\documentclass[twoside]{article}
\usepackage[accepted]{aistats2024}

\usepackage{lipsum}
\usepackage{amsfonts}
\usepackage{algorithm}
\usepackage{algpseudocode}
\usepackage{amsthm}
\usepackage{amsmath}
\usepackage{subcaption}
\usepackage{graphicx}
\usepackage{xcolor}
\usepackage{amsthm}
\usepackage{amsmath}
\usepackage{bbm}
\usepackage{amsthm}
\theoremstyle{plain}

\newtheorem*{theorem*}{Theorem}
\newtheorem*{prop*}{Proposition}

\usepackage{titlesec}
\usepackage{lipsum}

\usepackage{verbatim} 
 \usepackage{amsxtra}  
\newcounter{actr}
{\begin{list}{(\alph{actr})}{\usecounter{actr}}}{\end{list}}

\newcounter{ictr}
{\begin{list}{(\roman{ictr})}{\usecounter{ictr}}}{\end{list}}

\newtheorem{thm}{Theorem}

\newtheorem{prop}{Proposition}

\newenvironment{new-proof}[1]
{{\em Proof  #1: }}%
{ \noindent\qed }
\newcommand{\E}[1]{E\left[{#1}\right]}

\newcommand{\cE}{{\mathcal{E}}}

\newcommand{\cX}{{\mathcal{X}}}

\DeclareMathAlphabet{\mathbsf}{OT1}{cmss}{bx}{n}
\DeclareMathAlphabet{\mathssf}{OT1}{cmss}{m}{sl}

\DeclareSymbolFont{bsfletters}{OT1}{cmss}{bx}{n}  
\DeclareSymbolFont{ssfletters}{OT1}{cmss}{m}{n}
\DeclareMathSymbol{\bsfGamma}{0}{bsfletters}{'000}
\DeclareMathSymbol{\ssfGamma}{0}{ssfletters}{'000}
\DeclareMathSymbol{\bsfDelta}{0}{bsfletters}{'001}
\DeclareMathSymbol{\ssfDelta}{0}{ssfletters}{'001}
\DeclareMathSymbol{\bsfTheta}{0}{bsfletters}{'002}
\DeclareMathSymbol{\ssfTheta}{0}{ssfletters}{'002}
\DeclareMathSymbol{\bsfLambda}{0}{bsfletters}{'003}
\DeclareMathSymbol{\ssfLambda}{0}{ssfletters}{'003}
\DeclareMathSymbol{\bsfXi}{0}{bsfletters}{'004}
\DeclareMathSymbol{\ssfXi}{0}{ssfletters}{'004}
\DeclareMathSymbol{\bsfPi}{0}{bsfletters}{'005}
\DeclareMathSymbol{\ssfPi}{0}{ssfletters}{'005}
\DeclareMathSymbol{\bsfSigma}{0}{bsfletters}{'006}
\DeclareMathSymbol{\ssfSigma}{0}{ssfletters}{'006}
\DeclareMathSymbol{\bsfUpsilon}{0}{bsfletters}{'007}
\DeclareMathSymbol{\ssfUpsilon}{0}{ssfletters}{'007}
\DeclareMathSymbol{\bsfPhi}{0}{bsfletters}{'010}
\DeclareMathSymbol{\ssfPhi}{0}{ssfletters}{'010}
\DeclareMathSymbol{\bsfPsi}{0}{bsfletters}{'011}
\DeclareMathSymbol{\ssfPsi}{0}{ssfletters}{'011}
\DeclareMathSymbol{\bsfOmega}{0}{bsfletters}{'012}
\DeclareMathSymbol{\ssfOmega}{0}{ssfletters}{'012}

\usepackage{titlesec}

\usepackage{xr}
\usepackage[round]{natbib}

\newtheorem{rem}{Remark}

\usepackage{nicefrac} 

\usepackage{titlesec}
\usepackage{titling} 
\newcommand{\brN}{{\bar{N}}}
\newcommand{\bN}{\bar{N}}
\renewcommand{\E}{E}
\usepackage{lipsum}

\usepackage{paralist}
\usepackage{algorithm}
\usepackage{algpseudocode}
\usepackage{amsthm}
\usepackage{amsmath}
\usepackage{subcaption}
\usepackage{graphicx}
\usepackage{xcolor}
\usepackage{amsthm}
\usepackage{amsmath}
\usepackage{bbm}

\usepackage{lipsum}
\usepackage{amsfonts}
\usepackage{algorithm}
\usepackage{algpseudocode}
\usepackage{amsthm}
\usepackage{amsmath}
\usepackage{subcaption}
\usepackage{graphicx}
\usepackage{xcolor}
\usepackage{amsthm}
\usepackage{amsmath}
\usepackage{bbm}
\usepackage{amsthm}
\theoremstyle{plain}

\usepackage{amsfonts}       
\usepackage{amsthm}

\makeatletter

\usepackage{nicefrac} 

\usepackage{xr}
\usepackage[font=small]{caption}
\newcommand\blfootnote[1]{%
  \begingroup
  \renewcommand\thefootnote{}\footnote{#1}%
  \addtocounter{footnote}{-1}%
  \endgroup
}

\usepackage[round]{natbib}

\begin{document}


%
\runningtitle{Importance Matching Lemma for Lossy Compression with Side Information}

%

\twocolumn[

\aistatstitle{Importance Matching Lemma for Lossy Compression \\ with Side Information  }

\aistatsauthor{Buu Phan$^{*}$ \And Ashish Khisti$^{*}$ \And  Christos Louizos }

\aistatsaddress{ University of Toronto \And University of Toronto \\ Qualcomm AI Research$^{\dag}$   \And Qualcomm AI Research$^{\dag}$ } 
]
 
\begin{abstract}
  We propose two extensions to existing importance sampling based methods for lossy compression. First, we introduce an importance sampling based compression scheme that is a variant of  ordered random coding \citep{theis2022algorithms}  and is amenable to direct evaluation of the achievable compression rate for a  finite number of samples. Our second and major contribution is the \emph{importance matching lemma}, which is a finite proposal counterpart of the recently introduced {Poisson matching lemma} ~\citep{li2021unified}. By integrating  with deep learning, we provide a new coding scheme for distributed lossy compression with side information at the decoder.  We demonstrate the  effectiveness of the proposed scheme through experiments involving synthetic Gaussian sources, distributed image compression with MNIST and vertical federated learning with CIFAR-10.\looseness=-1

\end{abstract}

\section{INTRODUCTION}
  Lossy compression has become increasingly important in the field of machine learning, fueled by the expanding scale of data \citep{deng2009imagenet}, models \citep{ramesh2021zero,chatgpt}, and infrastructure\citep{tang2018communication}.  
 Furthermore, with the growing demand for decentralized and distributed learning, compression techniques that operate in multi-terminal settings must be considered.
We focus on a class of lossy compression techniques based on channel simulation, targeting one-shot or finite block length settings. The sender, upon observing $X\sim p_X(\cdot)$, communicates a noisy sample $Y\sim p_{Y|X}(.)$ to the decoder  at a rate of $R$ bits/sample. In practice, the  distribution $p_{Y|X}(.|x)$ is selected to satisfy a variety of constraints  e.g., fidelity constraint or distribution constraints. In a recent work, \cite{li2018strong}  propose a method based on the \emph{Poisson functional representation lemma} (PFRL) with  near optimal compression rate:
\begin{equation}
R\leq  I(X;Y) + \log (I(X;Y) + 1) + 5.
    \label{bound1}
\end{equation}
Here $I(X;Y)$, which denotes the mutual information between $X$ and $Y$ is well known to be a lower bound on the compression rate~(\cite{cuff2013distributed,bennett2002entanglement}). However PFRL, requires  an {\em infinite} number of samples to be generated  between the encoder and decoder. 
More practical approaches for channel simulation have been developed using importance sampling \citep{chatterjee2018sample} for a variety of applications e.g., neural compression \citep{flamich2020compressing,theis2022lossy}, federated learning \citep{isik2023communication, triastcyn2021dp}, differential privacy \citep{shah2022optimal}, and model compression \citep{havasi2019minimal}.  We will refer to these approaches as importance sampling based compression (ISC). In these methods, the output samples  follow a proxy distribution $\Tilde{p}_{Y|X}(\cdot)$ whose divergence w.r.t the target distribution $p_{Y|X}(\cdot)$  can be made arbitrarily small, provided that the number of samples is sufficiently large. Ordered random coding (ORC, \cite{theis2022algorithms}) is a recently proposed method in this family that also achieves near-optimal compression rate in ~\eqref{bound1}. The analysis of compression rate in ORC is  based on one-to-one comparison of the selected sample index with PFRL. 
\blfootnote{ $^{*}$ Equal Contribution.}
\blfootnote{ $^{\dag}$ Qualcomm AI Research is an initiative of Qualcomm Technologies, Inc. and/or its subsidiaries.}

\begin{figure*}[t]%
    \centering%
    \begin{subfigure}[t]{0.31\linewidth}%
        \includegraphics[width=\linewidth]{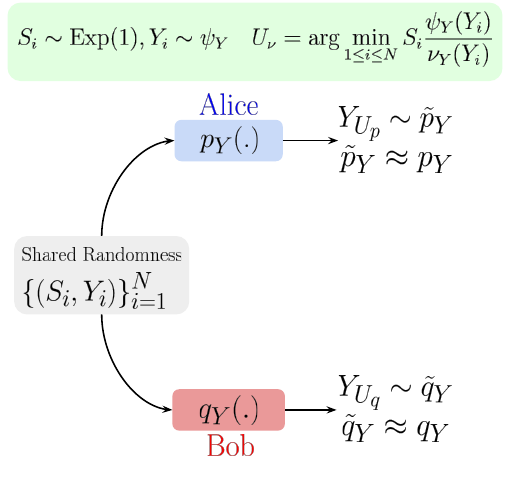}%
        \label{fig:minimal-example-pmp-ml:ml}%
    \end{subfigure}%
    \hspace{3mm}
    \begin{subfigure}[t]{0.25\linewidth}%
        \includegraphics[width=\linewidth,trim={0cm 0.5cm 1.cm 0cm}]{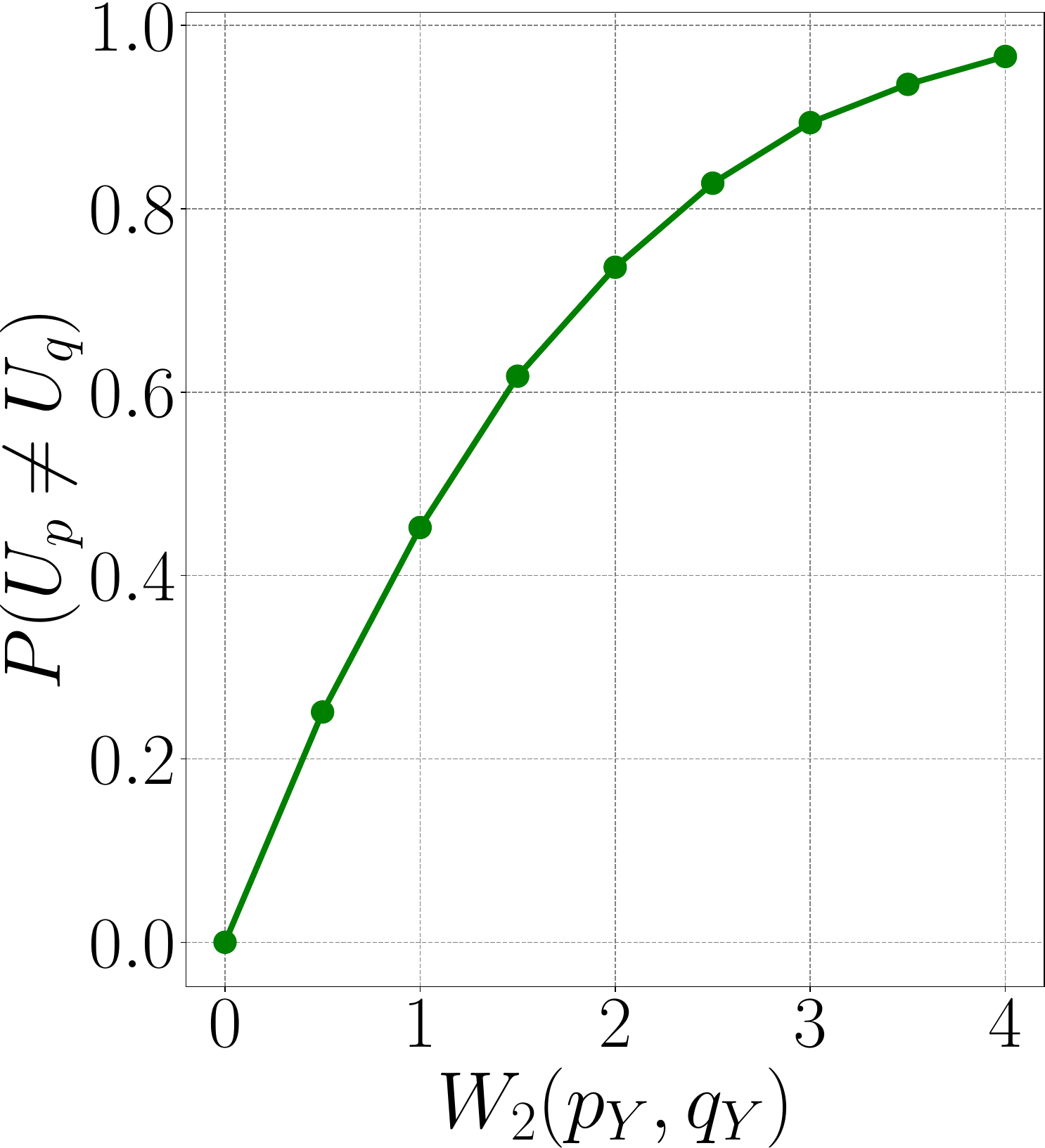}%
        \label{fig:minimal-example-pmp-ml:pmp}%
    \end{subfigure}%
    \hspace{6mm}
    \begin{subfigure}[t]{0.26\linewidth}%
        \includegraphics[width=\linewidth, trim={0.0cm 0.0cm -0.0cm 0.0cm}]{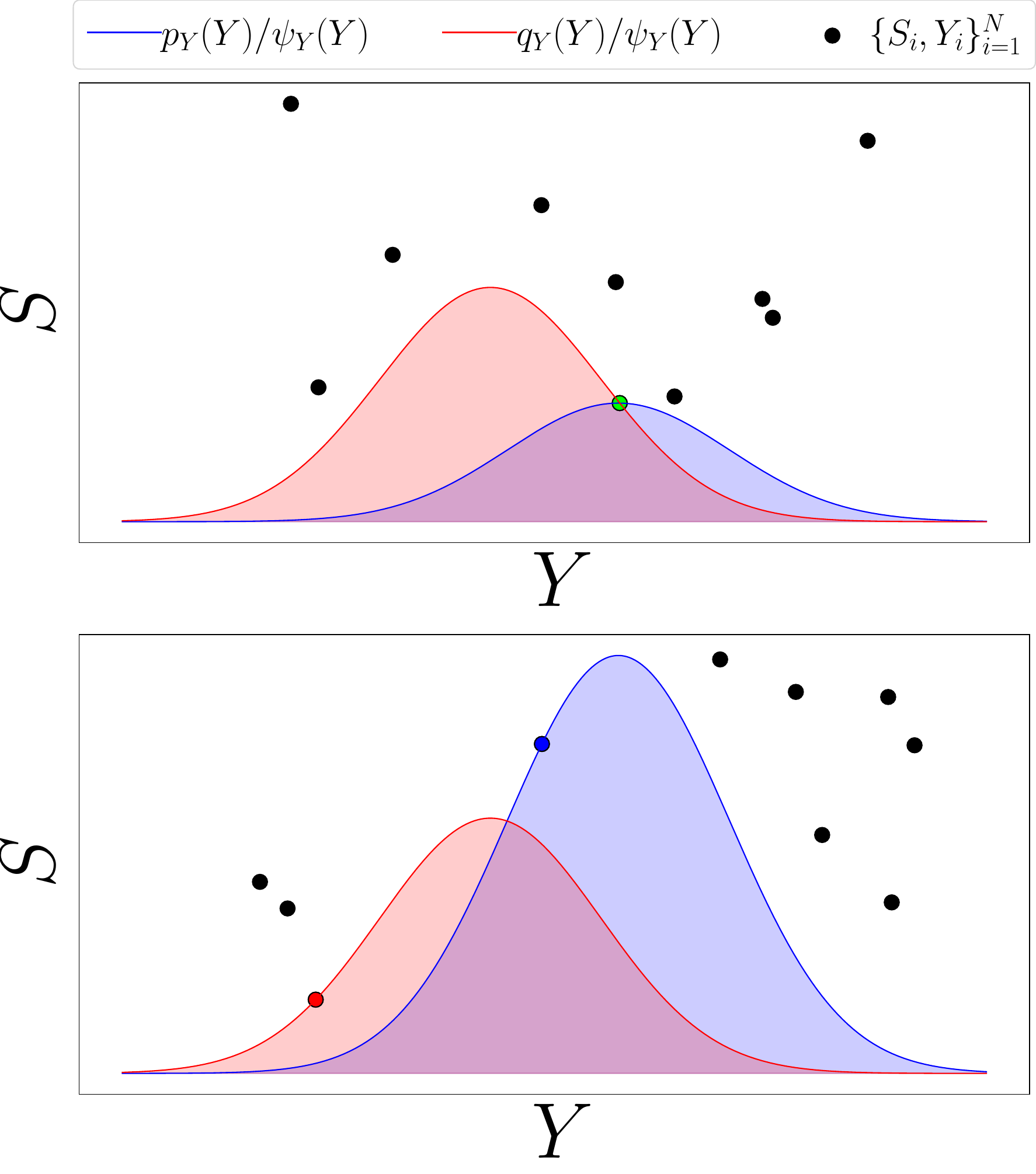}%
        \label{fig:minimal-example-pmp-ml:predictive-mixture-atypical}%
    \end{subfigure}%
    \caption{(Left) Overview of IML: Alice and Bob independently sample $Y_{U_p},Y_{U_q}$  by applying the Gumbel-max trick on the shared randomness. (Middle) The empirical mismatch probability with respect to the Wasserstein-2 distance $W_2(p_Y, q_Y)$, where $p_Y{=}\mathcal{N}(m,1)$, $q_Y{=}\mathcal{N}(-m,1)$ and $m \in [0,\infty)$. (Right) Mechanism of IML: each party scales their respective importance weights function $\frac{p_i(y)}{\psi_i(y)}$ and $\frac{q_i(y)}{\psi_i(y)}$ until one point $\{S_i,Y_i\}$ falls on the curve. Top and bottom figures show the matching and mismatching case respectively.} 
     \label{fig:importance_matching_lemma2}
\end{figure*}


To our knowledge ISC methods till date have not considered distributed source coding (DSC) which enables higher compression rates by exploiting the correlations between multiple sources \citep{el2011network}. We note that this is particularly relevant in many machine learning setups \citep{castiglia2022compressed, mital2022neural}. While the information-theoretic limits of DSC have been well studied in classical settings, practical implementations remain challenging.  First, the joint source distributions are often unknown and need to be learned using data. Second quantization based approaches for DSC  are challenging to implement in higher dimensions~\citep{zamir1998nested} and are mostly  studied in one-dimensional case~\citep{liu2006slepian,chen2010low, domanovitz2022data}. The extension of PFRL  to DSC settings has been  recently proposed in \citep{li2021unified} through the introduction of a new analysis technique called the Poisson matching lemma (PML).   It however requires an infinite number of samples to be generated and can be challenging to implement in practice.  In this work, as our main contribution, we introduce a new theoretical tool called the {\em importance matching lemma}, which enables us to extend ISC to DSC settings with provable guarantees. Our main contributions are:

\begin{compactitem}
    \item We  propose an ISC scheme, which we call {\em communication efficient importance sampling based compression} (CE-ISC). Our scheme is amenable to direct evaluation of the achievable compression rate. We also discuss a potential extension to the case of multiple importance sampling.

    \item We introduce a new analysis tool called the {\em importance matching lemma} (IML), which is the counterpart of  PML to importance sampling. This enables us to significantly expand the scope of ISC. We discuss in detail the application of ISC to DSC using IML.
    
    \item We conduct experimental studies on synthetic Gaussian sources, distributed compression involving the MNIST dataset, and a vertical federated learning setting with CIFAR-10 to demonstrate the effectiveness of our approach.  We propose a data-driven approach to implement the decoding rule, and make use of a feedback link from the decoder to  encoder to improve the rate-distortion performance. 
\end{compactitem}

A core technical challenge in our work involves analysis of self-normalized importance sampling, where  bounds on standard quantities (e.g., bias and variance) are considerably challenging to characterize~\citep{agapiou2017importance}. In our work, we are required to perform novel analysis of such methods for characterizing the compression rate in ISC and error probabilities associated with IML. 

\textbf{Related Work.}

\emph{Channel Simulation.} Our work falls in the class of ISC schemes, which include ORC \citep{theis2022algorithms} and minimum random coding (MRC) \citep{havasi2019minimal}.  Our proposed scheme is different from these techniques, is amenable to direct evaluation of the achievable rate and appears  compatible with multiple importance sampling~\citep{mis}. While ISC schemes are approximate sampling techniques with an upper bound on the total number of proposal samples, other methods such as
PFRL based sampling~\citep{li2018strong}, A* sampling~\citep{maddison2014sampling, flamich2022fast} and rejection sampling~\citep{harsha2007communication, flamich2023adaptive} are exact sampling techniques and may require arbitrarily large number of proposal samples in the worst case.

\emph{Distributed Source Coding (DSC).} To our knowledge the only channel simulation technique that extends to DSC is from \cite{li2021unified} discussed previously. Deep learning based DSC has been considered in some recent works (\cite{mital2022neural,whang2021neural,ozyilkan2023learned}), which provide empirical evidence that neural networks could learn the binning. In contrast, our approach is motivated by theoretical analysis of IML and integrates deep learning to incorporate complex joint distributions.  Traditional information theoretic approaches for  one-shot DSC \citep{verdu2012non,liu2015one,song2016likelihood}  do not appear amenable to practical implementations. Finally quantization based  approaches for DSC have several limitations as are already discussed before.

\section{COMMUNICATION-EFFICIENT IMPORTANCE SAMPLING}
We introduce our communication efficient ISC scheme, provide analysis of the achievable compression rate and discuss an extension to multiple importance sampling.

\subsection{Problem Setup}\label{setup}
We begin by introducing the setup of approximate channel simulation under communication constraints. Let $(X,Y) \in \mathcal{X}\times\mathcal{Y}$ be a pair of random variables distributed according to $p_{X,Y}$, with marginal distributions $p_X(\cdot)$ and $p_Y(\cdot)$, respectively. The encoder observes a realization $x$ of $X\sim p_X$. Additionally, there is a shared source of randomness denoted by $W \in \mathcal{W}$, which is available to both the encoder and the decoder.   We define the mappings $f$ and $g$ for the encoder and decoder respectively: 
\begin{align}
f &\colon \mathcal{X}\times \mathcal{W}\to \mathcal{M},\\
g &\colon \mathcal{M} \times \mathcal{W}\to \mathcal{Y}, \label{decoding-function}
\end{align}
where $\mathcal{M} \in \{0,1\}^{\star}$ denotes the (variable length) message that the encoder sends to the decoder and  $\ell(M)$ denotes the length of message $M$. A rate-divergence tuple $(R,\Tilde{\epsilon})$ is defined to be one-shot achievable if there exists $f,g$ satisfying:
\begin{align}
E[\ell(M)] \le R, \label{eq:l1}\\
D_\mathrm{TV}(\tilde{p}_{Y|X}(.|x), {p}_{Y|X}(.|x)) {\le} \Tilde{\epsilon}, \forall x \in {\mathcal X} \label{eq:l2}
\end{align}
where $\tilde{p}_{Y|X}(.|x)$ denotes the proxy distribution realized by our choice of $f$ and $g$, while $p_{Y|X}(.|x)$ is the desired target distribution.

\subsection{Coding Scheme}\label{coding_scheme}

The encoder and decoder generate  shared randomness $W=\{S_i,Y_i\}_{i=1}^N$ where $Y_i \sim p_Y(\cdot)$ and $S_i \sim \text{Exp}(1)$ are sampled in an i.i.d.\ fashion. The encoder, upon observing $X=x$ must select an index $i \in \{1,...,N\}$  
and communicate it such that both~\eqref{eq:l1} and~\eqref{eq:l2} are satisfied.
Specifically,  the  encoding function is described in the following steps:
\begin{compactenum}
\item\emph{Index Selection. } Upon observing $X{=}x$, the encoder selects index $U$ by:
\begin{equation}
    U = \arg \min_{1\le i \le N}S_{i} \frac{p_Y(Y_i)}{p_{Y|X}(Y_i|x)} \label{index_sel}
\end{equation}
which is also known as Gumbel-max trick \citep{maddison2014sampling}.
\item \emph{Index Reordering.} Instead of directly sending $U$ to the decoder, which will cost $\log(N)$ bits, the encoder will send its corresponding position $K$ in the sorted list: $S_{\pi(1)} {\le} S_{\pi(2)} {\le} \ldots {\le} S_{\pi(N)}$, such that $\pi(K) {=} U$. 
\item \emph{Entropy Coding.} Finally, the encoder convert $K$ into a bit string $M$ by entropy coding with a Zipf distribution \citep[Section 2]{li2018strong}.
\end{compactenum}

The decoding step $g$ is straightforward. As the sorted list is both known to the encoder and decoder,  the decoder can recover $K$ (and by so, $U$) losslessly from $M$, and outputs $Y_U$. 

Note that our index selection in step 1 is the same as in  importance sampling~\citep{havasi2019minimal}. In particular, given a common sequence of samples $Y_1,\ldots, Y_N$, our approach selects the sample $Y_i$ with probability:
\begin{align}\lambda_i = \left\{\frac{p_{Y|X}(Y_i|x)}{p_Y(Y_i)} \right\} / \left\{\sum_{i=1}^N \frac{p_{Y|X}(Y_i|x)}{p_Y(Y_i)}\right\}.\label{eq:lam-i} \end{align}
Conditioned on the common sequence $Y_1^N$, our setting reduces to the exponential functional representation lemma for discrete alphabets in~\citep[Def. 4.1]{li2017information}. Following their approach, we transmit the position of the selected index in the sorted list of $\{S_i\}$ in steps 2 \& 3.   Our construction is different from ORC~\citep[Sec. 3.4]{theis2022algorithms} that reorders (sorts) the exponential random variables before  the index selection operation in~\eqref{index_sel}. Interestingly  both schemes lead to the same output distribution when the samples are generated in an i.i.d.\ fashion.

Note that the proxy distribution generated by our scheme is:
\begin{align}
\tilde{p}_{Y|X}(y|x) = E_{Y_1,\ldots, Y_N} \left[\sum_{i=1}^N \lambda_i \cdot \delta(y-Y_i)\right]
\label{eq:psim}
\end{align}
where $\lambda_i$ are defined in~\eqref{eq:lam-i}. Following prior works~\citep{havasi2019minimal, theis2022algorithms,chatterjee2018sample} the output distribution can be  close to the target distribution $p_{Y|X}(\cdot)$ for sufficiently many samples. In particular, under the standard assumption:
\vspace{-0.5em}
\begin{align} 
\frac{p_{Y|X}(y|x)}{p_Y(y)} \le \omega,  \forall x, y, \label{eq:om-def}
\end{align} we can construct a $N_0(\epsilon)$ such that for $N \ge N_0(\epsilon)$, we have that
$D_\mathrm{TV}(\tilde{p}_{Y|X}(.|x), {p}_{Y|X}(.|x)) {\le} \epsilon$. A characterization of $N_0(\epsilon)$ is provided in the Section~\ref{sec:output} of the supplementary material for sake of completeness.
We next provide analysis of the achievable rate.

\begin{thm} \label{thm:rate} Given $(X,Y) \sim p_{X,Y}$,  and $N, K$   as in the scheme in Sec. \ref{coding_scheme},   we have that:
\begin{align}
E[\log K | X=x] \le E_{Y_1^N}\left[D({\bf \lambda}|| {\bf u})\right] + \delta \label{eq:bnd1}
\end{align}
where ${\bf \lambda} = (\lambda_1, \ldots, \lambda_N)$ is defined via~\eqref{eq:lam-i}, \textcolor{black}{$D({.}|| .)$ is the KL divergence}, ${\bf u} = \left(1/N,\ldots, 1/N\right)$ is associated with the uniform distribution and  $\delta = 1 + \log e/e$ is a constant. Furthermore,
    \begin{align}
H[K] {\le} I(X;Y) {+} \frac{\Delta}{N} {+} \log\left( I(X;Y) {+} \frac{\Delta}{N} {+}1\right){+} 4\label{eq:bnd2}
\end{align} 
where $\Delta:=\Delta(p_{X,Y})$ is a constant defined in the supplementary material via~\eqref{eq:Del-def} and~\eqref{eq:alpha2} that does not depend on $N$.
$\hfill\qed$
\end{thm}

We provide the proof of Theorem~\ref{thm:rate} in the Section~\ref{sec:thm-rate} in the supplementary material. While the upper bound in~\eqref{eq:bnd1} follows through connection with~\citep[Chapter 4]{li2017information}, the derivation of~\eqref{eq:bnd2} is complicated by the normalizing term in the denominator in~\eqref{eq:lam-i}. Theorem~\ref{thm:rate} demonstrates that the proposed scheme also achieves a near optimal compression rate and an additive penalty of at most $\Theta(1/N)$. We also provide an alternative bound to~\eqref{eq:bnd2} in the supplementary material in section~\ref{sec:alt-thm1}, which is simpler to derive but  involves a multiplicative penalty term.


\subsection{Beyond i.i.d.\ samples}\label{sec:mis}
Our discussion so far has assumed that the proposal samples $\{Y_i\}$ are generated in an i.i.d.\ fashion from a distribution $p_Y(\cdot)$. However in variance reduction schemes, such as multiple importance sampling~\citep{mis}, it is required that different samples be generated from different distributions.  As a simple example, suppose that  $Y_1, \ldots, Y_{\brN}$  are sampled i.i.d.\ from $p^{(1)}_Y(\cdot)$ and
$Y_{\brN+1}, \ldots, Y_{N}$ are sampled i.i.d.\ from $p^{(2)}_Y(\cdot)$ where $\brN=N/2$
and  $p^{(1)}_Y(\cdot)$ and $p^{(2)}_Y(\cdot)$ are selected to satisfy $p_Y(y)= \frac{1}{2}p^{(1)}_Y(y) + \frac{1}{2}p^{(2)}_Y(y)$. Given $X=x$, the probability that the output index $K=i$ in Multiple Importance Sampling (MIS) is proportional to $\lambda_i$ (see scheme N3 in ~\cite{mis}) in~\eqref{eq:lam-i}. As a result our proposed coding scheme in~Section~\ref{coding_scheme} can be immediately used as stated. In the rate analysis, note that the  upper bound in~\eqref{eq:bnd1} in Theorem~\ref{thm:rate} also applies with the difference that $Y_1^N$ are not i.i.d.\ but distributed according to either $p^{(1)}_Y(\cdot)$ or $p^{(2)}_Y(\cdot)$. We discuss further analysis of this specific case in Section~\ref{sec:MIS} in the supplementary material. We show that under a simplifying assumption that the denominator in~\eqref{eq:lam-i} equals its expectation, the distribution of the output samples equals the target distribution and the associated compression rate matches ORC. We also study a numerical example involving a Gaussian mixture model and demonstrate that MIS with our proposed compression scheme can achieve significantly lower bias and variance than the ORC scheme for a given number of samples. 

\section{IMPORTANCE MATCHING LEMMAS}

\subsection{Importance Matching Lemma}
\label{sec:IML}
The Poisson Matching Lemma (PML)~\citep{li2021unified}  enables the application of Poisson Functional Representation lemma (PFRL) to  a broad class of problems in multi-terminal source and channel coding with provable guarantees. In this section, we introduce the Importance Matching Lemma (IML), which enables application of importance sampling to such settings. We demonstrate the application of IML to a specific  setting of source coding with side information in the next section.  

Let $Y_1, \ldots, Y_N$ be sampled i.i.d.\ from distribution $p_Y(\cdot)$ and let $p_{Y|X}(\cdot|X=x)$ and $q_{Y|X}(\cdot|X=x)$ be two conditional distributions.  We note that such $p_Y(\cdot)$ can be replaced by any distribution over ${\mathcal Y}$ such that (\ref{eq:om-def}) is satisfied as our proof does not require any relation between $p_Y(\cdot)$ and $p_{Y|X}(\cdot)$ to hold. We generate two indices $U_p$ and $U_q$ as follows:

\begin{align}
U_p = \arg\min_{1\le i \le N}\frac{S_i}{\lambda_i^p}, \qquad  U_q = \arg\min_{1\le i \le N} \frac{S_i}{\lambda^q_i} \label{eq:U-def}
\end{align}
Where $S_1, \ldots, S_N$ are sampled i.i.d.\ $\mathrm{Exp}(1)$ and $\lambda_i^p$ and $\lambda_i^q$ are the importance weight counterparts of~\eqref{eq:lam-i}:
\begin{align}
\lambda^p_i {=} \frac{\frac{p_{Y|X}(Y_i|X=x)}{p_Y(Y_i)}}{\sum_{j=1}^N\frac{p_{Y|X}(Y_j|X=x)}{p_Y(Y_j)}},  
\lambda^q_i {=} \frac{\frac{q_{Y|X}(Y_i|X=x)}{p_Y(Y_i)}}{\sum_{j=1}^N\frac{q_{Y|X}(Y_j|X=x)}{p_Y(Y_j)}}. \label{eq:lam-i2}
\end{align}
The indices $U_p$ and $U_q$  selected via importance sampling have the same proposal distribution $p_Y(\cdot)$ but  different target distributions $p_{Y|X}$ and $q_{Y|X}$ respectively.  We bound the error event that $\{U_p \neq U_q\}$.

\begin{prop}
\label{prop:condPML}
Letting  $\Omega= \{y_1, \ldots, y_N\}$ denote the sequence of samples: 
\begin{align}
&\Pr(U_p \neq U_q | \Omega, X=x, U_p = k) \le \notag\\
&1 {-} \left(1 {+} \frac{{p_{Y|X}(y_k|x)}}{{q_{Y|X}(y_k|x)}} \frac{\left(\frac{1}{N}\sum_{j=1}^N\frac{q_{Y|X}(y_j|x)}{p_Y(y_j)}\right)}{\left(\frac{1}{N}\sum_{j=1}^N\frac{p_{Y|X}(y_j|x)}{p_Y(y_j)}\right)}\right)^{-1}. 
\label{eq:condPML}
\end{align}
\end{prop}
The proof of Prop.~\ref{prop:condPML} is Section~\ref{app:condPML} in the supplementary material. It exploits the fact that conditioned on the samples $Y_1^N$, the operation in~\eqref{eq:U-def} can be viewed as sampling over a discrete alphabet with probabilities given by $\lambda^p_i$ and $\lambda^q_i$ respectively and arguments as in~\citep{li2021unified} are applicable. Our main result in this section is the following where the conditioning on all $Y_i$, except $Y_k$ is removed.

\begin{thm}
\label{thm:PML2-1}
Define $\bN = N-1$, we have:
\begin{equation}\begin{aligned}
\Pr(U_p \neq U_q | Y_k=y_k, U_p = k, X=x) \le \\
1- \left(1 + \frac{{p_{Y|X}(y_k|x)}}{q_{Y|X}(y_k|x)}\mu_{y_k}(\bN)\right)^{-1} 
\label{eq:IML-1}
\end{aligned}\end{equation}
and $\mu_{y_k}(\bN)$ is defined via~\eqref{eq:nu-def-1}-\eqref{eq:l1-bnd2} in Section~\ref{app:PML2-1} in the supplementary material. Note that $\mu_{y_k}(\bN)$ scales as $\Theta(1)$ as $\bN {\rightarrow} \infty$ under some mild assumptions on the distributions (see Remark~\ref{rem:largeN} in Supplementary Material).  $\hfill\qed$
\end{thm}
The proof of Theorem~\ref{thm:PML2-1} is in Section~\ref{app:PML2-1} in the supplementary material. The main challenge is associated with the normalizing terms in~\eqref{eq:lam-i2}. We also provide an alternate bound in Section~\ref{app:alt-thm2} in the supplementary material that has a shorter derivation, but requires stronger assumptions on the distribution. 

Intuitively, IML bounds the mismatch between the sampled indices when different target distributions are used in importance sampling. This is further illustrated in Fig.~\ref{fig:importance_matching_lemma2}. As is the case with PML, it turns out that in practice we need a conditional version of IML. 



\subsection{Conditional Importance Matching Lemma.}\label{sec_cond_lemma}

Suppose that $(S_i, Y_i)_{i=1}^N$ be sampled i.i.d.\ as in Section~\ref{sec:IML}. Let $(X,Y, Z)$ be a triplet of random variables with a joint distribution $p_{X,Y, Z}(\cdot)$. Let $Q_{Y|Z}(\cdot)$ be an arbitrary conditional distribution satisfying $\frac{Q_{Y|Z}(y|z)}{P_Y(y)} {\le} \omega$ for all $(y,z)$. Given $X{=}x$ sampled independently of $(S_i, Y_i)_{i=1}^N$, suppose that we sample $Y=Y_{U_P}$ using importance sampling i.e., we select 
\begin{align}
U_P {=} \arg\min_{1\le i\le N} \frac{S_i}{\lambda^P_i}, \quad \lambda_i^P {=} \frac{\frac{p_{Y|X}(Y_i|X=x)}{p_Y(Y_i)}}{\sum_{j=1}^N \frac{p_{Y|X}(Y_j|X=x)}{p_Y(Y_j)}}.\label{eq:Up}
\end{align}
Next, given $X=x$ and $Y=y$ we generate a sample $Z \sim p_{Z|X,Y}(\cdot|X=x, Y=y)$ and note that
\begin{align}
Z \rightarrow (X,Y) \rightarrow (S_i, Y_i)_{i=1}^N
\label{eq:markov}
\end{align}
is satisfied by construction.  Given  a realization $Z=z$ we sample 
\begin{align}
U_Q = \arg\min_{1\le i\le N} \frac{S_i}{\lambda^Q_i}, \hspace{0.25em} \lambda_i^Q = \frac{\frac{Q_{Y|Z}(Y_i|Z=z)}{p_Y(Y_i)}}{\sum_{j=1}^N \frac{Q_{Y|Z}(Y_j|Z=z)}{p_Y(Y_j)}} \label{eq:Uq}. 
\end{align}

\begin{thm} The error probability satisfies:
\begin{equation}\begin{aligned}&\Pr\left(U_P \neq U_Q | U_P = k, X=x, Z=z, \Omega\right) \le \\ &1 {-} \left(1 {+} \frac{{p_{Y|X}(y_k|x)}}{{Q_{Y|Z}(y_k|z)}} \frac{\left(\frac{1}{N}\sum_{j=1}^N\frac{Q_{Y|Z}(y_j|z)}{p_Y(y_j)}\right)}{\left(\frac{1}{N}\sum_{j=1}^N\frac{p_{Y|X}(y_j|x)}{p_Y(y_j)}\right)}\right)^{-1},\label{eq:condbnd1_main} \end{aligned} \end{equation}
where  $\Omega=\{y_1,\ldots, y_N\}$,  and furthermore,
\begin{align}
&\Pr(U_p \neq U_q | Y_k=y_k, U_p = k, X=x, Z=z)  \le \notag \\ &1- \left(1 + \mu_{y_k}(\bN) \frac{{p_{Y|X}(y_k|x)}}{Q_{Y|Z}(y_k|z)}\right)^{-1}.
\label{eq:condbnd2_main}\end{align}
where $\mu_{y_k}(\bN)$ defined via~\eqref{eq:mu-def2-n}-\eqref{eq:l1-bnd3} in the supplementary material, scales as $\Theta(1)$ when $\bN\rightarrow\infty$ under some mild assumptions (c.f.~Remark~\ref{rem:2}).$\hfill\qed$
\label{thm:CML}
\end{thm}
The proof of Thm.~\ref{thm:CML} is in Section~\ref{app:CML} in the supplementary material.  As discussed in Remark~\ref{rem:2} there, under mild conditions on the distributions, we can exhibit an $N_1(\epsilon)$ such that
\begin{align}
\mu_y(N) \le 1 + \epsilon, \quad \forall N \ge N_1(\epsilon) \label{eq:N1-def}
\end{align}

In the special case $Z=X$,~\eqref{eq:condbnd1_main} reduces to~\eqref{eq:condPML} in Prop.~\ref{prop:condPML}. Likewise~\eqref{eq:condbnd2_main} reduces to~\eqref{eq:IML-1} in Theorem~\ref{thm:PML2-1}. The value of Theorem~\ref{thm:CML} is that it extends IML to any $Z$ that satisfies~\eqref{eq:markov}. \textcolor{black}{In other words,  the conditional version of IML is an extension of Theorem \ref{thm:PML2-1} where the decoder is revealed an observation $Z$ regarding the sample $Y{=}Y_{U_P}$ selected by the encoder, and improves the decoding rule (cf. (\ref{eq:Uq})) by making use of the posterior distribution $Q_{Y|Z}(\cdot)$.} We demonstrate an application of this result in the next section. 

\section{ LOSSY COMPRESSION WITH SIDE-INFORMATION}
\label{sec:SI}
\subsection{Problem Setup}
\label{ref:p-setup}
\begin{figure*}[t]%
    \centering%
    \begin{subfigure}[t]{0.53\linewidth}%
        \includegraphics[width=\linewidth]{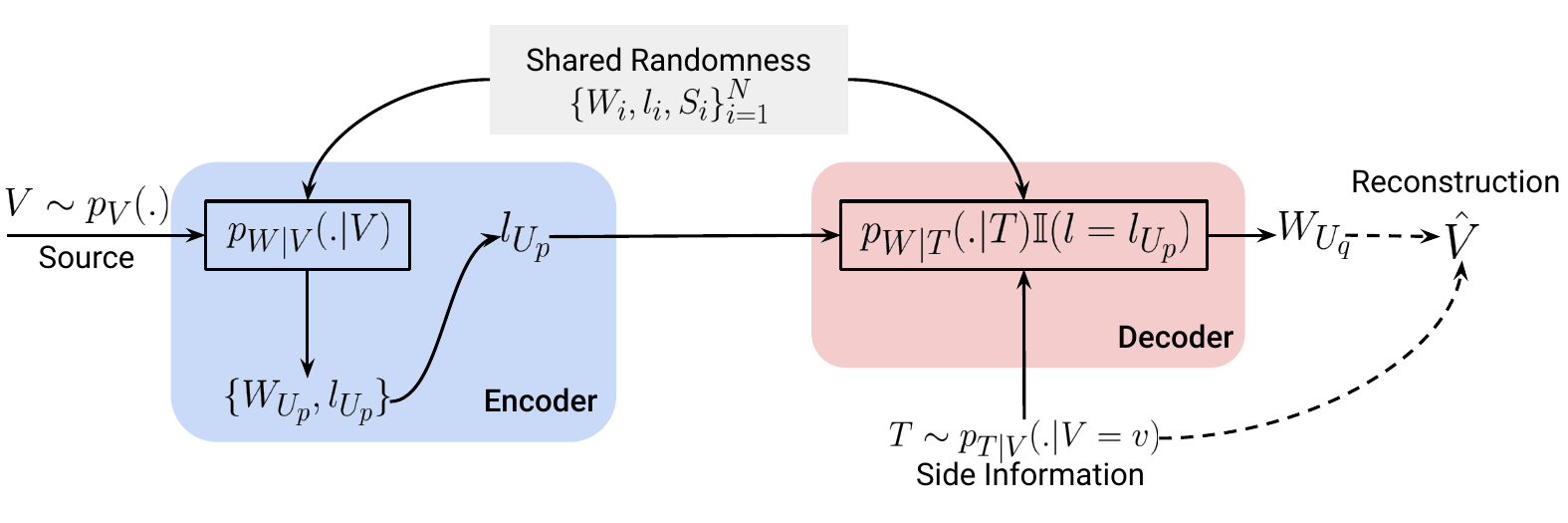}%
        \label{fig:minimal-example-pmp-ml:ml}%
    \end{subfigure}%
    \hspace{2mm}
    \begin{subfigure}[t]{0.44\linewidth}%
        \includegraphics[width=\linewidth]{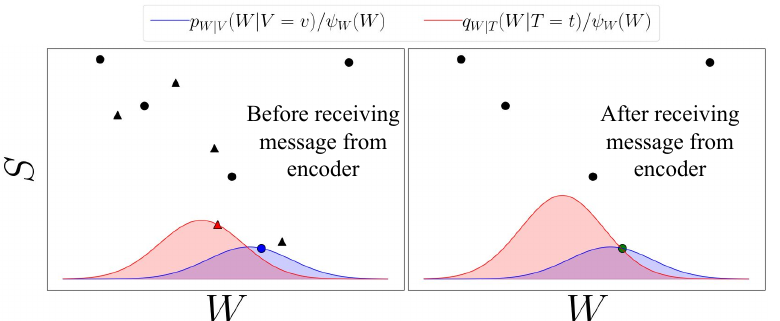}%
        \label{fig:minimal-example-pmp-ml:pmp}%
    \end{subfigure}%
    \caption{\textcolor{black}{(Left) Source coding with side information at the decoder with conditional IML. (Right) Decoding mechanism: the encoder scales $\frac{p_{W|V}(w|v)}{\psi_W(w)}$ and selects $W_{U_p}$(blue circle). Left sub-figure:  the decoder selects incorrect indices by purely scaling $\frac{p_{W|T}(w|t)}{\psi_W(w)}$ (equivalently, rate $R{=}0$). Right sub-figure: we generate extra one-bit information $l_i$ for each  codeword index by randomly marking it either a triangle ($l_i{=}0$) or a circle ($l_i{=}1$).   Upon receiving $l_{U_P}{=}1$ from the encoder,  the decoder  eliminates all indices marked by triangle  and correctly decode the index among the circles.}}
\label{fig:SI}
\end{figure*}

Just as the Poisson Matching Lemma (PML) has broad applications in multi-terminal source and channel coding settings, IML developed in Section~\ref{sec:IML}  can have analogous applications using importance sampling. We demonstrate the application of IML to the classical problem of lossy compression with side information at the decoder~\citep{wyner1976rate}. As illustrated in Figure \ref{fig:SI}(left), a source sample $V \sim p_V(\cdot)$ observed at the encoder, must be lossily compressed into a bit sequence of average rate $R$ bits/sample and sent to the decoder. Besides the information from the encoder, the decoder also has access to the side information $T\sim p_{T|V}(.|V=v)$, both of which are employed to output $W$ that is required to be approximately sampled from a conditional distribution $p_{W|V}(\cdot|v)$. In practice
this conditional distribution can be selected to satisfy
an average distortion constraint  $E[d(V,\hat{V})]$, where $\hat{V}$ is the final reconstruction expressed as a function of $W$ and $T$, i.e. $\hat{V}=\Tilde{g}(W,T)$.


We present a scheme based on importance sampling and present the error analysis by making use of IML in the previous section. 
\subsection{Coding Scheme}
Let $p_W(\cdot)$ be the marginal distribution of $W$ and $p_{W|T}(w|t) = \sum_{v} p_{W|V}(w|v)p_{V|T}(v|t)$ be the conditional of $W$ given $T$. Following the construction in Section~\ref{coding_scheme}, we sample  $W_1, \ldots, W_N$ i.i.d.\ from the distribution $p_W(\cdot)$. In addition, let $L>0$ be an integer and let $p_{\mathsf{l}}(\cdot)$ be uniform over the set of integers $\{1,2,\ldots, L\}$. We generate $l_1,\ldots,l_N$ i.i.d.\ from $p_\mathsf{l}(\cdot)$. Let us define $Y = (W,l)$ with $p_Y(y)= p_W(w)p_{\mathsf{l}}(l)$ and note that $Y_i = (W_i, l_i)$ is sampled i.i.d.\ from $p_Y(\cdot)$. Further let $X=V$ and  $p_{Y|X}(w,l|v) = p_{W|V}(w|v)p_{\mathsf{l}}(l)$ be the target distribution used at the encoder with the knowledge of $v$.  Finally we let $S_1,\ldots, S_N$ be a sequence of i.i.d.\ exponential random variables ${\mathrm{Exp}}(1)$ known to both the encoder and the decoder. Following~\eqref{eq:U-def}, the encoder selects an index $U_p$ given by:
\begin{align}
U_p {=} \arg\min_{1\le i \le N}\frac{S_i }{\frac{p_{Y|X}(Y_i|v)}{p_Y(Y_i)}}{=}
\arg\min_{1\le i \le N}\frac{S_i }{\frac{p_{W|V}(W_i|v)}{p_W(W_i)}}. \label{eq:Up-sel}
\end{align}
The encoder, in turn, transmits $l_{U_p} \in\{1,2,\ldots, L\}$ to the decoder using $\log L$ bits.  In defining the decoding rule, we let $Z = (T, l_{U_p})$. and note that $Z \rightarrow (X,Y
_{U_p})  \rightarrow (S_i, Y_i)_{i=1}^N$ is satisfied. Let \begin{align}Q_{Y|Z}(y|z) &= Q_{(W,l)|(T, l_{U_p})}(w,l|t,l_{U_p}) \\&= p_{W|T}(w|t){\mathbb {I}}({l=l_{U_p})} \label{eq:q-def}\end{align} be the distribution used at the decoder. Following~\eqref{eq:Up}, the decoder outputs an index $U_q$ given by:
\begin{align}
&U_q =  \arg\min_{1\le i \le N}\frac{S_i }{\frac{Q_{Y|Z}(Y_i|t,l_{U_p})}{p_Y(Y_i)}}\notag\\
&=\arg\min_{1\le i \le N}\frac{S_i }{\frac{p_{W|T}(W_i|t) {\mathbb I}(l_i= l_{U_p})}{p_W(W_i) p_{\mathsf l}(l_i)}} \label{eq:Uq-sel-wz}
\end{align} The decoder finally outputs $\hat{W}{=}W_{U_q}$ as the sample. 

Since the encoder selects the sample $W_{U_p}$ using importance sampling~\eqref{eq:Up-sel}, it follows from the discussion in Section~\ref{coding_scheme} that for sufficiently large $N$ the distribution $p_{W_{U_p}|V}(\cdot)$ can be  arbitrarily close to the target distribution $p_{W|V}(\cdot)$. The error probability can be bounded as below:
\begin{prop}
\label{prop:SI}For sufficiently large $N$,
\begin{align}
&\Pr(U_p \neq U_q) \le \notag\\  &E_{V,W,T}\left[1- \left(1+ (1+\epsilon)L^{-1} \textcolor{black}{2^{i(W;V|T)}} \right)^{-1} \right]
\label{eq:err-bnd-main}
\end{align}
where \textcolor{black}{$i_{W,V|T}(w;v|t) = \log \frac{p_{W|V}(w|v)}{p_{W|T}(w|t)}$} is the conditional information  density and recall $T \rightarrow V \rightarrow W$. $\hfill\qed$
\end{prop}
The proof of Prop.~\ref{prop:SI} is in Section~\ref{app:SI} in the supplementary material. 
Note that~\eqref{eq:err-bnd-main} provides a tradeoff between the compression rate $R {=} \log L$ and the error probability. The larger the value of $R$, the smaller will be the error probability. \textcolor{black}{ We illustrate how this decoding mechanism reduces the error in Figure \ref{fig:SI} (right) for the case  $R{=}0$ and $R{=}1$}. Similar to results in \cite{li2021unified}, we can extend this result to  bound the excess distortion probability $P_e{=}\Pr({d(V,\hat{V}) {>} D})$ where $\hat{V}{=}\Tilde{g}(W_{U_q}, T)$ is the reconstruction output by the decoder and $d(.,.)$ is the distortion  (Section ~\ref{app:excess_err} in the Supplementary).



\begin{rem}
\label{rem:N}
Note that the error probability directly depends on the compression rate $R=\log L$ and the conditional information density \textcolor{black}{$i(W;V|T)$}. Provided that $N$ is sufficiently large (which is necessary to track the target distribution in ISC)  it only exhibits a second-order dependence on $N$ via the $\epsilon$ factor.
\end{rem}
\begin{rem}
\label{rem:multiple}
In some of our experiments, we will consider compressing $k$ i.i.d.\ samples together. The error probability can be recovered by setting $L = 2^{k \cdot R}$ and \textcolor{black}{$i(W^k;V^k|T^k) = \sum_{i=1}^k i(W_i;V_i|T_i)$}. \\
\end{rem}

\subsection{Decision Feedback Based Scheme} 
\label{prac}
\begin{figure}[t]%
    \centering%
        \includegraphics[width=0.75\linewidth, trim={5.5cm 0.cm 4.5cm -0.5cm}]{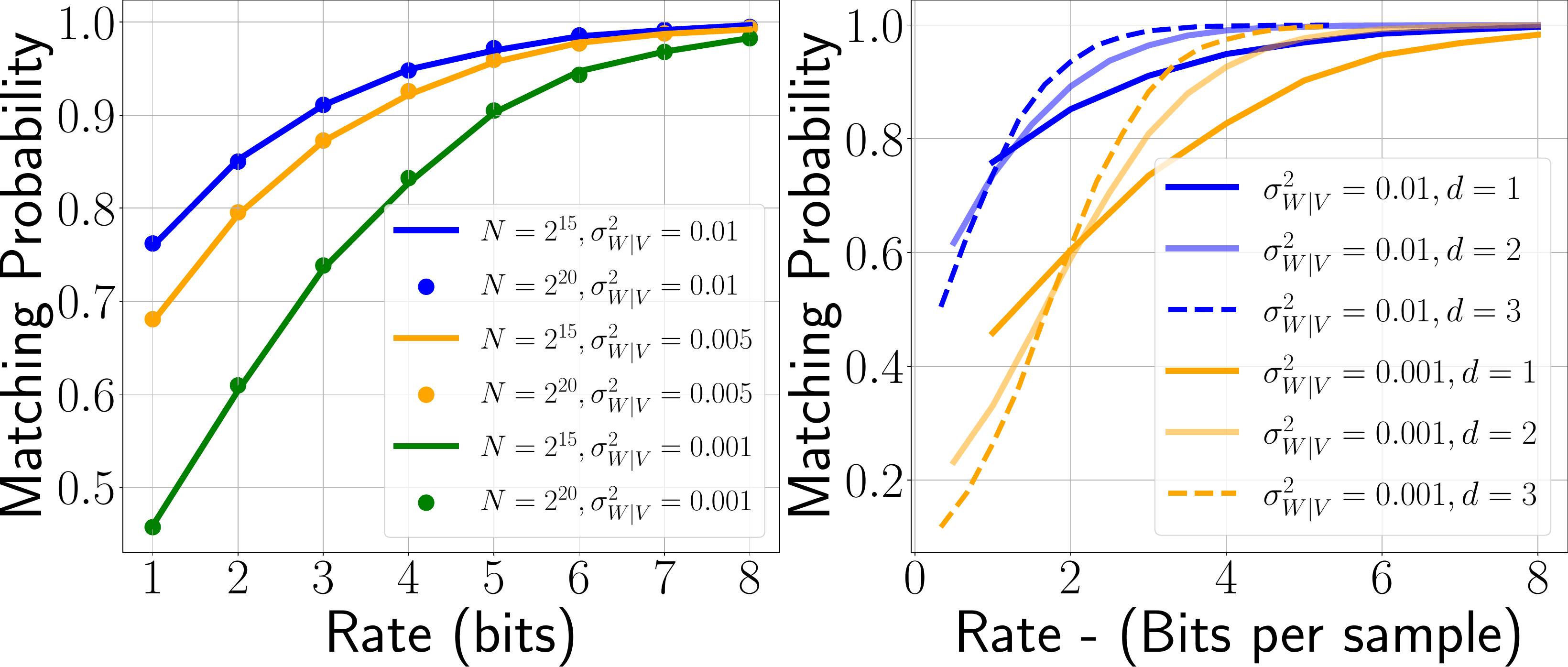}%
        \vspace{-1pt}
        \caption{Left: Empirical matching probabilities with different target distribution and number of proposals. Right: effects of compressing multiple samples jointly on the matching probability (Best view in screen).}%
     \vspace{-10pt}
     \label{fig:gaussian}
\end{figure}

In practice, a decoding error from ~\eqref{eq:err-bnd-main} can result in high average reconstruction distortion. This motivates us to use feedback communication to correct the errors and improve the rate-distortion performance as follows:
\begin{compactenum}
\item \emph{Index Selection. } The encoder communicates $\log_2(L)$ least significant bits (LSB) of the selected index $U_p$,  see (\ref{eq:Up-sel}), to the decoder.
\item \emph{Decoding and Feedback.}  The decoder outputs $U_q$ using (\ref{eq:Uq-sel-wz}) and  send $\log_2(N/L)$ most significant bits (MSB) to the encoder. 
\item \emph{Re-tranmission. } From the feedback MSB, if the index is correct, the encoder responds with an acknowledgment bit. Otherwise, it sends the MSB of its selection to the decoder.
\end{compactenum}
We verified in our experiments that the use of LSB instead of random bits in step $1$ did not have any noticeable difference.  While the feedback rate of $\log_2(N/L)$ bits guarantees that the encoder can perfectly locate the decoded index, we experimentally observe that we can reduce it by sending a hashed value of MSB, while tolerating a slight increase in distortion. \textcolor{black}{Note that this small hash of the decoded index requires significantly fewer bits than full side information (e.g., by $> 500$ times in Sec \ref{sec:mnist_exp}).}  

Our rate-distortion analysis uses the total length of the messages in both index selection and re-transmission (including any acknowledgement messages) for computing the rate. We do not include the rate of the feedback message, however. This  can be justified if there is an asymmetric cost in communication in the forward and reverse directions, e.g., wireless channels. \textcolor{black}{For details about the rate distortion analysis of this scheme, see Section \ref{feedback_rate} in the supplementary material.} 

\section{EXPERIMENTS}\label{sec:exp}
We experimentally study ISC schemes for the setup in Section~\ref{sec:SI} with different datasets as discussed below. For real-world datasets, we note that the decoding step requires computing $\log \frac{p_W(W_i)}{p_{W|T}(W_i|t)}$, which can be learned by training a neural estimator \citep{hermans2020likelihood}. Further details are available in Section \ref{app:add_exps} in the supplementary.

\subsection{Synthetic Gaussian Source}\label{sec:expgauss}

For the setup in Section~\ref{ref:p-setup} we assume that the source $V{\sim} \mathcal{N}(0,\sigma_V^2{=}1.0)$ and the side information $T = V + \zeta$ where $\zeta {\sim} \mathcal{N}(0, 0.01)$, i.e $p_{T|V}(.|v){=}\mathcal{N}(v, \sigma_{T|V}^2= 0.01)$. Furthermore, the encoder and decoder have access to the shared randomness $(S_i,Y_i,\ell_i)^N_i$ as described previously. The decoder must ideally output $W \sim p_{W|V}$, where $p_{W|V}(\cdot|v) {=}{\mathcal N}(v, \sigma_{W|V}^2)$. The encoder follows (\ref{eq:Up-sel}) to select the index $U_P$ and transmit $l_{U_p}$ to the decoder, using $\log_2(L)$ bits. Upon receiving $l_{U_p}$, the decoder selects its index following (\ref{eq:Uq-sel-wz}) and outputs $\hat{W}{=}W_{U_q}$. Finally, note that in this scenario, a closed-form solution for $p_{W|T}$ exists, and a refined reconstruction $\hat{V}{=}\Tilde{g}(\hat{W}, T)$ can be generated with inverse variance weighting (see Section \ref{app:add_exps} in the supplementary).
\begin{figure}[t]%
    \centering%
    
    \begin{subfigure}[t]{0.39\linewidth}%
        \includegraphics[width=\linewidth]{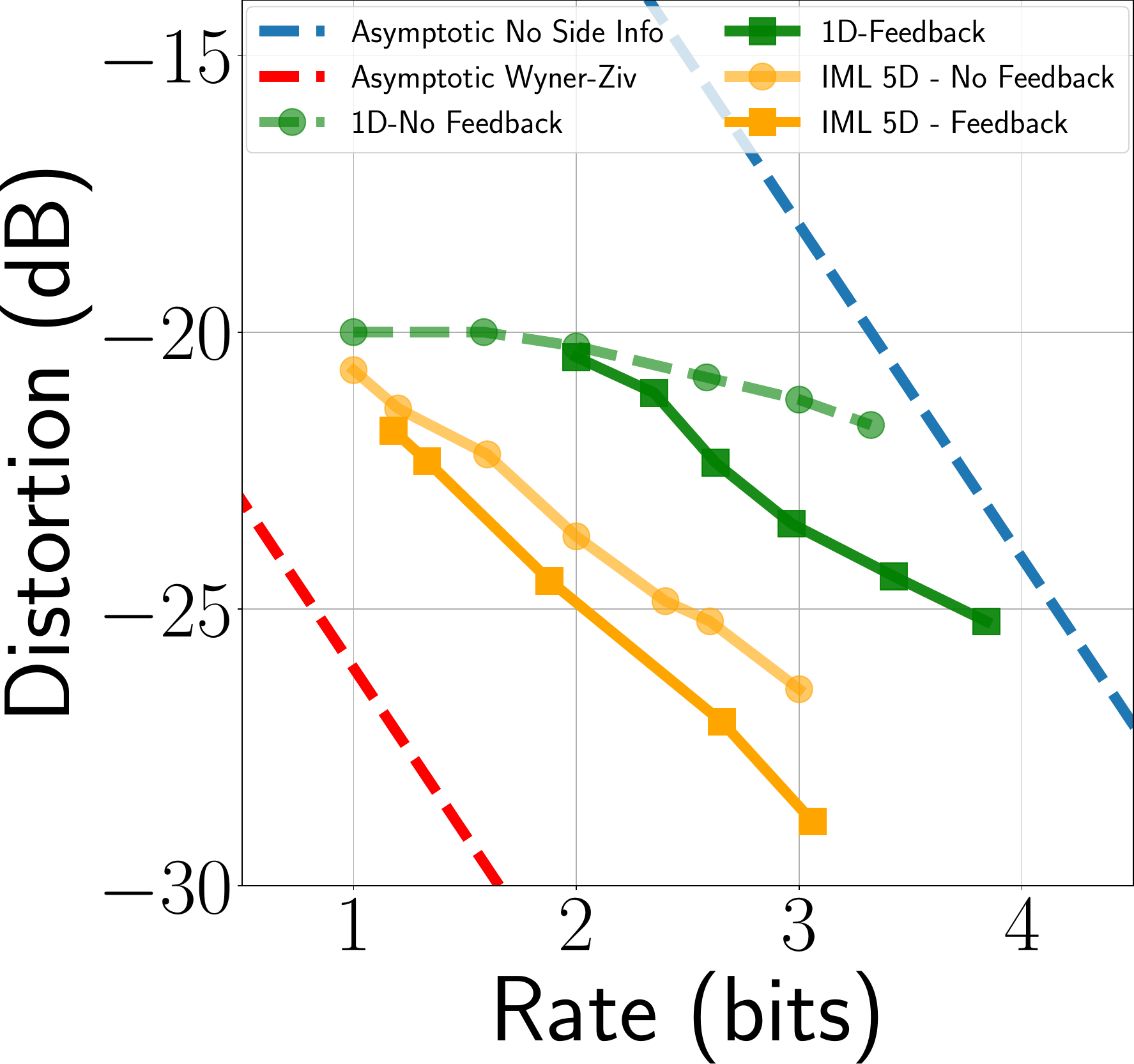}%
        \vspace{-3pt}
        \caption{Rate-Distortion Performance. }%
        \label{fig:gauss2}%
    \end{subfigure}%
    \hspace{2mm}
    \begin{subfigure}[t]{0.4\linewidth}%
        \includegraphics[width=\linewidth]{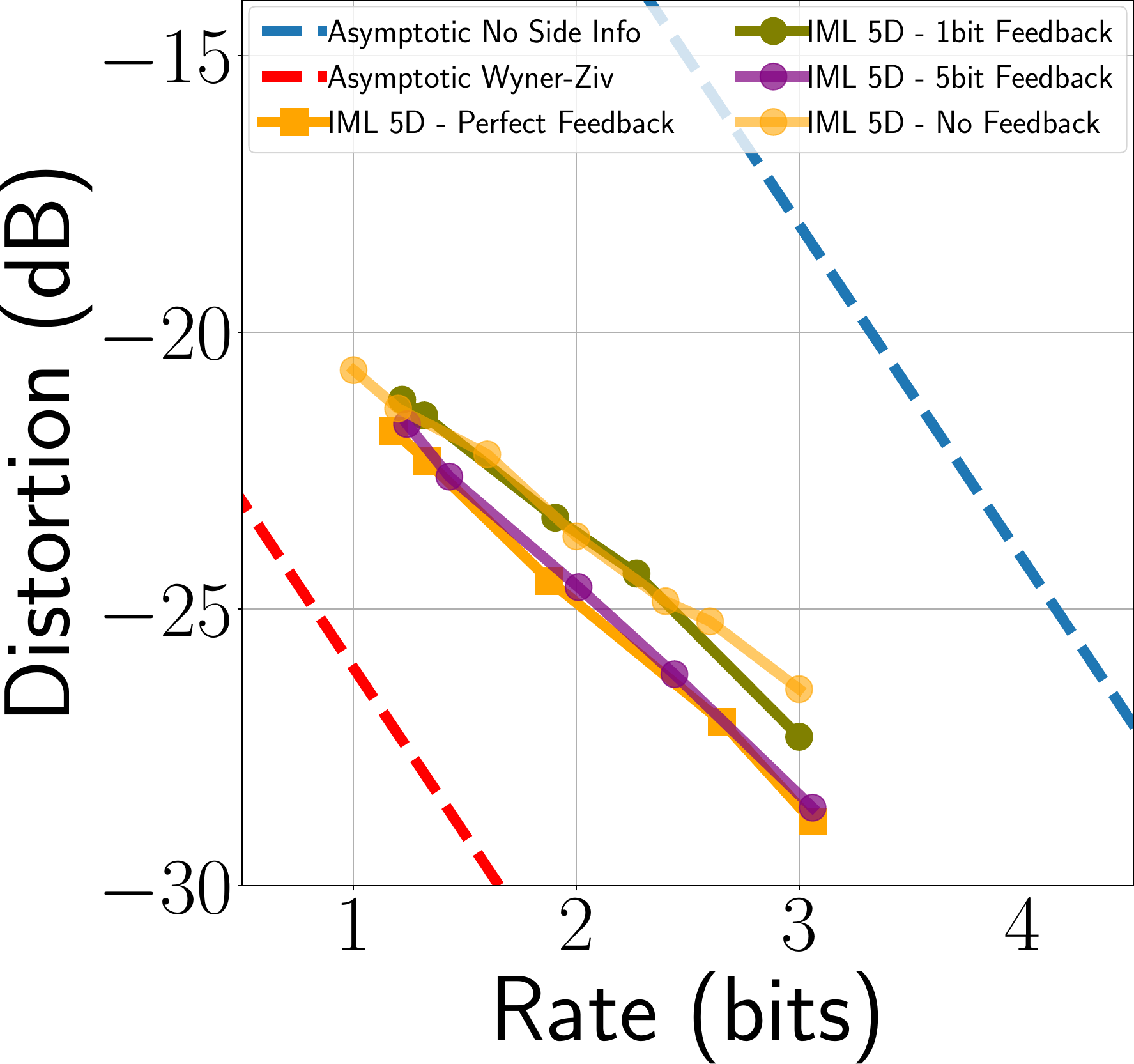}%
        \vspace{-3pt}
        \caption{Comparison of Different Feedback Rates.}%
        \label{fig:feedback}%
    \end{subfigure}%
    \vspace{-6pt}
    \caption{Analysis and rate-distortion performance of different IML schemes. (Best view in screen) }
    \vspace{-6pt}
     \label{fig:gaussanalysis}
\end{figure}

Figure \ref{fig:gaussian} illustrates the empirical matching probability: $p_m = \Pr(U_p = U_q)$ in our setup. In the left figure, we plot $p_m$ as a function of rate for different choices of $\sigma^2_{W|V}$ and different number of proposal candidates $N$. We note that provided  $N$ is sufficiently large (which is required for guaranteeing $\tilde{p}_{W|V}\approx p_{W|V}$ in all ISC schemes), it has a negligible effect on $p_m$ as noted in Remark~\ref{rem:largeN}. We also note that consistent with the theoretical analysis (1) increasing the rate for a fixed $\sigma^2_{W|V}$ increases $p_m$ and (2) increasing $\sigma^2_{W|V}$ with a fixed rate increases $p_m$. Finally in the right sub-figure in Fig.~\ref{fig:gaussian}, we also demonstrate the behaviour of $p_m$ when compressing multiple (say $k$) independent source samples. In the regime where $p_m$ (and correspondingly $R$) is large, which is of practical interest, we observe that compressing $k>1$ independent samples improves $p_m$. This effect is also reflected in our theoretical analysis in~
Remark~\ref{rem:multiple}. In particular if we approximate \textcolor{black}{$\sum_{i=1}^k i(W_i;V_i|T_i)$} by the expectation, \textcolor{black}{$k I(W;V|T)$}, then the error probability is decreasing in $k$ 
provided \textcolor{black}{$R> I(W;V|T)$}.
 



Figure \ref{fig:gauss2} presents the rate-distortion (RD) trade-off with and without feedback when compressing either a single sample or $5$ samples together  (which we will refer to as 1D and 5D respectively).  First, note that  there is a substantial improvement in the RD trade-off in the 5D case. It is worth mentioning that each operating point in the figure is such that compressing multiple samples is beneficial (see Fig.~\ref{fig:gaussian}). Secondly, when considering the 1D Gaussian case, feedback demonstrates a significant enhancement in performance, albeit at the cost of an additional acknowledgment bit required in the re-transmission step. The overhead of the acknowledgment bit gets amortized over $5$ samples in the 5D case, resulting in the overhead of $0.2$ bits/sample. As such when comparing the 1D and 5D compression schemes with feedback, the consistent improvement in the latter can be attributed to both the reduced overhead in the acknowledgment bit, as well as improved matching probability. Note that the feedback communication here is perfect as described previously in Section ~\ref{prac}. 


We show the effects of different feedback rates on the RD tradeoffs in Figure \ref{fig:feedback} for the 5D case. Note that $1$ and $5$ bits are not sufficient to recover the index since we use $N{=}2^{27}$ and the maximum value of $L$ is $2^{15}$. Nevertheless, they can correct most of the decoding error as the result shows. With  $1$ bit of feedback, we can offset the penalty of the overhead in re-transmission and with $5$ bit feedback, the performance remains close to that achieved with full feedback (at least 12  bits). This demonstrates that limited number of feedback bits may also be sufficient in practice. 

\subsection{Distributed Image Compression} \label{sec:mnist_exp}


\begin{figure}[t]%
    \centering%
    \vspace{-0.1cm}
    \begin{subfigure}[t]{0.42\linewidth}%
        \centering
    \includegraphics[width=1.0\textwidth]{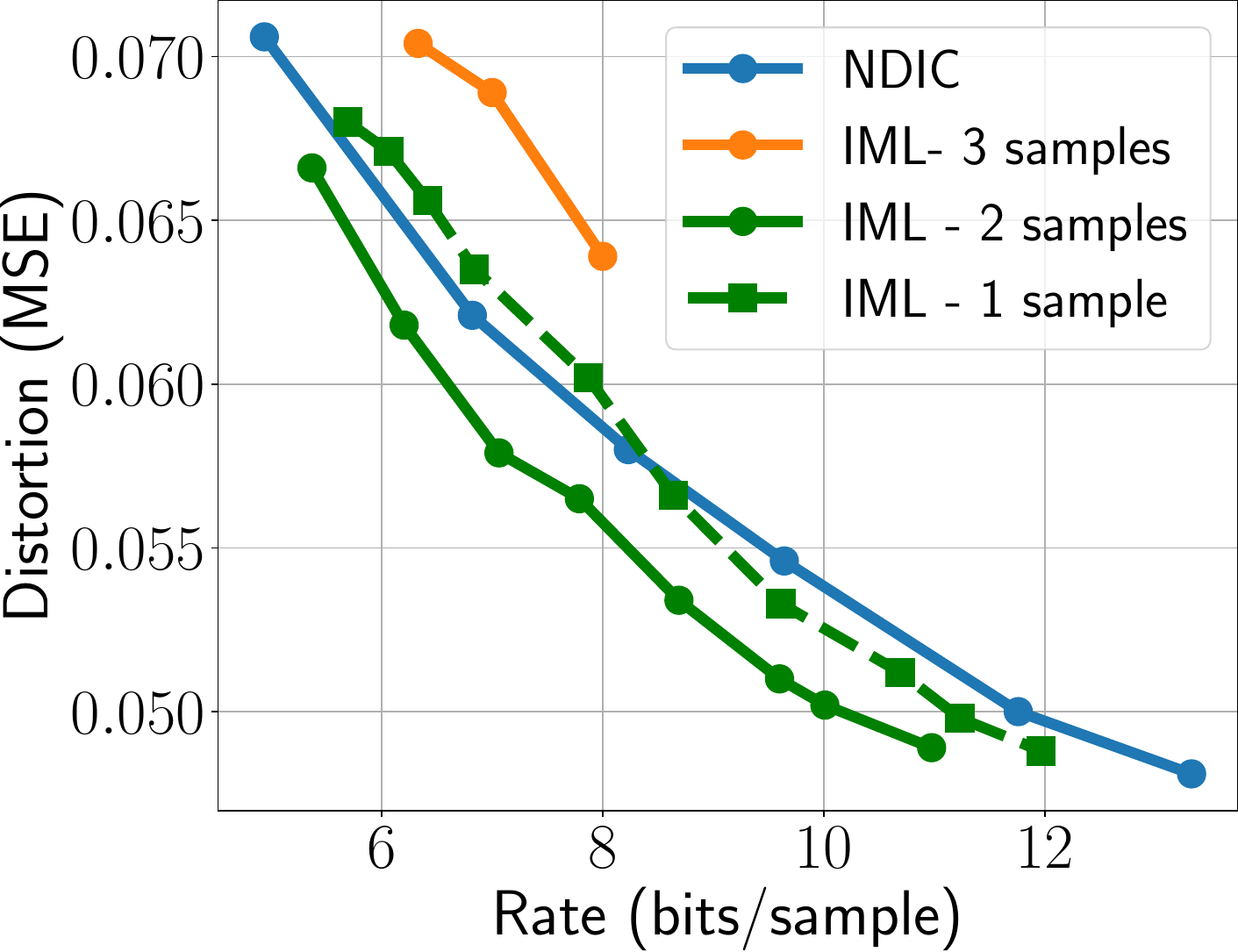}
        \caption{Rate-Distortion Performance. }
        \label{mnista}
    \end{subfigure}%
    \hspace{1mm}
    \begin{subfigure}[t]{0.4\linewidth}%
        \includegraphics[width=\linewidth]{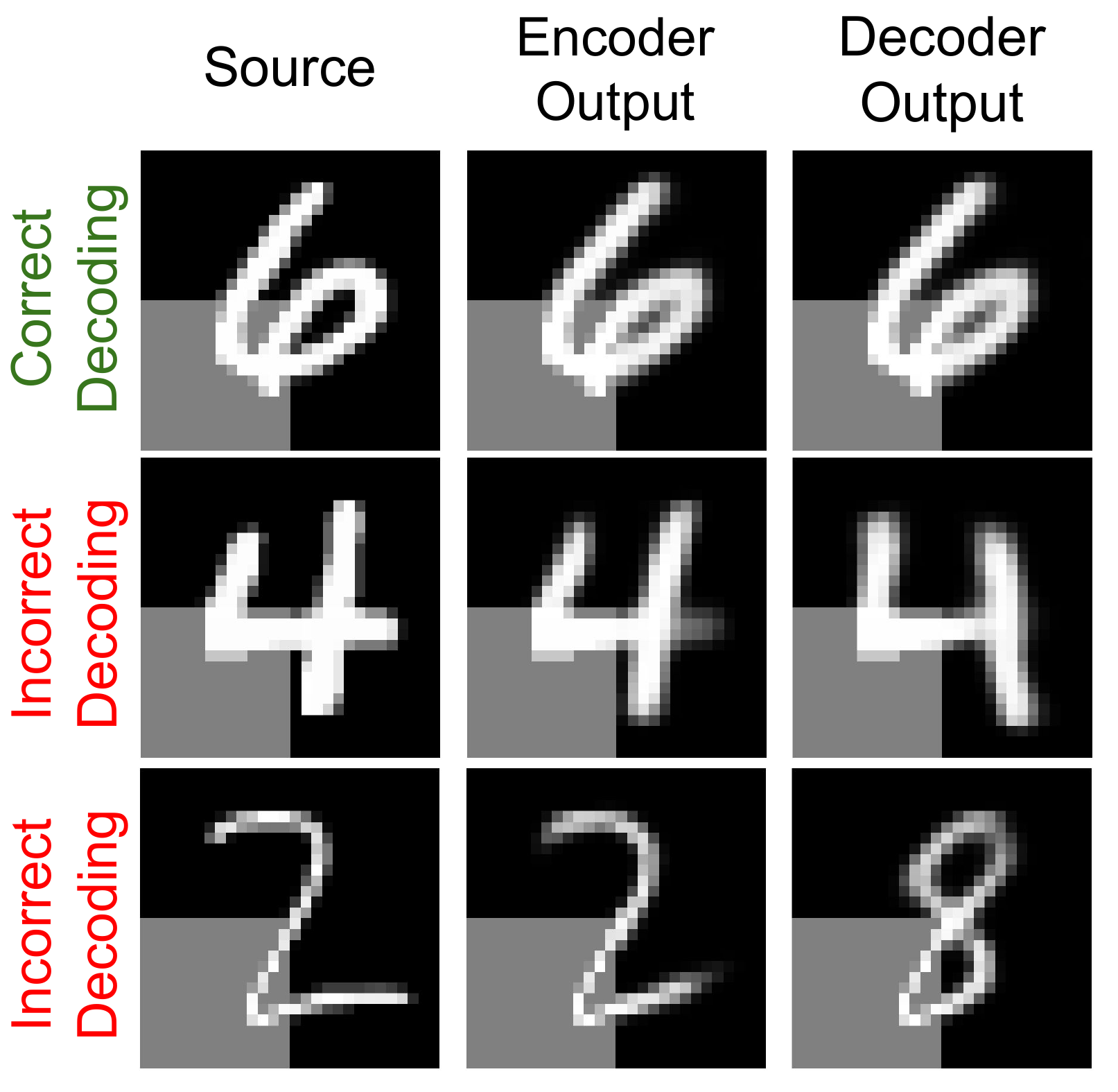}%
        \caption{Decoded Examples (No Feedback).}%
        \label{mnistb}%
    \end{subfigure}%
    \vspace{-5pt}
    \caption{Distributed Image Compression with MNIST. In (a), the orange curve is IML without feedback and the performance is restricted to 8 bits due to limited computing resources. In (b), the gray area denotes the side-information sent to the decoder, which is 0 at the encoder side.}
   \vspace{-12pt}
     \label{fig:mnist}
\end{figure}

We validate the efficacy of our method through a distributed image compression setting \citep{whang2021neural, mital2022neural}. We consider the MNIST dataset where the side information is the cropped bottom-left quadrant of the image, see Figure \ref{mnistb}, while the source image to be reconstructed is the remaining.

Directly sending the noisy source image as in the Gaussian case will incur high complexity. Instead, we rely on the learned compression approach, where a $\beta$-VAE \citep{higgins2016beta} is trained to first project the source image to the embedding vector of size $4$. This vector, together with the side information, is then fed into a neural network to reconstruct the source image, where the whole process is trained end-to-end. Once the $\beta$-VAE converges, we train a neural estimator (details in  Section \ref{app:nce} in the supplementary) for the decoding process at test time, where the proposal distribution of the 4D embedding is the prior distribution used in $\beta$-VAE. At test time, we deploy the feedback process described in Section \ref{prac}. Note that the variance of the target distribution is dependent on the input and we vary the $\beta$ factor to obtain different rate-distortion tradeoffs. Also, we use a fixed transmission size of 5 bits per feedback, irrespective of the number of samples to be compressed. 


We compare our approach with NDIC \citep{mital2022neural}, a method that enhances compression rates in this task by modeling the common information where we also set its bottleneck dimension to 4. Figure \ref{mnista} shows that our approach achieves comparable performance with a single sample (4D vector) while consistently outperforming NDIC when jointly compressing two samples (8D vector), as in the Gaussian case.  It is also worth highlighting that due to the high dimensionality of the side information ($14\times14$), employing classical 1D binning is not straightforward. Unlike NDIC which heavily relies on the learning capability of neural networks for encoding and decoding, our scheme explicitly exploits the statistical correlation between the source and side information and gives stronger RD performance. Figure \ref{mnistb} shows examples of when the decoder selects the correct/incorrect index, showing that the neural estimator selects messages that are semantically related to the side information. \textcolor{black}{Finally, we provide additional experiment with different feedback rates in Section \ref{app:add_exps} in the supplementary material.}
\subsection{Vertical Federated Learning} 
\begin{figure}[!t]%
    \centering%
    \vspace{-0.32cm}
    \begin{subfigure}[t]{0.4\linewidth}%
        \includegraphics[width=\linewidth,trim={0cm 1.2cm 0.cm 0cm}]{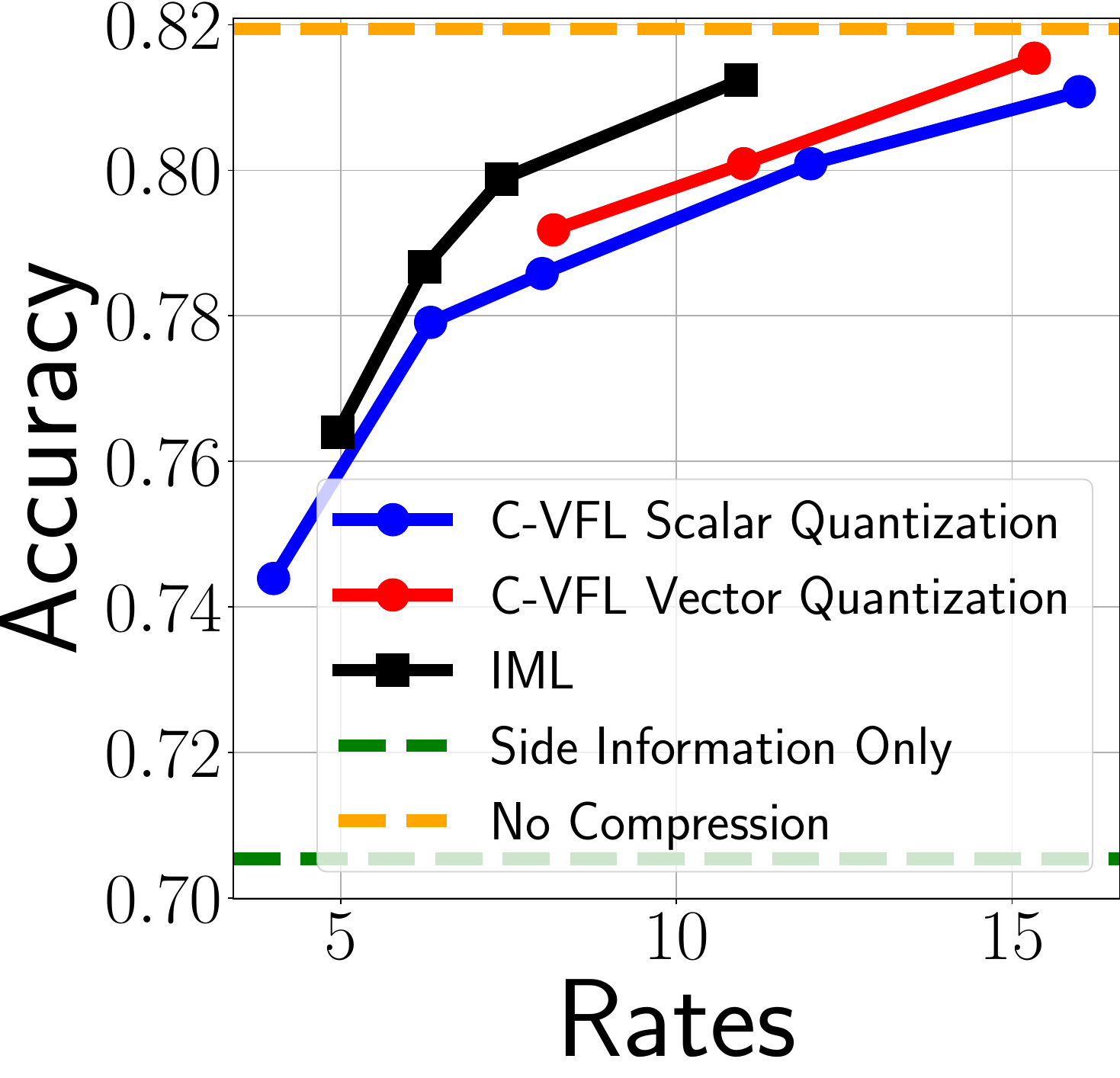}%
        \label{fig:minimal-example-pmp-ml:ml}%
    \end{subfigure}%
    \hspace{2mm}
    \begin{subfigure}[t]{0.48\linewidth}%
        \includegraphics[width=\linewidth,,trim={0cm -0.8cm 0.cm 0cm}]{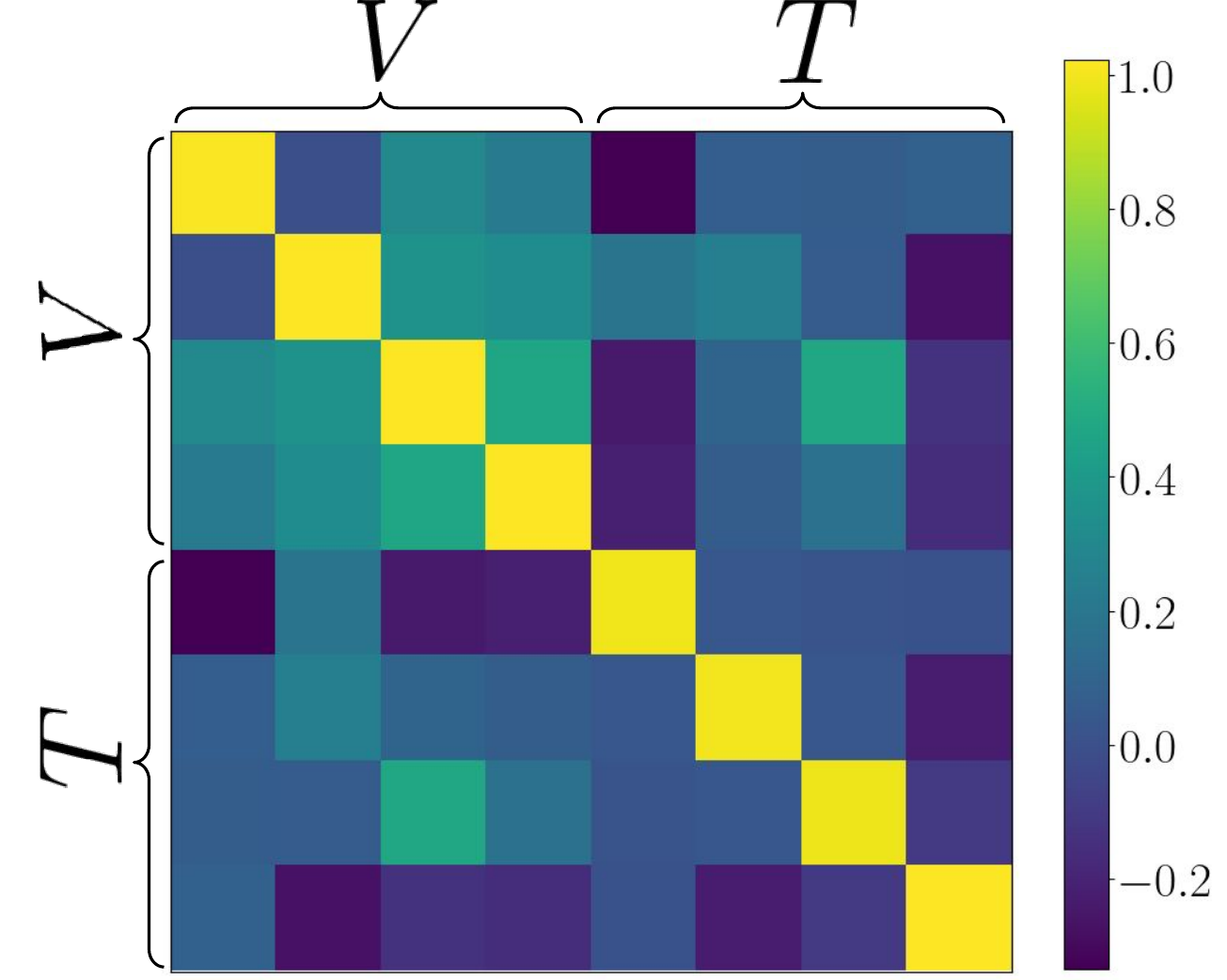}%
        \label{fig:minimal-example-pmp-ml:pmp}%
    \end{subfigure}%
    \caption{VFL with CIFAR-10. Left: Rate-Accuracy performance. Right: Covariance matrix between the source and side information's embeddings ($V$ and $T$ respectively).}%
    \vspace{-12pt}
     \label{fig:cifar}
\end{figure}
\subsubsection{CIFAR-10 Dataset.}
We demonstrate the applicability of our method to the compressed-vertical federated learning setting proposed by \cite{castiglia2022compressed}, whose work we will refer to as C-VFL. Specifically, given a pre-trained model (e.g. the neural networks at the server and each party), we want to exploit the statistical correlation between the features or learned embeddings to efficiently compress and therefore save communication costs from each party to the server during inference, while also minimizing the accuracy drop. We adapt the setup and network architecture in C-VFL for CIFAR-10 to the two-party scenario, where each party is assigned a non-overlapping quadrant of an input image. Each party then transforms their given quadrant into an embedding of dimension 4.   Assuming perfect transmission, one party's embedding is compressed losslessly (32 bits per dimension) and treated as side information, while the other party's embedding is  compressed in a lossy manner. Note that we exclude the possibility of splitting images into half since the side information alone in this case achieves near-optimal accuracy, rendering the source information unnecessary, and vice versa. This lets us evaluate the effectiveness of different compression methods when one party's information alone is insufficient for optimal accuracy.  





Following our CE-IS approach, we compress the embedding by communicating its noisy version. Here, each of the 4 dimensions is perturbed with independent Gaussian noise with zero mean and the same variance\footnote{For each dimension, before compressing, we shift and scale their values to zero-mean and unit variance. This helps us avoid searching the variance for each dimension.}, whose value is varied to obtain different rate-accuracy tradeoffs. To further the efficiency, similar to the MNIST experiment, we exploit the correlation from the side information by training a neural estimator (see Supplementary, Section~\ref{app:nce}) and use the feedback scheme to communicate the perturbed embedding. Note that we employ 4-bit feedback, which is sufficient for locating the index in this experiment.


We compare our method with scalar and vector quantization baselines, proposed by \cite{castiglia2022compressed}. In scalar quantization, each dimension of the 4D embeddings is discretized, while in vector quantization, a 2D lattice is constructed for every 2 dimensions. In Figure \ref{fig:cifar} (left), we observe that our method outperforms the baselines, achieving near-optimal accuracy of 81.24\% with ${\sim}11$ bits, while both vector and scalar quantization requires up to ${\sim}15$ bits for similar accuracy. This improvement can be attributed to the utilization of the correlation structure between the source and side information, as depicted in the covariance matrix in the right figure. These results further support the effectiveness of our method, even when the objective of the task considered is not related to source reconstruction. 



\subsubsection{UCI Breast Cancer Dataset.}
\begin{figure}[!t]%
    \centering%
    \vspace{-0.32cm}
    \begin{subfigure}[t]{0.4\linewidth}%
        \includegraphics[width=\linewidth]{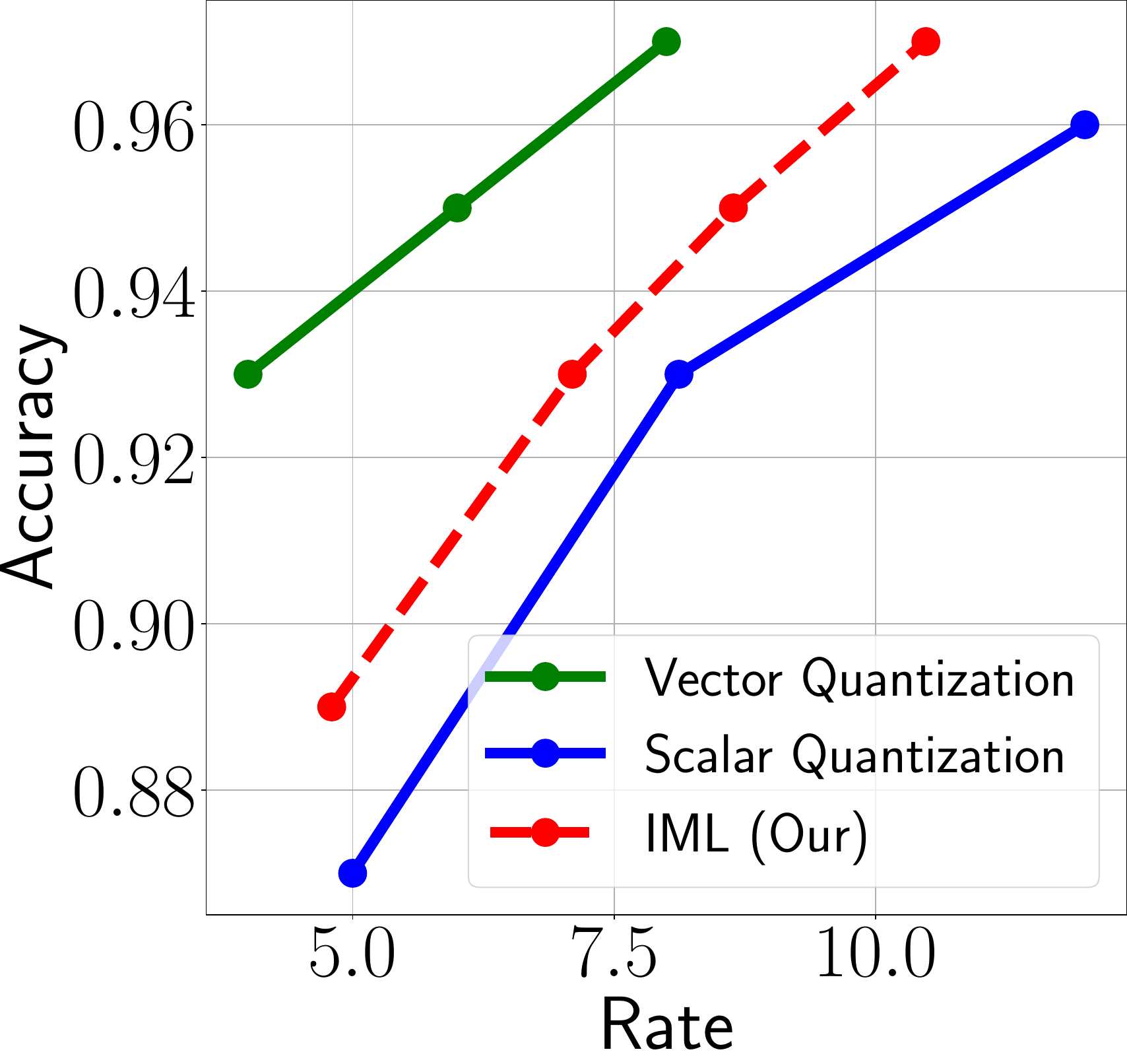}%
    \end{subfigure}%
    \hspace{2mm}
    \begin{subfigure}[t]{0.48\linewidth}%
        \includegraphics[width=\linewidth]{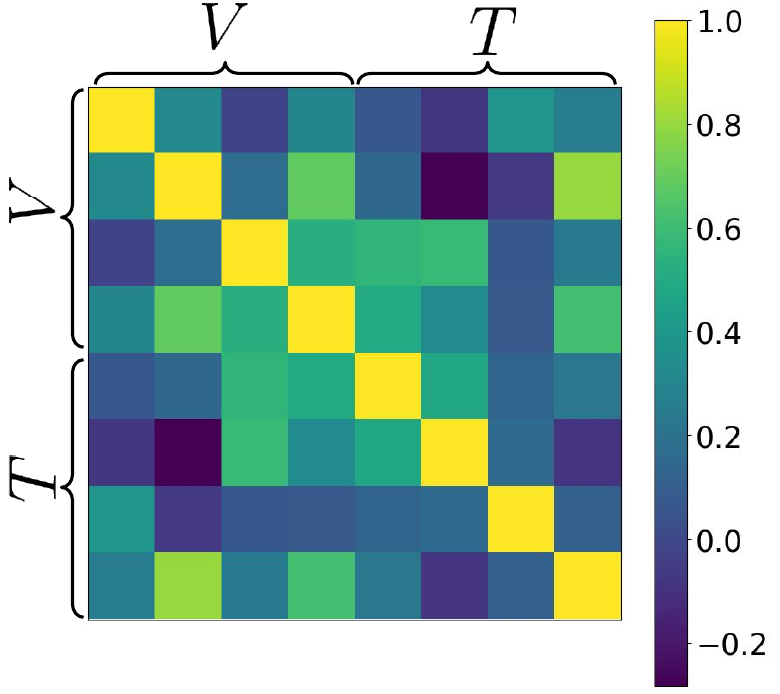}
    \end{subfigure}%
    \caption{VFL with Breast Cancer Dataset. Left: Rate-Accuracy performance. Right: Covariance matrix between the source and side information's features.}%
    \vspace{-10pt}
     \label{fig:extra_vfl}
\end{figure}

\textcolor{black}{We compare our IML method with scalar quantization on the Breast Cancer dataset \citep{misc_breast_cancer_wisconsin_(diagnostic)_17}. For this task, we use the following features at the sender: ``mean texture'', ``mean area'', ``mean smoothness'', ``mean concavity'', and the following as side information, ``mean symmetry'',	``mean fractal dimension'',	``texture error'',	``area error''. Here, we directly compress the features at the sender and the combined input features (size of 8) will be fed into a neural network consisting of 2 hidden layers of size 8 (with ReLU activation) and an output layer of size 1. The results are shown in Figure \ref{fig:extra_vfl}, which shows that IML consistently outperforms scalar quantization. On the other hand, our scheme with the current hyperparameter tuning did not outperform vector quantization in this experiment. Results are averaged over 10 runs.}


\section{CONCLUSIONS} 
We introduce a new one-shot ISC scheme with theoretical guarantees for lossy compression with side information at the decoder.  Different from the previous work by \citep{li2021unified}, we introduce the importance matching lemma that quantifies the influence of the number of proposals $N$ on the mismatch probability.  We also present a detailed study of synthetic Gaussian sources to validate our theoretical results. On the practical side, we present an algorithm that uses neural networks to enable the extension of IML to complex probability distributions. We then demonstrated its effectiveness in the task of distributed image compression with MNIST and vertical federated learning with CIFAR-10.

For future research, an important direction is to further scale and extend our approach to other DSC and machine learning settings. We note that it may be possible to extend this method to higher dimensional source models, by employing similar techniques proposed by \cite{havasi2019minimal} for model compression. \textcolor{black}{Specifically, one can split source vectors into $k$ smaller parts and transmit them separately.  This can reduce the proposals by roughly a factor of $2^{O(k)}$ but increase the decoding error probability. }Finally, another avenue for exploration is the elimination of feedback for latency reduction.

\bibliography{ref}
\section*{Checklist}

The checklist follows the references. For each question, choose your answer from the three possible options: Yes, No, Not Applicable.  You are encouraged to include a justification to your answer, either by referencing the appropriate section of your paper or providing a brief inline description (1-2 sentences). 
Please do not modify the questions.  Note that the Checklist section does not count towards the page limit. Not including the checklist in the first submission won't result in desk rejection, although in such case we will ask you to upload it during the author response period and include it in camera ready (if accepted).


 \begin{enumerate}

 \item For all models and algorithms presented, check if you include:
 \begin{enumerate}
   \item A clear description of the mathematical setting, assumptions, algorithm, and/or model. [Yes]
   \item An analysis of the properties and complexity (time, space, sample size) of any algorithm. [Yes]
   \item (Optional) Anonymized source code, with specification of all dependencies, including external libraries. [Not Applicable]
 \end{enumerate}

 \item For any theoretical claim, check if you include:
 \begin{enumerate}
   \item Statements of the full set of assumptions of all theoretical results. [Yes]
   \item Complete proofs of all theoretical results. [Yes]
   \item Clear explanations of any assumptions. [Yes]     
 \end{enumerate}

 \item For all figures and tables that present empirical results, check if you include:
 \begin{enumerate}
   \item The code, data, and instructions needed to reproduce the main experimental results (either in the supplemental material or as a URL). [Yes]
   \item All the training details (e.g., data splits, hyperparameters, how they were chosen). [Yes]
         \item A clear definition of the specific measure or statistics and error bars (e.g., with respect to the random seed after running experiments multiple times). [Yes]
         \item A description of the computing infrastructure used. (e.g., type of GPUs, internal cluster, or cloud provider). [Yes]
 \end{enumerate}

 \item If you are using existing assets (e.g., code, data, models) or curating/releasing new assets, check if you include:
 \begin{enumerate}
   \item Citations of the creator If your work uses existing assets. [Yes]
   \item The license information of the assets, if applicable. [Not Applicable]
   \item New assets either in the supplemental material or as a URL, if applicable. [Not Applicable]
   \item Information about consent from data providers/curators. [Not Applicable]
   \item Discussion of sensible content if applicable, e.g., personally identifiable information or offensive content. [Not Applicable]
 \end{enumerate}

 \item If you used crowdsourcing or conducted research with human subjects, check if you include:
 \begin{enumerate}
   \item The full text of instructions given to participants and screenshots. [Not Applicable]
   \item Descriptions of potential participant risks, with links to Institutional Review Board (IRB) approvals if applicable. [Not Applicable]
   \item The estimated hourly wage paid to participants and the total amount spent on participant compensation. [Not Applicable]
 \end{enumerate}

 \end{enumerate}


\onecolumn 

\titlespacing*{\section}
{0pt}{5.5ex plus 1ex minus .2ex}{4.3ex plus .2ex}
\titlespacing*{\subsection}
{0pt}{5.5ex plus 1ex minus .2ex}{4.3ex plus .2ex}

\aistatstitle{Supplementary Material: Importance Matching Lemma for Lossy Compression  with Side Information  }

\section{Output Distribution of Importance Sampling}
\label{sec:output}

The result in this section has already been shown in prior works~\citep{havasi2019minimal,theis2022algorithms,chatterjee2018sample}. In particular, following~\citep[Corollary 3.2]{theis2022algorithms}, we have that, for each $x$, if we set:
\begin{equation}
    N = 2^{ [D_{\mathrm{KL}}(p_{Y|X}(\cdot|x) \,||\, p_Y(\cdot))+t]}
\end{equation}
for any $t \ge 2\log(e)/(e)$, then $D_\mathrm{TV}(\tilde{p}_{Y|X}(.|x), {p}_{Y|X}(.|x)) {\le} 4\epsilon$, where recall that $\tilde{p}_{Y|X}(.|x)$ is the output distribution of the importance sampling scheme as defined in~\eqref{eq:psim} in the main paper. Here $\epsilon$ is given by:
\begin{align}
\epsilon = 2^{-t/8} {+} \sqrt{2} \exp\left(-\frac{1}{4B^2}\left(t/2{-} \frac{\log e}{e}\right)^2\right)
\end{align}
and $B = \log \omega$, where $\omega$ is defined in~\eqref{eq:om-def} in the main paper. Thus for any given $\epsilon>0$, we can construct a $t(\epsilon)$ such that selecting $N \ge N(x,\epsilon)$, where 
$$N(x,\epsilon)=2^{ [D_{\mathrm{KL}}(p_{Y|X}(\cdot|x) \,||\, p_Y(\cdot))+t(\epsilon)]}$$
guarantees that $D_\mathrm{TV}(\tilde{p}_{Y|X}(.|x), {p}_{Y|X}(.|x)) {\le} 4\epsilon$. Finally since this bound must hold for every $x$ it suffices to take $N_0(\epsilon) = \max_x N(x, \epsilon)$.

\section{Proof of Theorem~\ref{thm:rate}}
\label{sec:thm-rate}

For convenience we re-state the Theorem below:

\begin{theorem*}[Restatement of Theorem~\ref{thm:rate} in main paper] Given $(X,Y) \sim p_{X,Y}$,  and $N, K$ are defined as in the scheme in Sec. \ref{coding_scheme} in the main paper, then  we have that:
\begin{align}
E[\log K | X=x] \le E_{Y_1^N}\left[D({\bf \lambda}|| {\bf u})\right] + \delta \label{eq:bnd1-n}
\end{align}
where ${\bf \lambda} = (\lambda_1, \ldots, \lambda_N)$ is defined in~\eqref{eq:lam-i} (in the main paper),  ${\bf u} = \left(1/N,\ldots, 1/N\right)$ is associated with the uniform distribution and  $\delta = 1 + \log e/e$ is a constant. Furthermore,
    \begin{align}
H[K] {\le} I(X;Y) {+} \frac{\Delta}{N} {+} \log( I(X;Y) {+} \frac{\Delta}{N} {+}1){+} 4 ,\label{eq:bnd2-n}
\end{align} 
where $\Delta:=\Delta(p_{X,Y})$ is a constant defined via~\eqref{eq:Del-def} and~\eqref{eq:alpha2} (below) that depends on the distributions $p_{Y|X}(\cdot|x)$, $p_Y(\cdot)$ and $\omega$ in~\eqref{eq:om-def}, but not on $N$.
\end{theorem*}
Here we introduce:
\begin{align}
\Delta = 6{(\omega -1) \log \omega} + E_{X \sim p_X(\cdot)}\left[\alpha(p_{Y}(\cdot), p_{Y|X}(\cdot|X=x))\right] 
\label{eq:Del-def}
\end{align}
and
\begin{align}
\alpha(p_Y(\cdot),p_{Y|X}(\cdot|x)) &= 2(\omega-1) + 2\sqrt{\omega-1}(d_3(p_Y(\cdot)||p_{Y|X}(\cdot|x))-d_2^2(p_Y(\cdot)||p_{Y|X}(\cdot|x)))^{\frac{1}{2}} \notag \\&+ 4\omega \cdot d_2(p_Y(\cdot)||p_{Y|X}(\cdot|x)) \label{eq:alpha2},
\end{align}
and finally
\begin{align}
d_{N+1}(p_Y(\cdot), p_{Y|X}(\cdot|X=x)) = E_{Y \sim p_Y(\cdot)}\left[\frac{p^N_Y(\cdot)}{p^N_{Y|X}(\cdot|X=x)}\right], \label{eq:d3-def}
\end{align}for each $N\ge 1$.

\subsection{Proof of~\eqref{eq:bnd1-n} (Eq.~\eqref{eq:bnd1} in main paper)}
\label{sec:part1}
We start with the proof of~\eqref{eq:bnd1-n}. Following the description of the coding scheme in Section~\ref{coding_scheme} in the main paper, we note the following: conditioned on $Y_1^n=y_1^n$, our construction is equivalent to channel simulation over a discrete alphabet
\begin{align}
\Omega= \{y_1, \ldots, y_N\}
\end{align}
where the probability of selecting $y_i$ must equal:
\begin{align}
\lambda_i = \left\{\frac{p_{Y|X}(y_i|x)}{p_Y(y_i)} \right\}/ \left\{\sum_{i=1}^N \frac{p_{Y|X}(y_i|x)}{p_Y(y_i)}\right\}.\label{eq:lam-i-n} 
\end{align}
The index selection rule i.e.,~\eqref{index_sel} in Section~\ref{coding_scheme} in the main text and the associated compression scheme is equivalent to the construction of the Exponential Function Representation Lemma~\citep[Chapter 4]{li2017information} which is summarized below.
\begin{enumerate}
\item {\bf Input}: $\Omega=\{y_1,\ldots, y_N\}$ is a discrete alphabet, $\mu(\cdot)$ is a proposal distribution over $\Omega$ known to both the encoder and the decoder while $\nu(\cdot)$ is a target distribution only known to the encoder. 
\item {\bf Step 1}: The encoder and decoder sample $S_1, \ldots, S_N$ i.i.d. from ${\mathrm Exp}(1)$ distribution using shared randomness.
\item {\bf Step 2}: The encoder and decoder compute $\phi_i = \frac{S_i}{\mu(y_i)}$ and sort them so that:
$$\phi_{\pi_1} \le \phi_{\pi_2} \ldots \le \phi_{\pi_N}.$$
\item {\bf Step 3}: Given $\nu(\cdot)$, the encoder computes 
$$I = \arg\min_{1\le i\le N} \frac{S_i}{\nu(y_i)}$$
and transmits index $K$ such that $I = \pi_K$.
\item {\bf Analysis}: Following the analysis in~\citep[Chapter 4]{li2017information} we can show that with the selected index $I$ we have $y_I \sim \nu(\cdot)$ and furthermore
\begin{align}
E[\log K] \le D(\nu(\cdot) || \mu(\cdot)) + \underbrace{\frac{\log e}{e} +1}_{=\delta}. \label{eq:efrl}
\end{align}
  \end{enumerate}
We provide a proof of~\eqref{eq:efrl} for completeness in Section~\ref{app:efrl} in this document.

Note that our proposed scheme is equivalent to the exponential functional representation lemma over the discrete alphabet $\Omega$ where the proposal distribution $\mu(\cdot) = {\bf u}$ is the uniform distribution over all $N$ samples and the target distribution $\nu(\cdot) = (\lambda_1,\ldots, \lambda_N)$ is based on~\eqref{eq:lam-i-n}. It follows that the transmitted index $K$ satisfies:
\begin{align}
E[\log K | Y_1^N=y_1^n, X=x]  \le D( {\bf \lambda} || {\bf u})  + \delta \label{eq:delta}
\end{align}
where  ${\bf \lambda} = \left(\lambda_1, \ldots, \lambda_N\right)$,
 with $\lambda_i$ defined in~\eqref{eq:lam-i-n} and ${\bf u} = (1/N, 1/N, \ldots, 1/N)$ is the uniform distribution.  Taking expectation w.r.t.\ $Y_1^N$  completes the proof.
 
 \subsection{Proof of~\eqref{eq:bnd2-n} above (Eq.~\eqref{eq:bnd2} in main paper)}
 \label{sec:thm1-bnd}
We will provide some intuition behind the proof under certain heuristic assumptions. First note that:
\begin{align}
&E_{Y_1^N}\left[\sum_{i=1}^N \lambda_i \log \frac{\lambda_i}{u_i}\right] \\
&= E_{Y_1^N}\left[\sum_{i=1}^N \lambda_i \log (N \lambda_i) \right] \\
&= \sum_{i=1}^N E_{Y_1^N}\left[\lambda_i \log (N \lambda_i) \right] \\
&= N E_{Y_1^N} [\lambda_1 \log (N \lambda_1)] =  E_{Y_1^N} [N \lambda_1 \log (N \lambda_1)]  \label{eq:expectaton}
\end{align}
The last step follows from symmetry since $Y_1, \ldots Y_N$  are sampled i.i.d.\ from $p_Y(\cdot).$ 

Next observe that for each  $i=1,2,\ldots N$ we have that
$$E_{Y_i \sim P_Y(\cdot)}\left[\frac{P_{Y|X}(Y_i|x)}{P_Y(Y_i)}\right] = \int_y P_{Y}(y) \frac{P_{Y|X}(y|x)}{P_Y(y)} dy = \int_y P_{Y|X}(y|x) =1.$$
Also since $Y_1, \ldots Y_N$ are sampled i.i.d.\, it follows by law of large numbers that $\frac{1}{N}\sum_{i=1}^N \frac{P_{Y|X}(Y_i|x)}{P_Y(Y_i)} \rightarrow 1$ as $N \rightarrow \infty$.

 Note the followings heuristic approximation:
\begin{align}
N \lambda_1 =\frac{\frac{p_{Y|X}(Y_1|x)}{p_Y(Y_1)} }{\frac{1}{N}\sum_{i=1}^N \frac{p_{Y|X}(Y_i|x)}{p_Y(Y_i)}} \approx \frac{\frac{p_{Y|X}(Y_1|x)}{p_Y(Y_1)} }{\frac{1}{N}\sum_{i=2}^N \frac{p_{Y|X}(Y_i|x)}{p_Y(Y_i)}}  \approx \frac{N}{N-1}\frac{p_{Y|X}(Y_i|x)}{p_Y(Y_i)}
\end{align}
In turn, for large $N$ by assuming $\frac{N}{N-1} \approx 1$ we have that 
\begin{align}
E_{Y_1^N} [N \lambda_1 \log (N \lambda_1)] \approx E_{Y_1}\left[ \frac{p_{Y|X}(Y_1|x)}{p_Y(Y_1)} \log \frac{p_{Y|X}(Y_1|x)}{p_Y(Y_1)} \right] = D(P_{Y|X}(\cdot|x) || P_Y(\cdot))
\end{align}
Thus it shows that under the above approximations $E[\log K]$ is upper bounded by $E_{X}[D(P_{Y|X}(\cdot|x) || P_Y(\cdot))] = I(X;Y) +\delta$. Finally as in~\citep{li2018strong} the upper bound on $E[\log K]$ can be converted into an upper bound on $H(K)$ using the maximum entropy theorem:
\begin{align}
H(K) \le E[\log K] + \log(E[\log K]+1)+1, \label{eq:max-ent}
\end{align}
 which will complete the proof.

 In establishing~\eqref{eq:bnd2-n} we formalize the heuristic argument  by adding a penalty term that scales as $O(1/N)$. In particular we will show that:
\begin{prop}
For any $N \ge1$ we have that:
\begin{align}
\label{eq:X-exp}
E_{Y_1^N}\left[\sum_{i=1}^N \lambda_i \log \frac{\lambda_i}{u_i}\right] \le D(p_{Y|X}(\cdot|x)||p_Y(\cdot)) + 
\frac{6(\omega-1)\log \omega}{N} + \frac{\alpha(p_Y(\cdot),p_{Y|X}(\cdot))}{N}
\end{align}
\label{prop:1}\qed
\end{prop}

Note that upon taking expectation with respect to $X$ on both sides in~\eqref{eq:X-exp} and using~\eqref{eq:delta} we have that:
\begin{align}
E[\log K] \le I(X;Y) + \frac{\Delta}{N} + \delta
\end{align}
where
\begin{align}
\Delta = 6{(\omega-1) \log \omega} + E_{X \sim p_X(\cdot)}\left[\alpha(p_{Y}(\cdot), p_{Y|X}(\cdot|X=x))\right] 
\label{eq:Del-def2}
\end{align}

Finally using the maximum entropy theorem as in~\citep{li2018strong} we can show that:

\begin{align}
H(K) \le I(X;Y) + \frac{\Delta}{N} + \log\left( I(X;Y) + \frac{\Delta}{N} +1\right)+ 4, \label{eq:fin-bnd}
\end{align} 
which completes the proof of~\eqref{eq:bnd2-n}.

It thus remains to provide a proof of Prop.~\ref{prop:1}. The proof is rather long and the main challenge is to handle the normalizing term in the expression for $\lambda_i$ carefully. It is presented in Section~\ref{sec:prop1-proof} in this document.

\section{Alternative Bound for Eq.~\eqref{eq:bnd2} in Theorem~\ref{thm:rate} in the main paper}
\label{sec:alt-thm1}

We establish the following alternate upper bound on $H(K)$, which is the counterpart of~\eqref{eq:bnd2-n}.
For any $\epsilon >0$, we have:
\begin{align}
\label{eq:altn}
H(K) \le  \alpha_N(\epsilon) I(X;Y) + \beta_N(\epsilon) + \log\left( \alpha_N(\epsilon) I(X;Y) + \beta_N(\epsilon)+1\right)+4
\end{align}
where
\begin{align}
\alpha_N(\epsilon)=\frac{N}{(N-1)(1-\epsilon)}
\end{align}
and 
\begin{align}
\beta_N(\epsilon)=\frac{N}{(N-1)(1-\epsilon)}  \log \frac{N}{(N-1)(1-\epsilon)}   + N\log N \exp\left(-2(N-1) \epsilon^2/\omega^2\right) 
\end{align}
Note that the upper bound in~\eqref{eq:altn} involves a multiplicative constant for $I(X;Y)$ and appears weaker than the bound in~\eqref{eq:bnd2} in main paper. However by setting $\epsilon \rightarrow 0$ and $N \rightarrow \infty$ such that $N\epsilon^2 \rightarrow \infty$ we can have $\alpha_N(\epsilon) \rightarrow 1$ and $\beta_N(\epsilon) \rightarrow 0$, so that we can also attain the same rate as in Theorem~\ref{thm:rate} when $N \rightarrow\infty$. 

The key step in the following proposition:

\begin{prop}
We have that for any $\epsilon > 0$
\begin{equation}
\begin{aligned}
E_{Y_1^N} \left[\sum_{i=1}^N \lambda_i \log \frac{\lambda_i}{u_i}\right] &\le \alpha_N(\epsilon) D(p_{Y|X}(\cdot|x) || p_Y(\cdot))  +\beta_N(\epsilon).
\label{eq:bound2}
\end{aligned}
\end{equation}
\label{prop:entbnd}
\end{prop}
The proof of Prop.~\ref{prop:entbnd} is relegated to Section~\ref{app:entbnd} in this document. Note it follows from Prop.~\ref{prop:entbnd},
\begin{align}
E[\log K | X=x] \le \alpha_N(\epsilon) D(p_{Y|X}(\cdot|x) || p_Y(\cdot))  +\beta_N(\epsilon),
\end{align}
and thus we have:
\begin{align}
E[\log K] \le \alpha_N(\epsilon) I(X;Y)  +\beta_N(\epsilon). \label{eq:bnd-3}
\end{align}
The upper bound in~\eqref{eq:bnd-3} leads to an upper bound on $H(K)$ as in the previous section. That argument is similar and will not be repeated. 

\section{Multiple Importance Sampling}
\label{sec:MIS}

\subsection{Analysis}
We consider the setting discussed in Section~\ref{sec:mis} in the main text. We generate our samples as follows:
\begin{itemize}
\item $Y_1, \ldots, Y_{{\bar{N}}}$ are sampled i.i.d.\ from $p^{(1)}_Y(\cdot)$
\item $Y_{{\bar{N}}+1}, \ldots, Y_{N}$ are sampled i.i.d.\ from $p^{(2)}_Y(\cdot)$
\end{itemize}
where we select $p^{(1)}_Y(\cdot)$ and $p^{(2)}_Y(\cdot)$ to satisfy: $p_Y(y)= \frac{1}{2}p^{(1)}_Y(y) + \frac{1}{2}p^{(2)}_Y(y)$.

Given $X=x$, the index $K$ in Multiple Importance Sampling (MIS) is selected~\cite{mis} using the following probability distribution:
\begin{align}
\Pr(K=i) = \lambda_i = \frac{\frac{p_{Y|X}(Y_i|X=x)}{p_Y(Y_i)}}{\sum_{j=1}^N\frac{p_{Y|X}(Y_j|X=x)}{p_Y(Y_j)}}
\end{align}

We perform approximate analysis assuming that $N$ is sufficiently large and that we can approximate
\begin{align}
{\sum_{j=1}^N\frac{p_{Y|X}(Y_j|X=x)}{p_Y(Y_j)}}  \approx N,
\end{align}
so that
\begin{align}
\lambda_i \approx \frac{1}{N}\frac{p_{Y|X}(Y_i|X=x)}{p_Y(Y_i)}.  \label{eq:lam-approx}
\end{align}
The above approximation is justified by noting that for $i=1,2,\ldots, {\bar{N}}$ we have that:
\begin{align}
&\frac{1}{N}\sum_{i=1}^{{\bar{N}}} E_{Y_i, Y_{i+{\bar{N}}}}\left[\frac{p_{Y|X}(Y_i|X=x)}{p_Y(Y_i)} + \frac{p_{Y|X}(Y_{i+{\bar{N}}}|X=x)}{p_Y(Y_{i+{\bar{N}}})}\right]\notag\\
& = \frac{1}{2}\int_{y} \frac{p_{Y|X}(y|x)}{p_Y(y)} \left(p^{(1)}(y)+p^{(2)}(y) \right) dy =1.
\end{align}

We argue that under the simplifying assumption~\eqref{eq:lam-approx}, the proxy distribution is close to the target distribution. Indeed, note that from~\eqref{eq:psim} in the main text
\begin{align}
\tilde{p}_{Y|X}(y|x) &= E_{Y_1,\ldots, Y_N} \left[\sum_{i=1}^N \lambda_i \cdot \delta(y-Y_i)\right] \\
&\approx \frac{1}{N}E_{Y_1,\ldots, Y_N} \left[\sum_{i=1}^N \frac{p_{Y|X}(Y_i|X=x)}{p_Y(Y_i)} \cdot \delta(y-Y_i)\right]\\
&=\frac{1}{N}\sum_{i=1}^N \int_{y_i} \frac{p_{Y|X}(y_i|X=x)}{p_Y(y_i)}p_Y(y_i)\delta(y-y_i)\\
&=p_{Y|X}(y|X=x).
\end{align}
Thus the distribution of the output samples will be close to the target distribution. For the approximate rate analysis, 
we note that the proof in Section~\ref{sec:part1} in this document still applies and we have:
\begin{align}
E[\log K |  X=x]  \le E_{Y_1,\ldots, Y_N}[D( {\bf \lambda} || {\bf u})]  + \delta 
\end{align}
where again we have $$\lambda_i = \frac{\frac{p_{Y|X}(Y_i|X=x)}{p_Y(Y_i)}}{\sum_{j=1}^N\frac{p_{Y|X}(Y_j|X=x)}{p_Y(Y_j)}},\qquad i=1,2,\ldots, N$$
and $\bold{u}$ denotes the uniform distribution.  
Now note that:
\begin{align}
&E_{Y_1,\ldots, Y_N}[D( {\bf \lambda} || {\bf u})] = E_{Y_1,\ldots, Y_N}\left[\sum_{i=1}^{\bar{N}} \left(\lambda_i \log N \lambda_i + \lambda_{i+{\bar{N}}} \log N \lambda_{i+{\bar{N}}} \right) \right] \\
&\approx E_{Y_i,Y_{i+{\bar{N}}}}\left[ \frac{1}{2} \frac{p_{Y|X}(Y_i|X=x)}{p_Y(Y_i)}\log \frac{p_{Y|X}(Y_i|X=x)}{p_Y(Y_i)} \right. \notag\\&\qquad+ \left. \frac{1}{2} \frac{p_{Y|X}(Y_{i+{\bar{N}}}|X=x)}{p_Y(Y_{i+{\bar{N}}})}\log \frac{p_{Y|X}(Y_{i+{\bar{N}}}|X=x)}{p_Y(Y_{i+{\bar{N}}})} \right] \\
&= D(p_{Y|X}(\cdot|X=x) || p_{Y}(\cdot))
\end{align}
Thus to the first order approximation in $N$, the MIS scheme achieves the same compression rate as the IS scheme. 

\subsection{Numerical Example}
\begin{figure}[t!]%
    \centering%
        \includegraphics[width=1.0\linewidth]{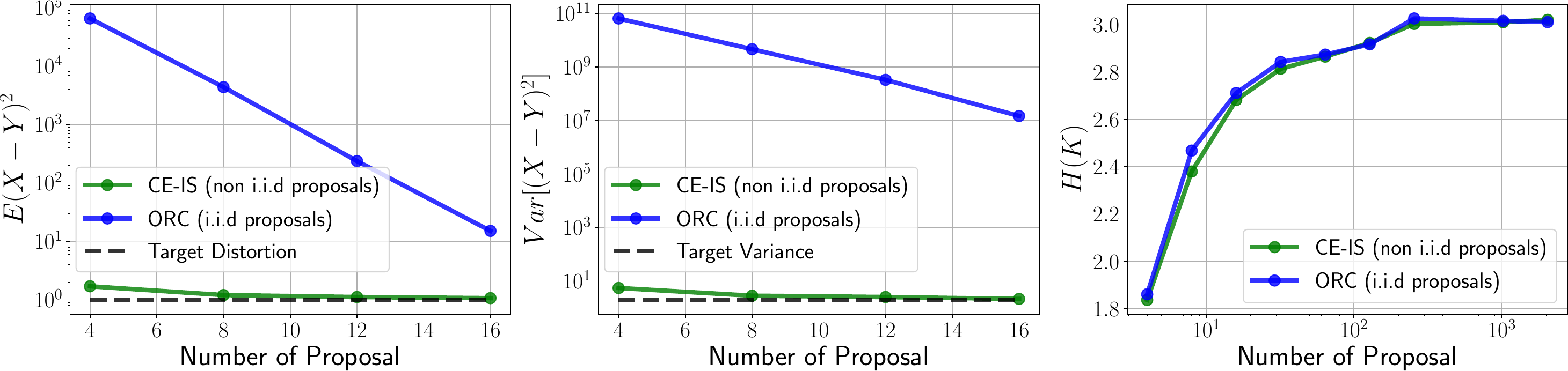}%
        
        \caption{Multiple Importance Sampling. From left to right: expected distortion, distortion variance, and compression rate as a function of the number of proposals, $N$. MIS exhibits faster convergence to the target levels while maintaining a comparable compression rate to ORC. We set $m=512$ and $D=1$.}%
        \label{fig:mis}%
    \label{fig:importance_matching_lemma}%
\end{figure}

We demonstrate the advantage of our proposed over ORC through a numerical example. We assume that the source sample $X$ is a mixture of two Gaussians: $p_X(x) = \frac{1}{2}p_X^{(1)}(x)+ \frac{1}{2}p_X^{(2)}(x)$, where we have $p_X^{(1)}(x) = {\mathcal N}(m,1)$ and $p_X^{(2)}(x) = {\mathcal N}(-m,1)$. The conditional distribution of $Y$ given $X$ is given by $Y = X+\zeta$, where $\zeta \sim {\mathcal N}(0,D)$. Note that we can also express the marginal distribution of $Y$ as $p_Y(y) = \frac{1}{2}p_Y^{(1)}(y)+ \frac{1}{2}p_Y^{(2)}(y)$ where $p_Y^{(1)}(y) = {\mathcal N}(m, 1+D)$ and $p_Y^{(2)}(y) = {\mathcal N}(-m, 1+D)$. In this example, we set $m=512$, $D=1$ and average our results over $2^{20}$ simulations.

In Figure \ref{fig:mis}, given a source sample $X$ at the encoder and output $Y$ at the decoder, we compute $E[(Y-X)^2]$, $\mathrm{Var}((Y-X)^2)$ as well as the compression rate for our proposed scheme (computed from the index histogram) and ORC (with i.i.d.\ samples from $p_Y(\cdot)$) for a different number of candidate proposals. Here, $E[(Y-X)^2]$ is the expected distortion, and $\mathrm{Var}((Y-X)^2)$ is the distortion variance and is equal to $D$ and $2D^2$ respectively when $\Tilde{p}_{Y|X}(.|x) = p_{Y|X} (.|x)$ for all $x$, which we refer to as target distortion and target variance. We observe that while both schemes achieve a similar compression rate, our proposed scheme outperforms ORC in other metrics indicating that it more closely approximates the target distribution $p_{Y|X}(.|x)$. This occurs because when a small number of i.i.d proposals  $N$ is considered, there is a high probability that all the proposals are sampled from different modes of the source variable X. This probability is $2^{-(N+1)}$ and can result a significant distortion in the output. The extent of this distortion is primarily determined by the distance between the two modes $p^{(1)}_X(.)$ and $p^{(2)}_X(.)$ of $p_X(.)$, which is $4m^2$ in this example. On the other hand, such probability is $0$ in our MIS scheme, enabling us to achieve a better convergence rate in this example.

\begin{figure}[t!]%
    \centering%
        \includegraphics[width=0.8\linewidth]{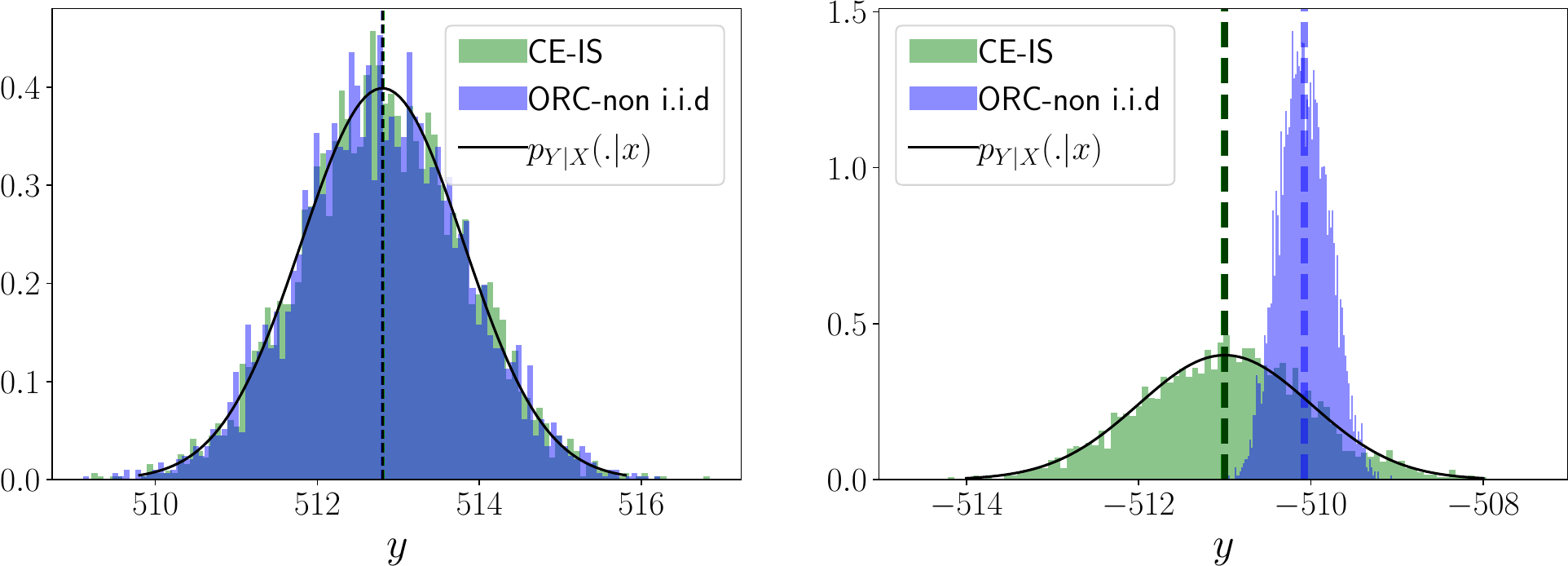}%
        
        \caption{ORC does not simulate the target distribution properly in the case of non-i.i.d proposals. (Left) When $X=x$ is from the positive mode, i.e. $x=512.88$ in this case, ORC can simulate the $\tilde{p}_{Y|X}(.|x)$ approximately close to the target $p_{Y|X}(.|x)$. (Right), $X=x$ is from the negative mode, i.e. $x=-509.93$, when ORC is unable to simulate $\tilde{p}_{Y|X}(.|x)$ accurately. Our method CE-IS can simulate accurately in both cases. We set $m=512$ and $D=1$.}%
        \label{fig:mis_orc}%
    \label{fig:importance_matching_lemma}%
\end{figure}

In Figure ~\ref{fig:mis_orc}, we show that when the proposals $\{Y_i\}^N_{1}$ are non-i.i.d, ORC is unable to simulate the target distribution $p_{Y|X}(.|x)$ properly. In this example, for each figure, we fix a source sample $X=x$ and plot the histogram of the obtained samples $Y\sim \Tilde{p}_{Y|X}(.|x)$, from multiple sets of non-i.i.d proposals. We set $N=512$ in this case and refer to $p^{(1)}_X(.) = \mathcal{N}(m, 1)$ and $p^{(2)}_X(.) = \mathcal{N}(-m, 1)$ as the positive and negative mode respectively. Figure ~\ref{fig:mis_orc} plots and compares the simulated histogram of our method CE-IS and ORC in the case where $X$ is from the postive mode (left) and negative mode (right). Our first observation is that our proposed scheme CE-IS is able to simulate the distribution accurately in both cases. On the other hand, while ORC seems to simulate accurately the distribution when $x>0$, or from the positive mode (left figure), its simulated distribution when $x<0$ (negative mode) is very different from the target distribution $p_{Y|X}(.|x)$ (right figure). This is because ORC requires the exponential variables $S_1,...,S_{2N}$ to be sorted before performing index selection, which only works when the proposals are i.i.d. In the right figure, when the proposals are non-i.i.d and $x < 0$, ORC will ignore the first half of the proposals (since $\log\frac{p_Y(Y_i)}{p_{Y|X}(Y_i|x)}$ is extremely large due to large $m$) to select samples in the second half of the proposals $Y_{N+1}...Y_{2N}$. As a result, it loses the exponential race property to simulate proper distribution. When $x>0$, on the other hand, ORC does not need to ignore the first half and as a result, can simulate approximately accurate $\tilde{p}_{Y|X}(.|x)$.





\section{Proof of Prop.~\ref{prop:condPML} in the Main Paper}
\label{app:condPML}
We restate the result for convenience.
\begin{prop*}
\label{prop:condPML-n}
Let  $\Omega= \{y_1, \ldots, y_N\}$ denote the sequence of samples. 
\begin{align}
\Pr(U_p \neq U_q | \Omega, X=x, U_p = k) \le 
1 - \left(1 + \frac{{p_{Y|X}(y_k|x)}}{{q_{Y|X}(y_k|x)}} \frac{\left(\frac{1}{N}\sum_{j=1}^N\frac{q_{Y|X}(y_j|x)}{p_Y(y_j)}\right)}{\left(\frac{1}{N}\sum_{j=1}^N\frac{p_{Y|X}(y_j|x)}{p_Y(y_j)}\right)}\right)^{-1}.
\label{eq:condPML-n}
\end{align}
\end{prop*}

The proof follows by observing that given $Y_1^N = y_1^N$, the sampling procedures in~\eqref{eq:U-def} in the main paper, is equivalent to the Poisson Matching Lemma applied to a discrete alphabet $\Omega = \{y_1,\ldots, y_N\}$ with target distributions given by
\begin{align}
\lambda^p_i = \frac{\frac{p_{Y|X}(y_i|X=x)}{p_Y(y_i)}}{\sum_{j=1}^N\frac{p_{Y|X}(y_j|X=x)}{p_Y(y_j)}}, \quad 
\lambda^q_i = \frac{\frac{q_{Y|X}(y_i|X=x)}{p_Y(y_i)}}{\sum_{j=1}^N\frac{q_{Y|X}(y_j|X=x)}{p_Y(y_j)}}. 
\end{align}

We provide the analysis for completeness below. Let us define:
\begin{align}
\tilde{S}_p = \min_{1\le i \le N} \frac{S_i}{\lambda_i^p} \sim \mathrm{Exp}(\sum_i \lambda_i^p = 1).
\end{align}

We further condition on $U_p=k$ and $\tilde{S}_p = s$. It follows that for each $j \neq k$ we have that:
\begin{align}
\frac{S_j}{\lambda_j^p} \ge s, 
\end{align}
and thus $S_j - \lambda_j^p \cdot s$ is an $\mathrm{Exponential}(1)$ random variable. Since this holds for each $s$, it follows tht $S_j - \lambda_j^p \cdot \tilde{S}_p$ is also an  $\mathrm{Exponential}(1)$  and is independent of the variable $\tilde{S}_p$.

We consider the following events:

\begin{align}
\Pr(U_p \neq U_q | \Omega, U_p = k) &= \Pr\left(\min_{j\neq k} \frac{S_j}{\lambda_j^q} \le \frac{S_k}{\lambda_k^q}~\bigg|~\Omega, U_p=k \right)\\
&= \Pr\left(\min_{j\neq k} \frac{S_j}{\lambda_j^q} \le \tilde{S}_p \frac{\lambda_k^p}{\lambda_k^q}~\bigg|~\Omega, U_p=k  \right)\\
&\le \Pr\left(\min_{j\neq k} \frac{S_j - \lambda_j^p  \cdot \tilde{S}_p}{\lambda_j^q} \le \tilde{S}_p \frac{\lambda_k^p}{\lambda_k^q}~\bigg|~\Omega, U_p=k  \right)\\
&= \frac{ \sum_{j\neq k} \lambda_j^q }{\sum_{j\neq k} \lambda_j^q + \frac{\lambda_k^q}{\lambda_k^p}} \\
&\le \frac{1}{1+ \frac{\lambda_k^q}{\lambda_k^p}} = 1-\left(1 + \frac{\lambda_k^p}{\lambda_k^q}\right)^{-1} \\
&= 1 - \left(1 + \frac{\frac{p_{Y|X}(y_k|x)}{p_{Y}(y_k) }}{\frac{q_{Y|X}(y_k|x)}{p_Y(y_k)}} \frac{\left(\sum_{j=1}^N\frac{q_{Y|X}(y_j|x)}{p_Y(y_j)}\right)}{\left(\sum_{j=1}^N\frac{p_{Y|X}(y_j|x)}{p_Y(y_j)}\right)}\right)^{-1}\\
&= 1 - \left(1 + \frac{{p_{Y|X}(y_k|x)}}{{q_{Y|X}(y_k|x)}} \frac{\left(\frac{1}{N}\sum_{j=1}^N\frac{q_{Y|X}(y_j|x)}{p_Y(y_j)}\right)}{\left(\frac{1}{N}\sum_{j=1}^N\frac{p_{Y|X}(y_j|x)}{p_Y(y_j)}\right)}\right)^{-1}
\end{align}

\section{Proof of Theorem~\ref{thm:PML2-1} in Main Paper}
\label{app:PML2-1}
In what follows, we will set $k=1$ without loss of generality. The argument can be easily extended to any $k$. We rewrite the Theorem below for sake of convenience.
\begin{theorem*}[Expanded version of Theorem~\ref{thm:PML2-1} in main paper.]

Define $\bN = N-1$, we have:
\begin{align}
\Pr(U_p \neq U_q | Y_1=y_1, U_p = 1) \le
1- \left(1 + \frac{{\lambda(y_1)}}{\beta(y_1)}\mu_{y_1}(\bN)\right)^{-1}, 
\label{eq:IML-1-n}
\end{align}

where 

\begin{align}
\mu_{y_1}(\bN)= \left(\frac{\frac{\beta(y_1)} {\bN}+1}{\frac{\lambda(y_1)}{\bN}+1}\right) + \frac{1}{\bN} \left(1+\frac{\lambda(y_1)}{\bN}\right)K(\bN)+ \frac{2\omega}{\bN} \left(1+\frac{\lambda(y_1)}{\bN}\right)L(\bN),
\label{eq:nu-def-1}
\end{align}


\begin{align}
K(\bN) =  4 \frac{(\omega-1)}{ (1+ \frac{\lambda(y_1)}{\bN})^2}\left(1+ \frac{N+1}{\bN}\omega\right)\sqrt{2+ 4\left(\frac{1+ \frac{\beta(y_1)}{\bN}}{1+ \frac{2\lambda(y_1)}{\bN}}\right)^2\left\{\left(1+ \frac{N+1}{\bN}\omega\right)^2 + \frac{(\omega-1)}{\bN}\right\}} \label{eq:k1-bnd2}
\end{align}
and
\begin{align}
L(\bN)=  \sqrt{\omega-1}\sqrt{(d_5(p_Y(\cdot)||p_{Y|X}(\cdot|X=x))-d_3^2(p_Y(\cdot)||p_{Y|X}(\cdot|x)))} + (\omega-1) d_3(p_Y(\cdot)||p_{Y|X}(\cdot|x))  \label{eq:l1-bnd2}
\end{align}
 in~\eqref{eq:k1-bnd2} and~\eqref{eq:l1-bnd2} respectively scale as $\Theta(1)$ as $\bN \rightarrow \infty$. Also, we use  $\lambda(y_1) = \frac{p_{Y|X}(y_1|x)}{p_Y(y_1)}$, and  $\beta(y_1) = \frac{q_{Y|X}(y_1|x)}{p_Y(y_1)}$. 
\end{theorem*}
\begin{rm}
\label{rem:largeN}
Upon examining~\eqref{eq:k1-bnd2} and \eqref{eq:l1-bnd2}, note that as $\bN \rightarrow \infty$, the dominating term in $K(\bN)$ scales as $\omega^3$,  while $L(\bN)$ is upper-bounded by $\omega \cdot d_5(p_Y(\cdot)|| p_{Y|X}(\cdot|x))$, regardless of $N$. Under the  assumption that $d_5(p_Y(\cdot)|| p_{Y|X}(\cdot|x)) < \infty$, see~\eqref{eq:d3-def}, we observe that for any $\epsilon>0$ there is a sufficiently large $N_1(\epsilon)$, we will have that  $\mu_{y_1}(N) \le1+\epsilon$ for any $N \ge N_1(\epsilon)$.  In turn~\eqref{eq:IML-1-n} recovers the bound in the Poisson Matching Lemma in~\citep{li2021unified}.
\end{rm}

To proceed with the proof, observe that:
\begin{align}
&\Pr(U_p \neq U_q | Y_1=y_1, U_p = 1) \\
&=1 - E_{Y_2^N}\left[\left(1 + \frac{{p_{Y|X}(y_1|x)}}{{q_{Y|x}(y_1|x)}} \frac{\left(\frac{1}{N}\sum_{j=1}^N\frac{q_{Y|X}(Y_j|x)}{p_Y(Y_j)}\right)}{\left(\frac{1}{N}\sum_{j=1}^N\frac{p_{Y|X}(Y_j|x)}{p_Y(Y_j)}\right)}\right)^{-1}\bigg| Y_1=y_1, U_p=1\right] \\
&\le 1- \left(1 + \frac{{p_{Y|X}(y_1|x)}}{{q_{Y|X}(y_1|x)}} E_{Y_2^N}\left[\frac{\left(\frac{1}{N}\sum_{j=1}^N\frac{q_{Y|X}(Y_j|x)}{p_Y(Y_j)}\right)}{\left(\frac{1}{N}\sum_{j=1}^N\frac{p_{Y|X}(Y_j|x)}{p_Y(Y_j)}\right)} \bigg| Y_1=y_1, U_p=1\right]\right)^{-1}, \label{eq:t-n-1}
\end{align}
where the last step is a consequence of Jensen's inequality since the function $f(t) =1/t$ is a convex function so that $E[f(t)] \ge f(E[t])$ holds. We thus need to upper bound the expectation above.
\begin{align}
&E_{Y_2^N}\left[\frac{\left(\sum_{j=1}^N\frac{q_{Y|X}(Y_j|x)}{p_Y(Y_j)}\right)}{\left(\sum_{j=1}^N\frac{p_{Y|X}(Y_j|x)}{p_Y(Y_j)}\right)} \bigg| Y_1=y_1, U_p=1\right]\\
&=\int_{y_2^N} \frac{\left(\sum_{j=1}^N\frac{q_{Y|X}(y_j|x)}{p_Y(y_j)}\right)}{\left(\sum_{j=1}^N\frac{p_{Y|X}(y_j|x)}{p_Y(y_j)}\right)}p_{Y_2^N|\{Y_1, U_p\}}(y_2,\ldots, y_N|Y_1=y_1, U_p=1)dy_2\ldots dy_N\\
&=\int_{y_2^N} \frac{\sum_{j=1}^N\ \beta(y_j) }{\sum_{j=1}^N \lambda(y_j) }p_{Y_2^N|\{Y_1, U_p\}}(y_2,\ldots, y_N|Y_1=y_1, U_p=1)dy_2\ldots dy_N\\
\end{align}

where we define  \begin{align}
\beta(y_j) =\frac{q_{Y|X}(y_j|x)}{p_Y(y_j)} \qquad \lambda(y_j) = \frac{p_{Y|X}(y_j|x)}{p_Y(y_j)} \label{eq:beta-def}
\end{align}

Now consider the joint density function $p_{Y_2^N|\{Y_1, U_p\}}(y_2,\ldots, y_N|Y_1=y_1, U_p=1)$:

\begin{align}
&p_{Y_2^N|\{Y_1, U_p\}}(y_2,\ldots, y_N|Y_1=y_1, U_p=1) \\ &= \frac{\Pr(U_p=1|Y_1=y_1, \ldots, Y_N=y_N) p_{Y_2^N|Y_1}(y_2^N |Y_1=y_1) }{\Pr(U_p=1|Y_1=y_1)}\\
&= \frac{\Pr(U_p=1|Y_1=y_1, \ldots, Y_N=y_N) \prod_{j=2}^N p_{Y_j}(y_j) }{\Pr(U_p=1|Y_1=y_1)}\\
&=\frac{\lambda(y_1)}{\sum_{j=1}^N \lambda(y_j)}\frac{\prod_{j=2}^N p_{Y_j}(y_j) }{\Pr(U_p=1|Y_1=y_1)}
\end{align}

Next we consider the probability $\Pr(U_p=1|Y_1=y_1)$.
\begin{align}
\Pr(U_p=1|Y_1=y_1) &= E_{Y_2^n}[\Pr(U_p=1|Y_1=y_1, Y_2=y_2, \ldots, Y_N=y_N) ] \\
&= E_{Y_2^N}\left[\frac{\lambda(y_1)}{\lambda(y_1) + \sum_{j=2}^N \lambda(y_j)} \right] \\
&\ge \frac{\lambda(y_1)}{\lambda(y_1) + E_{Y_2^N} [\sum_{j=2}^N \lambda(y_j)] }\\
&=\frac{\lambda(y_1)}{\lambda(y_1) + (N-1)}
\end{align}
where we have again used the convexity of $f(t) = 1/t$ and applied Jensen's inequality.

Thus we can express the following:
\begin{align}
&E_{Y_2^N}\left[\frac{\left(\sum_{j=1}^N\frac{q_{Y|X}(Y_j|x)}{p_Y(Y_j)}\right)}{\left(\sum_{j=1}^N\frac{p_{Y|X}(Y_j|x)}{p_Y(Y_j)}\right)} \bigg| Y_1=y_1, U_p=1\right]\\
&\le (N-1+\lambda(y_1)) \int_{y_2^N} \frac{\sum_{j=1}^N\ \beta(y_j) }{(\sum_{j=1}^N \lambda(y_j) )^2 }\prod_{j=2}^n p_{Y}(y_j)dy_2\ldots dy_N\\
&= (N-1+\lambda(y_1)) E_{Y_2^N} \left[ \frac{\sum_{j=1}^N\ \beta(Y_j) }{(\sum_{j=1}^N \lambda(Y_j) )^2  }\bigg| Y_1=y_1\right]  \label{eq:t-simp}
\end{align}

We now establish the following bound:
\begin{prop}
For any $N \ge 1$ with $\bN = N-1$, we have that:
\begin{align}
&\bN  E_{Y_2^N}\left[\frac{\sum_{i=1}^N \beta(Y_i)}{\left(\sum_{i=1}^N \lambda(Y_i)\right)^2} \Bigg| Y_1=y_1\right] \notag  \le
\frac{\frac{\beta(y_1)}{\bN} + 1}{\left(1+ \frac{\lambda(y_1)}{\bN}\right)^2} + \frac{1}{\bN} K(\bN) + \frac{2\omega}{\bN}L(\bN)
\end{align} and in turn using~\eqref{eq:t-simp}, we have
\begin{align}
E_{Y_2^N}\left[\frac{\left(\sum_{j=1}^N\frac{q_{Y|X}(Y_j|x)}{p_Y(Y_j)}\right)}{\left(\sum_{j=1}^N\frac{p_{Y|X}(Y_j|x)}{p_Y(Y_j)}\right)} \bigg| Y_1=y_1, U_p=1\right]
\le \frac{\beta(y_1) + \bN}{\lambda(y_1)+\bN} + \frac{1 + \frac{\lambda(y_1)}{\bN}}{\bN} K_1(\bN) + \frac{2\omega(1+\frac{\lambda(y_1)}{\bN})}{\bN} L(\bN). \label{eq:t-n-2}
\end{align}
where  $K(\bN)$ is given in~\eqref{eq:k1-bnd2} and $L(\bN)$ is given in~\eqref{eq:l1-bnd2}.
\label{prop:lem2}
\end{prop}

The proof of Prop.~\ref{prop:lem2} is in rather long and relegated to Section~\ref{app:lem2}. Note that substituting~\eqref{eq:t-n-2} into~\eqref{eq:t-n-1}
\begin{align}
\Pr(U_p \neq U_q | Y_1=y_1, U_p = 1) \le
1- \left(1 + \frac{{\lambda(y_1)}}{\beta(y_1)}\left(\frac{\frac{\beta(y_1)} {\bN}+1}{\frac{\lambda(y_1)}{\bN}+1} + \frac{1 + \frac{\lambda(y_1)}{\bN}}{\bN} K(\bN) + \frac{2\omega(1+\frac{\lambda(y_1)}{\bN})}{\bN}L(\bN)\right)\right)^{-1},
\end{align}
which completes the proof.

\section{Alternative Upper Bound in~\eqref{eq:IML-1} in Theorem~\ref{thm:PML2-1} in Main Paper}
\label{app:alt-thm2}
We establish the following upper bound:
\begin{align}
\Pr(U_p \neq U_q | Y_1=y_1, U_p = 1) \le  1- \left(1 + \frac{{\lambda(y_1)}}{\beta(y_1)}   \mu'_{y_1}(N) \right)^{-1}.
\label{final-bnda1}
\end{align}
where $\lambda(y_1)$ and $\beta(y_1)$ are defined in~\eqref{eq:beta-def} and we have
\begin{align}
 \mu'_{y_1}(N)= (N-1+\lambda(y_1))\left(\frac{\beta(y_1) + (N-1)(1+\epsilon)}{(\lambda(y_1) + (N-1)(1-\epsilon))^2} + \frac{N \omega }{\lambda(y_1)^2} 2 e^{-(N-1)\epsilon^2/\omega^2}\right)
\end{align} Note that this upper bound requires that $\lambda(y_1) > 0$, which is a stronger condition than the condition in Theorem~\ref{thm:PML2-1}. It can also be seen that for sufficiently large $N$, we can choose $\epsilon$ to be arbitrarily small and recover PML.

In order to establish~\eqref{final-bnda1}, we will show the following:
\begin{prop}
For any $0 < \epsilon < 1$ we have that:
\begin{align}
&E_{Y_2^N} \left[ \frac{\sum_{j=1}^N\ \beta(Y_j) }{(\sum_{j=1}^N \lambda(Y_j) )^2  }\bigg| Y_1=y_1\right] \le \frac{\beta(y_1) + (N-1)(1+\epsilon)}{(\lambda(y_1) + (N-1)(1-\epsilon))^2} + \frac{N \omega }{\lambda(y_1)^2} 2 e^{-(N-1)\epsilon^2/\omega^2} \label{eq:hoeff-1}
\end{align}
and in turn using~\eqref{eq:t-simp}, we have
\begin{align}
E_{Y_2^N}\left[\frac{\left(\sum_{j=1}^N\frac{q_{Y|X}(Y_j|x)}{p_Y(Y_j)}\right)}{\left(\sum_{j=1}^N\frac{p_{Y|X}(Y_j|x)}{p_Y(Y_j)}\right)} \bigg| Y_1=y_1, U_p=1\right]
\le \mu'_{y_1} (N)
\label{eq:hoeff-2}
\end{align}
$\hfill\Box$
\label{prop:hoeff}
\end{prop}
Note that the bound in~\eqref{final-bnda1} follows directly by substituting~\eqref{eq:hoeff-2} into~\eqref{eq:t-n-1}. We relegate the proof of Prop.~\ref{prop:hoeff} to Section~\ref{app:hoeff} in this document.

\section{Proof of Theorem~\ref{thm:CML} in Main Paper}
\label{app:CML}
We state an expanded version of Theorem~\ref{thm:CML} in the main paper. We assume that $k=1$ without loss of genrality.
\begin{theorem*}
Let $\Omega=\{y_1,\ldots, y_N\}$. The error probability satisfies:
\begin{equation}\begin{aligned}Pr\left(U_P \neq U_Q | U_P = 1, X=x, Z=z, \Omega\right) \le  1 - \left(1 + \frac{{p_{Y|X}(y_1|x)}}{{Q_{Y|Z}(y_1|z)}} \frac{\left(\frac{1}{N}\sum_{j=1}^N\frac{Q_{Y|Z}(y_j|z)}{p_Y(y_j)}\right)}{\left(\frac{1}{N}\sum_{j=1}^N\frac{p_{Y|X}(y_j|x)}{p_Y(y_j)}\right)}\right)^{-1}, \label{eq:condbnd1}\end{aligned}\end{equation}
and furthermore,
\begin{equation}\begin{aligned}
Pr(U_p \neq U_q | Y_1=y_1, U_p = k, X=x, Z=z)  \le  1- \left(1 + \mu_{y_1}(\bN) \frac{{p_{Y|X}(y_1|x)}}{Q_{Y|Z}(y_1|z)}\right)^{-1}.
\end{aligned}\label{eq:condbnd2}\end{equation}
where
\begin{align}
\mu_{y_1}(N)= \left(\frac{\frac{\beta(y_1)} {\bN}+1}{\frac{\lambda(y_1)}{\bN}+1}\right) + \frac{1}{\bN} \left(1+ \frac{\lambda(y_1)}{\bN}\right)K(\bN) + \frac{2\omega}{\bN}\left(1+ \frac{\lambda(y_1)}{\bN}\right)L(\bN)), \label{eq:mu-def2-n} 
\end{align}
where
$\bN=N-1$, $\beta(y_1)=\frac{Q_{Y|Z}(y_1|z)}{p_Y(y_1)}$ and $\lambda(y_1)=\frac{p_{Y|X}(y_1|x)}{p_Y(y_1)}$  
and
\begin{align}
K(\bN) =  4 \frac{(\omega-1)}{ (1+ \frac{\lambda(y_1)}{\bN})^2}\left(1+ \frac{N+1}{\bN}\omega\right)\sqrt{2+ 4\left(\frac{1+ \frac{\beta(y_1)}{\bN}}{1+ \frac{2\lambda(y_1)}{\bN}}\right)^2\left\{\left(1+ \frac{N+1}{\bN}\omega\right)^2 + \frac{(\omega-1)}{\bN}\right\}}, \label{eq:k1-bnd3}
\end{align}
\begin{align}
L(\bN)&=  \sqrt{\omega-1}\sqrt{(d_5(p_Y(\cdot)||p_{Y|X}(\cdot|X=x))-d_3^2(p_Y(\cdot)||p_{Y|X}(\cdot|x)))} + (\omega-1) d_3(p_Y(\cdot)||p_{Y|X}(\cdot|x)).  \label{eq:l1-bnd3}
\end{align}

\end{theorem*}

\begin{rm}
\label{rem:2}
Upon examining~\eqref{eq:mu-def2-n}-\eqref{eq:l1-bnd3}, and assuming that $d_5(p_Y(\cdot)|| p_{Y|X}(\cdot|x)) < \infty$, see~\eqref{eq:d3-def}, it is clear that there exists an $N_1(\epsilon)$ such that
\begin{align}
\mu(N) \le 1 + \epsilon, \quad \forall  N \ge N_1(\epsilon) \label{eq:N-def2}
\end{align}
\end{rm}

We define $\Omega = \{y_1^N\}$ and consider:
\begin{align} 
&\Pr\left(U_P \neq U_Q | U_P = k, X=x, Z=z, \Omega\right)\\
&= \Pr\left(\min_{j\neq k} \frac{S_j}{\lambda_j^q} \le \frac{S_k}{\lambda_k^q}~\bigg|~\Omega, U_p=k, X=x, Z=z \right)\\
&=\Pr\left(\min_{j\neq k} \frac{S_j}{\lambda_j^q} \le \tilde{S}_P\frac{\lambda_k^p}{\lambda_k^q}~\bigg|~\Omega, U_p=k, X=x, Z=z \right) \\
&\le \Pr\left(\min_{j\neq k} \frac{S_j -\lambda_j^p \tilde{S}_P}{\lambda_j^q} \le \tilde{S}_P\frac{\lambda_k^p}{\lambda_k^q}~\bigg|~\Omega, U_p=k, X=x, Z=z \right)\\
&=\Pr\left(\min_{j\neq k} \frac{S_j -\lambda_j^p \tilde{S}_P}{\lambda_j^q} \le \tilde{S}_P\frac{\lambda_k^p}{\lambda_k^q}~\bigg|~\Omega, U_p=k, X=x \right)\label{eq:indep}\\
&\le \frac{1}{1+ \frac{\lambda_k^q}{\lambda_k^p}} = 1-\left(1 + \frac{\lambda_k^p}{\lambda_k^q}\right)^{-1} \\
&= 1 - \left(1 + \frac{\frac{p_{Y|X}(y_k|x)}{p_{Y}(y_k) }}{\frac{Q_{Y|Z}(y_k|z)}{p_Y(y_k)}} \frac{\left(\sum_{j=1}^N\frac{Q_{Y|Z}(y_j|z)}{p_Y(y_j)}\right)}{\left(\sum_{j=1}^N\frac{p_{Y|X}(y_j|x)}{p_Y(y_j)}\right)}\right)^{-1}\\
&= 1 - \left(1 + \frac{{p_{Y|X}(y_k|x)}}{{Q_{Y|Z}(y_k|z)}} \frac{\left(\frac{1}{N}\sum_{j=1}^N\frac{Q_{Y|Z}(y_j|z)}{p_Y(y_j)}\right)}{\left(\frac{1}{N}\sum_{j=1}^N\frac{p_{Y|X}(y_j|x)}{p_Y(y_j)}\right)}\right)^{-1}
\end{align}
where~\eqref{eq:indep} follows from the Markov condition~\eqref{eq:markov} and the subsequent stepsilon follows the the analysis done previously.

We will assume that $k=1$ without loss of generality. Taking expectation with respect to $\Omega_2 = \{Y_2^N\}$ we have that:

\begin{align}
&\Pr\left(U_P \neq U_Q | U_P = 1, X=x, Z=z, Y=y\right)\\
&=E_{Y_2^N}\left[1 - \left(1 + \frac{{p_{Y|X}(Y_1|x)}}{{Q_{Y|Z}(Y_1|z)}} \frac{\left(\frac{1}{N}\sum_{j=1}^N\frac{Q_{Y|Z}(Y_j|z)}{p_Y(Y_j)}\right)}{\left(\frac{1}{N}\sum_{j=1}^N\frac{p_{Y|X}(Y_j|x)}{p_Y(Y_j)}\right)}\right)^{-1} \Bigg| 
U_P = 1, X=x, Z=z, Y=y \right] \\
&\le 1 - \left(1 + \frac{{p_{Y|X}(Y_1|x)}}{{Q_{Y|Z}(Y_1|z)}} \cdot E_{Y_2^N | \{U_p, X, Y, Z\}}\left[\frac{\left(\frac{1}{N}\sum_{j=1}^N\frac{Q_{Y|Z}(Y_j|z)}{p_Y(Y_j)}\right)}{\left(\frac{1}{N}\sum_{j=1}^N\frac{p_{Y|X}(Y_j|x)}{p_Y(Y_j)}\right)}\Bigg|  
U_P = 1, X=x, Z=z, Y=y\right]\right)^{-1}  \\
&= 1 - \left(1 + \frac{{p_{Y|X}(Y_1|x)}}{{Q_{Y|Z}(Y_1|z)}} \cdot E_{Y_2^N | \{U_p, X, Y\}}\left[\frac{\left(\frac{1}{N}\sum_{j=1}^N\frac{Q_{Y|Z}(Y_j|z)}{p_Y(Y_j)}\right)}{\left(\frac{1}{N}\sum_{j=1}^N\frac{p_{Y|X}(Y_j|x)}{p_Y(Y_j)}\right)}\Bigg|  
U_P = 1, X=x, Y=y\right]\right)^{-1} 
\end{align}

By defining $\beta(Y_j) = \frac{Q_{Y|Z}(Y_j|z)}{p_Y(Y_j)}$ and $\lambda(Y_j) = \frac{p_{Y|X}(Y_j|x)}{p_Y(Y_j)}$ and leading to the same sequence of stepsilon that lead to~\eqref{eq:t-n-2} in Prop.~\ref{prop:lem2} in this document, we can complete the proof.

\section{Proof of Prop.~\ref{prop:SI} in the Main document}
\label{app:SI}

\begin{prop*}
\begin{align}
&\Pr(U_p \neq U_q) \le E_{V,W,T}\left[1- \left(1+ (1+\epsilon)L^{-1} \textcolor{black}{2^{i(W;V|T)}} \right)^{-1} \right]
\label{eq:err-bnd}
\end{align}
where \textcolor{black}{$i_{W,V|T}(w;v|t) = \log \frac{p_{W|V}(w|v)}{p_{W|T}(w|t)}$} is the conditional information  density. 
\end{prop*}

By application of the conditional IML in Theorem~\ref{thm:CML}  in the main document, and assuming that $N$ is sufficiently large, as stated in~\eqref{eq:N1-def}, it follows that:
\begin{align}
&\Pr(U_p\neq U_q | U_p=k, X=v, Y_k = (w,l), Z=(t,l)) \notag \\
&\le 1- \left(1+ (1+\epsilon)\frac{p_{Y|X}(w, l | v)}{Q_{Y|Z}(w,l|t,l)}\right)^{-1}\\
&=1- \left(1+ (1+\epsilon)\frac{p_{W|V}(w | v)p_{{\mathsf l}}(l)}{p_{W|T}(w|t) }\right)^{-1}\\
&= 1- \left(1+ (1+\epsilon)L^{-1} \textcolor{black}{2^{i_{W,V|T}(w;v|t)}} \right)^{-1}
\end{align}
where \textcolor{black}{$i_{W,V|T}(w;v|t) = \log \frac{p_{W|V}(w|v)}{p_{W|T}(w|t)}$} is the conditional information  density. It thus follows that
\begin{align}
&\Pr(U_p \neq U_q) \le E_{V,W,T}\left[1- \left(1+ (1+\epsilon)L^{-1} 2^{i(W;T|V)} \right)^{-1} \right].
\label{eq:err-bnd}
\end{align}

\section{Bound on the Probability of Excess Distortion}
\label{app:excess_err}
We introduce and prove the following bound on the probability of excess distortion.
\begin{prop*}
For a large enough $N$, the probability of excess distortion $\Pr({d(V,\hat{V}) > D})$, where $\hat{V}=\tilde{g}(W_{U_q}, T)$ is the reconstruction output by the decoder, can be bound as follow.
\begin{align}
&P_e=\Pr({d(V,\hat{V}) > D}) \le E_{W,V,T}\left\{1 {-} \mathbb{I}(d(V, \tilde{g}(W,T)) {\leq} D) \left(1+ (1+\epsilon)L^{-1} 2^{i(W;T|V)} \right)^{-1} \right\}
\label{eq:excess_err}
\end{align}
where $i_{W,T|V}(w;t|v) = \log \frac{p_{W|V}(w|v)}{p_{W|T}(w|t)}$ is the conditional information  density, $d(.,.)$ is the distortion measure. 
\end{prop*}

\proofname: We recall that for sufficiently large $N$, $p_{W_{U_p}|V}$ can be arbitrarily close to $p_{W|V}$. As such:
\begin{align}
P_e  &\overset{}{=} 1 - \Pr({d(V,\hat{V}) \leq D}) 
\\&\overset{(a)}{\le} 1 - \Pr({d(V,\hat{V}) \leq D}, W_{U_p} = W_{U_q}) 
\\ &= 1 - \Pr({d(V,\tilde{g}(W_{U_q}, T)) \leq D}, W_{U_p} = W_{U_q}) 
\\ &\overset{(b)}= 1 - \Pr({d(V,\tilde{g}(W_{U_p}, T)) \leq D}, W_{U_p} = W_{U_q}) 
\\ &\overset{(c)}= 1 - E_{W,V,T}\{\Pr({d(V,\tilde{g}(W_{U_p}, T)) \leq D}, W_{U_p} {=} W_{U_q}| W_{U_p}{=}W,V,T)  \}\\ &\overset{(d)}= 1 - E_{W,V,T}\{\Pr({d(V,\tilde{g}(W, T)) \leq D}|W_{U_p}{=}W_{U_q}{=}W,V,T) \Pr\{W_{U_p} = W_{U_q}| W_{U_p}=W,V,T)\}  \} 
\\ &\overset{(e)}= 1 - E_{W,V,T}\{\mathbb{I}(d(V, \tilde{g}(W,T)) {\leq} D) \Pr\{W_{U_p} {=} W_{U_q}| W_{U_p}{=}W,V,T)\}  \} 
 \\ &\overset{(f)}{\le} E_{W,V,T}\{1 {-} \mathbb{I}(d(V, \tilde{g}(W,T)) {\leq} D) \Pr \{U_p{=}U_q | W_{U_p}{=}W,V,T \}\}
 \\ &\overset{(g)}{\le} E_{W,V,T}\left\{1 {-} \mathbb{I}(d(V, \tilde{g}(W,T)) {\leq} D) \left(1+ (1+\epsilon)L^{-1} \textcolor{black}{2^{i(W;V|T)}} \right)^{-1} \right\}
\end{align}
where (a) is by marginalization, (b) is by $W_{U_p}=W_{U_q}$, (c) is by the law of iterated expectation , (d) is by chain rule for joint probability, (e) the event ${d(V,\tilde{g}(W, T)) \leq D}$ is a function of the conditioned random variables and therefore the probability $\Pr({d(V,\tilde{g}(W, T)) \leq D}|W_{U_p}{=}W_{U_q}{=}W,V,T)$ becomes the indicator function  $\mathbb{I}(d(V, \tilde{g}(W,T)) {\leq} D)$, (f) the event $\{W_{U_p}=W_{U_q}\} = \{U_p=U_q\} \cup \{U_p\neq U_q \text{ and } W_{U_p} = W_{U_q}\}$. For (g), we note that: 
\begin{align}
    \Pr \{U_p{=}U_q | W_{U_p}{=}w,v,t \} &= E_{\mathsf{k},\mathsf{l}} [\Pr \{U_p{=}U_q| W_{U_p}{=}w,v,t, U_p= \mathsf{k}, l_{U_p}= \mathsf{l} \} ] \\
    &= E_{\mathsf{k},\mathsf{l}} [\Pr \{U_p{=}U_q|  U_p= \mathsf{k}, Y_\mathsf{k} = (W_{U_p}, l_{U_p}) = (w,\mathsf{l}), Z=(t,\mathsf{l}) \} ] \\
    &= E_{\mathsf{k},\mathsf{l}} [1 - \Pr \{U_p{\neq}U_q|  U_p= \mathsf{k}, Y_\mathsf{k} = (W_{U_p}, l_{U_p}) = (w,\mathsf{l}), Z=(t,\mathsf{l}) \} ] \\
    &\geq E_{\mathsf{k},\mathsf{l}}\left[\left(1+ (1+\epsilon)L^{-1} \textcolor{black}{2^{i(w;v|t)}} \right)^{-1}\right] \\
    &= \left(1+ (1+\epsilon)L^{-1} \textcolor{black}{2^{i(w;v|t)}} \right)^{-1}
\end{align}
where the first line (equality) is by the law of iterated expectation over  $(U_{p}=\mathsf{k}, l_{U_p}=\mathsf{l})$; in the second line (equality) we rewrite it in the form of Thm.~\ref{thm:CML} (conditional importance matching lemma); the third line (equality) is because the two event $\{U_p{=}U_q\}$ and $\{U_p{\neq}U_q\}$ are complementary; the fourth line (inequality) is by applying Prop.~\ref{prop:SI}; the final line (equality) is because the term inside the bracket does not depend on $\mathsf{k}$ and $\mathsf{l}$. Following this, we arrive at (g) above.

\section{Rate Distortion Analysis for Feedback Scheme}\label{feedback_rate}
\textcolor{black}{We present the rate-distortion analysis for lossy compression with side information with feedback. Recall that in the feedback scheme, after sending the LSB of size $\log_2(L)$ of $U_p$ to the decoder, the encoder will outputs an acknowledgement bit of $1$ if the feedback signal indicates that the decoder outputs the same index, i.e. $U_q=U_p$. On the other hand, if the signal indicates $U_q \neq U_p$, the encoder outputs the MSB of its selection $U_p$ to the decoder. This means that the encoder message will be $\log_2(L)+1$ if the decoder outputs the correct index in the first try and $\log_2(N)$ otherwise.}

Assuming perfect feedback, the output distribution between the encoder and decoder is the same, i.e. $\Tilde{p}_{W|V}$, since the decoder always recover the correct $W_{U_p}$. Here, we note that the  output distribution $\tilde{p}_{W|V}$ correspond to the number of samples $N$, and achieve the expected distortion $D=E[d(V,\hat{V})]$. Since each $L$ yields different matching probability, the rate $R(D)$ in this case is:
\begin{equation}
    R(D) = \min_L R(D,L)
\end{equation}
where:
\begin{align}
    R(D,L) &= [\log_2(L)+1][1 - \Pr(U_p \neq U_q)] + \log_2(N) \Pr(U_p \neq U_q) \\
     &= \log_2(L) + 1 + (\log_2N - \log_2L -1)\Pr(U_p \neq U_q) \\
     &\leq \log_2(2L)  + (\log_2\frac{N}{2L})E_{V,W,T}\left[1- \left(1+ (1+\epsilon)L^{-1} 2^{i(W;V|T)} \right)^{-1} \right] \label{feedback_upbound}
\end{align}


\section{Proof of Equation~\eqref{eq:efrl} in Section~\ref{sec:part1} in this document}
\label{app:efrl}
We will show that:
\begin{align}
E[\log K] \le D(\nu(\cdot) || \mu(\cdot)) + \underbrace{\frac{\log e}{e} +1}_{=\delta}. \label{eq:efrl-1}
\end{align}
Our proof directly follows~\citep[Chapter 4]{li2017information} with a change in notation. We note that in~\cite{li2017information},
$\nu(\cdot)=p_{Y|X}(\cdot|x)$ and $\mu= p_Y(\cdot)$. However $p_{Y|X}(\cdot|x)$ and $p_Y(\cdot)$ are treated as arbitrary distributions in the proof and  fact that $p_Y(\cdot)$ is related to $p_{Y|X}(\cdot|x)$ through marginalization is not required.  We thus provide the proof using the notation in the present paper.

Without loss of generality we assume $\Omega = \{1,2,\ldots, N\}$.

We introduce:
\begin{align}
I = \arg\min_{y\in \Omega} \frac{S_y}{\nu(y)} \qquad \Theta = \min_{y\in \Omega} \frac{S_y}{\nu(y)} 
\end{align}

It follows from the exponential-race property~\citep{maddison2014sampling} that  (1) $Y = I \sim \nu(\cdot)$  (2) $\Theta \sim {\mathrm{Exp}}(1)$ and (3) (1) $Y$ and $\Theta$ are mutually independent of each other.

Note note with $\phi_{y} = \frac{S_y}{\mu(y)}$,  and since $K$ is the index of $\phi_Y$ in $\{\phi_y\}_{y\in {\mathcal Y}}$ sorted in ascending order:
\begin{align}
K = \left\{y' : \phi_{y'} < \phi_y \right\} + 1
\end{align}

\begin{align}
E[\log K] &= \sum_{y\in \Omega}\nu(y) E[\log K | Y=y] \\
&=\sum_{y\in \Omega}\nu(y) E_\Theta [\log K | Y=y, \Theta=\theta] \\
&=\sum_{y\in \Omega}\nu(y) \int_{\theta=0}^\infty  e^{-\theta}E[\log K|Y=y,\Theta=\theta] d\theta. \label{eq:tm1-2}
\end{align}
Now consider:
\begin{align}
E[\log K|Y=y,\Theta=\theta] &= E\left[\log \big|y'\neq y: \phi_{y'} < \phi_y \big| +1 | Y=y, \Theta=\theta\right]  \\
&\le \log E\left[\big|y'\neq y: \phi_{y'} < \phi_y \big| +1 | Y=y, \Theta=\theta\right]\\
&= \log E\left[\Big|y' \neq y: \phi_{y'} < \theta \frac{\nu(y)}{\mu(y)} \Big| +1 \Big| Y=y, \Theta=\theta\right]\label{eq:subt}\\
&=\log E\left[\Big|y'\neq y : \phi_{y'}  < \theta \frac{\nu(y)}{\mu(y)} \Big| +1 \Big| Y=y, \frac{S_{y'}}{\nu(y')} \ge \theta, \forall y' \in \Omega \right]\label{eq:condn}\\
&=\log E\left[\sum_{y' \neq y} {\mathbb I}\left\{\phi_{y'}  < \theta \frac{\nu(y)}{\mu(y)} \right\} +1 \Bigg| Y=y, \frac{S_{y'}}{\nu(y')} \ge \theta, \forall y' \in \Omega \right]\\
&=\log\left(\sum_{y' \neq y} \Pr\left(\phi_{y'}\le \theta \frac{\nu(y)}{\mu(y)} \Big| Y=y, \frac{S_{y'}}{\nu(y')} \ge \theta, \forall y' \in \Omega \right) + 1\right)\\
&=\log\left(\sum_{y' \neq y} \Pr\left(S_{y'}\le \theta \frac{\nu(y)}{\mu(y)}\mu(y') \Big| Y=y, \frac{S_{y'}}{\nu(y')} \ge \theta, \forall y' \in \Omega \right) + 1\right) \label{eq:tm2-2a}\\
&=\log\left(\sum_{y' \neq y} \Pr\left(S_{y'}\le \theta \frac{\nu(y)}{\mu(y)}\mu(y') \Big|  \frac{S_{y'}}{\nu(y')} \ge \theta \right) + 1\right)  \label{eq:tm2-2}
\end{align}
where~\eqref{eq:subt} uses the fact that ${\phi_y}{\mu(y)} = \theta \nu(y) = S_y$ and~\eqref{eq:condn} follows form the fact that the event $\{Y=y, \Theta =\theta\}$ is equivalent to the event $\{Y=y, \frac{S_{y'}}{\mu(y')} \ge \theta, \forall y' \in \Omega\}$ by definition of $\Theta$. Eq.~\eqref{eq:tm2-2} follows from the fact that the distribution of $S_{y'}$ only depends on the event  $\{\frac{S_{y'}}{\nu(y')} \ge \theta\}$ given the conditioning in~\eqref{eq:tm2-2a}.

Next we define $r(y) = \nu(y)/\mu(y)$ for each $y\in \Omega$ and use the fact that $S_{y'} \sim {\mathrm{Exp}}(1)$ so that:
\begin{align}
\Pr\left(S_{y'}\le \theta \frac{\nu(y)}{\mu(y)}\mu(y') \Big|  \frac{S_{y'}}{\nu(y')} \ge \theta \right) &\le {\mathbb I}(r(y')\le r(y))\left(1- \exp(-\theta \mu(y')(r(y)-r(y')))\right)\\
&\le {\mathbb I}(r(y')\le r(y)) \theta \mu(y')(r(y)-r(y'))\\
&\le \theta \mu(y') r(y).
\end{align}
Thus using~\eqref{eq:tm2-2}, we have:
\begin{align}
\log\left(\sum_{y' \neq y} \Pr\left(S_{y'}\le \theta \frac{\nu(y)}{\mu(y)}\mu(y') \Big|  \frac{S_{y'}}{\nu(y')} \ge \theta \right) + 1\right)&\le \log(\theta r(y)\sum_{y'} \mu(y')+1) \\
&= \log(\theta r(y)+1)
\end{align}
Using~\eqref{eq:tm1-2} we have that:
\begin{align}
E[\log K] &\le \sum_{y\in \Omega} \nu(y)\int_{\theta \ge 0} e^{-\theta} \log(\theta r(y)+1) d\theta \\
&\le \sum_{y\in \Omega} \nu(y)\log(r(y)+1) \label{eq:jn-2}\\
&=\sum_{y\in \Omega: r(y)\ge 1} \nu(y)\log(r(y)+1) + \sum_{y\in \Omega: r(y)\le 1} \nu(y)\log(r(y)+1) \\
&\le \sum_{y\in \Omega: r(y)\ge 1} \nu(y)\log(r(y)+1) +  \sum_{y\in \Omega: r(y)\le 1} \nu(y) \\
&\le \sum_{y\in \Omega: r(y)\ge 1} \nu(y)(\log r(y)+1) +  \sum_{y\in \Omega: r(y)\le 1} \nu(y) \\
&= \sum_{y\in \Omega: r(y)\ge 1} \nu(y)\log r(y) + 1 \\
&= \sum_{y\in \Omega} \nu(y) \log r(y) - \sum_{y\in \Omega: r(y)<1} \nu(y)\log r(y) + 1\\
&=D(\nu(\cdot)||\mu(\cdot)) - \sum_{y\in \Omega: r(y)<1} \nu(y)\log r(y) + 1 \\
&\le D(\nu(\cdot)||\mu(\cdot)) +\frac{\log e}{e} + 1
\end{align}
where we use Jensen's inequality in~\eqref{eq:jn-2} and the following inequality in~\citep[Appendix A]{harsha2007communication}:
For any two distributions $P(\cdot)$ and $Q(\cdot)$ on $\cX$ and any $\cX' \subset \cX$
\begin{align}
-\sum_{x \in \cX'} P(x) \log\frac{P(x)}{Q(x)} \le \frac{\log e}{e}.
\end{align}
This completes the proof.

\section{Proof of Prop.~\ref{prop:1} in Section~\ref{sec:thm1-bnd} in this document}
\label{sec:prop1-proof}
For simplicity in notation we will use $p_i = p_Y(Y_i)$ and $q_i = p_Y(Y_i|X=x)$. Thus the objective we need to simplify reduces to
\begin{align}
&E_{Y_1^N}\left[\sum_{i=1}^N \frac{\frac{q_i}{p_i}\log\left(N\frac{\frac{q_i}{p_i}}{\sum_{j=1}^N \frac{q_j}{p_j}}\right)}{\sum_{j=1}^N \frac{q_j}{p_j}}\right] \notag\\ &= E_{Y_1^N}\left[\sum_{i=1}^N \frac{\frac{q_i}{p_i}\log\left(\frac{N}{\sum_{j=1}^N \frac{q_j}{p_j}}\right)}{\sum_{j=1}^N \frac{q_j}{p_j}}\right] + E_{Y_1^N}\left[\sum_{i=1}^N \frac{\frac{q_i}{p_i}\log\left(\frac{q_i}{p_i}\right)}{\sum_{j=1}^N \frac{q_j}{p_j}}\right]  \\
&=E_{Y_1^N}\left[\log\left(\frac{N}{\sum_{j=1}^N \frac{q_j}{p_j}}\right)\right] + E_{Y_1^N}\left[\sum_{i=1}^N \frac{\frac{q_i}{p_i}\log\left(\frac{q_i}{p_i}\right)}{\sum_{j=1}^N \frac{q_j}{p_j}}\right] \label{eq:decomp}
\end{align}
We will establish an upper bound on each of the two terms in~\eqref{eq:decomp} separately. Consider the first term:
\begin{align}
E_{Y_1^N}\left[\log\left(\frac{N}{\sum_{j=1}^N \frac{q_j}{p_j}}\right)\right] \le \log \left(E_{Y_1^N}\left[\frac{N}{\sum_{j=1}^N \frac{q_j}{p_j}}\right]\right) \label{eq:jen1}
\end{align}
which follows from Jensen's inequality. 
Now consider:
\begin{align}
&E_{Y_1^N}\left[\frac{N}{\sum_{i=1}^N \frac{q_i}{p_i}}\right] - 1 = E_{Y_1^N}\left[\frac{1}{\frac{1}{N}\sum_{i=1}^N \frac{q_i}{p_i}}-1\right]  \\
&=E_{Y_1^N}\left[\frac{1}{\frac{1}{N}\sum_{i=1}^N \frac{q_i}{p_i}}\left(1- {\frac{1}{N}\sum_{i=1}^N \frac{q_i}{p_i}}\right)\right] \\
&=E_{Y_1^N}\left[\left(\frac{1}{\frac{1}{N}\sum_{i=1}^N \frac{q_i}{p_i}}-1\right)\left(1- {\frac{1}{N}\sum_{i=1}^N \frac{q_i}{p_i}}\right)\right] \label{eq:zero-mean}\\
&=E_{Y_1^N}\left[\left(\frac{1}{\frac{1}{N}\sum_{i=1}^N \frac{q_i}{p_i}}\right)\left(1- {\frac{1}{N}\sum_{i=1}^N \frac{q_i}{p_i}}\right)^2\right] \\
\end{align}
where~\eqref{eq:zero-mean} follows from the fact that $E_{Y_1^N}\left[1- {\frac{1}{N}\sum_{i=1}^N \frac{q_i}{p_i}}\right]=0$.
Next observe that:
\begin{align}&E_{Y_1^N}\left[\left(\frac{1}{\frac{1}{N}\sum_{i=1}^N \frac{q_i}{p_i}}\right)\left(1- {\frac{1}{N}\sum_{i=1}^N \frac{q_i}{p_i}}\right)^2\right] \notag \\
&= E_{Y_1^N}\left[\frac{1}{\frac{1}{N}\sum_{i=1}^N \frac{q_i}{p_i}}\left(1-\frac{1}{N}\sum_{i=1}^N \frac{q_i}{p_i}\right)^2{\mathbb I}\left(\frac{1}{N}\sum_{i=1}^N \frac{q_i}{p_i} \ge \frac{1}{2}\right)\right] \notag\\
&+ E_{Y_1^N}\left[\frac{1}{\frac{1}{N}\sum_{i=1}^N \frac{q_i}{p_i}}\left(1-\frac{1}{N}\sum_{i=1}^N \frac{q_i}{p_i}\right)^2{\mathbb I}\left(\frac{1}{N}\sum_{i=1}^N \frac{q_i}{p_i} < \frac{1}{2}\right)\right]
\end{align}
and by using the triangular inequality, we have that:
\begin{align}
 &\bigg| E_{Y_1^N}\left[\frac{N}{\sum_{i=1}^N \frac{q_i}{p_i}}\right] - 1 \bigg| \notag\\
 &\le 
    \Bigg|\E_{Y_1^N}\left[\frac{1}{\frac{1}{N}\sum_{i=1}^N \frac{q_i}{p_i}}\left(1-\frac{1}{N}\sum_{i=1}^N \frac{q_i}{p_i}\right)^2{\mathbb I}\left(\frac{1}{N}\sum_{i=1}^N \frac{q_i}{p_i} \ge \frac{1}{2}\right) \right]\Bigg| \notag\\
&+ \Bigg|\E_{Y_1^N}\left[\frac{1}{\frac{1}{N}\sum_{i=1}^N \frac{q_i}{p_i}}\left(1-\frac{1}{N}\sum_{i=1}^N \frac{q_i}{p_i}\right)^2{\mathbb I}\left(\frac{1}{N}\sum_{i=1}^N \frac{q_i}{p_i} < \frac{1}{2}\right)\right]\Bigg| \label{eq:t1-terms}
\end{align}
We consider each of the two terms in~\eqref{eq:t1-terms} separately. For the first term:
\begin{align}
&\E_{Y_1^N}\left[\frac{1}{\frac{1}{N}\sum_{i=1}^N \frac{q_i}{p_i}}\left(1-\frac{1}{N}\sum_{i=1}^N \frac{q_i}{p_i}\right)^2{\mathbb I}\left(\frac{1}{N}\sum_{i=1}^N \frac{q_i}{p_i} \ge \frac{1}{2}\right) \right]  \notag\\ &\le 2\E_{Y_1^N}\left[\left(1-\frac{1}{N}\sum_{i=1}^N \frac{q_i}{p_i}\right)^2{\mathbb I}\left(\frac{1}{N}\sum_{i=1}^N \frac{q_i}{p_i} \ge \frac{1}{2}\right) \right] \\
&\le
2\E_{Y_1^N}\left[\left(1-\frac{1}{N}\sum_{i=1}^N \frac{q_i}{p_i}\right)^2 \right] \\
&= \frac{2}{N}\left(\E_{Y\sim p(Y)}\left[\frac{q^2(Y)}{p^2(Y)}\right]-1\right) = \frac{2}{N}\left(d_2(q||p)-1\right)
\end{align}
where $d_2(q||p) = E_{q}\left[\frac{q}{p}\right]$.  We next consider the second term in~\eqref{eq:t1-terms}.
\begin{align}
&\E_{Y_1^N}\left[\frac{1}{\frac{1}{N}\sum_{i=1}^N \frac{q_i}{p_i}}\left(1-\frac{1}{N}\sum_{i=1}^N \frac{q_i}{p_i}\right)^2{\mathbb I}\left(\frac{1}{N}\sum_{i=1}^N \frac{q_i}{p_i} < \frac{1}{2}\right)\right] \notag\\
&\le
\E_{Y_1^N}\left[\frac{1}{\frac{1}{N}\sum_{i=1}^N \frac{q_i}{p_i}}{\mathbb I}\left(\frac{1}{N}\sum_{i=1}^N \frac{q_i}{p_i} < \frac{1}{2}\right)\right] 
\end{align}
We will next use the following key inequality established at the end of this section:
\begin{align}
\frac{1}{\frac{1}{N} \sum_{i=1}^N \frac{q_i}{p_i}} \le \frac{1}{N}\sum_{i=1}^N \frac{p_i}{q_i} \label{eq:inv_ineq}
\end{align}
Also using $\E_{p}\left[\frac{p}{q}\right] = d_2(p||q)$, we have 
\begin{align}
&\E_{Y_1^N}\left[\frac{1}{\frac{1}{N}\sum_{i=1}^N \frac{q_i}{p_i}}{\mathbb I}\left(\frac{1}{N}\sum_{i=1}^N \frac{q_i}{p_i} < \frac{1}{2}\right)\right] \le \E_{Y_1^N}\left[\left(\frac{1}{N}\sum_{i=1}^N \frac{p_i}{q_i}\right){\mathbb I}\left(\frac{1}{N}\sum_{i=1}^N \frac{q_i}{p_i} < \frac{1}{2}\right)\right] \\
&=\E_{Y_1^N}\left[\left(\frac{1}{N}\sum_{i=1}^N \frac{p_i}{q_i} - d_2(p||q)\right){\mathbb I}\left(\frac{1}{N}\sum_{i=1}^N \frac{q_i}{p_i} < \frac{1}{2}\right)\right] + d_2(p||q)\Pr\left(\frac{1}{N}\sum_{i=1}^N \frac{q_i}{p_i} <\frac{1}{2}\right)\\
&\le \sqrt{\E_{Y_1^N}\left[\left(\frac{1}{N}\sum_{i=1}^N \frac{p_i}{q_i} - d_2(p||q)\right)^2\right]}\sqrt{\E_{Y_1^N}\left[
{\mathbb I}\left(\frac{1}{N}\sum_{i=1}^N \frac{q_i}{p_i} < \frac{1}{2}\right)\right]}+ d_2(p||q)\Pr\left(\frac{1}{N}\sum_{i=1}^N \frac{q_i}{p_i} <\frac{1}{2}\right)\label{eq:cheb1}\\
&=\sqrt{\E_{Y_1^N}\left[\left(\frac{1}{N}\sum_{i=1}^N \frac{p_i}{q_i} - d_2(p||q)\right)^2\right]}\sqrt{\Pr\left(\frac{1}{N}\sum_{i=1}^N \frac{q_i}{p_i} < \frac{1}{2}\right)}+ d_2(p||q)\Pr\left(\frac{1}{N}\sum_{i=1}^N \frac{q_i}{p_i} <\frac{1}{2}\right)
\end{align}where~\eqref{eq:cheb1} follows from the Cauchy-Schwartz inequality: $E[X\cdot Y]\le \sqrt{E[X^2]E[Y^2]}$.
Using Chebyshev's Inequality, we have the following:
\begin{align}
\Pr\left(\frac{1}{N}\sum_{i=1}^N \frac{q_i}{p_i} <\frac{1}{2}\right) &\le     
\Pr\left(\Bigg|\frac{1}{N}\sum_{i=1}^N \frac{q_i}{p_i}-1 \Bigg|\ge \frac{1}{2}\right)\\
&\le \frac{\frac{1}{N}E_{p}\left[\left(\frac{q}{p}-1\right)^2\right]}{1/4}\\
&= \frac{4}{N}\left(d_2(q||p)-1\right).\label{eq:chebyshev}
\end{align}
Finally note that:
\begin{align}
\E_{Y_1^N}\left[\left(\frac{1}{N}\sum_{i=1}^N \frac{p_i}{q_i} - d_2(p||q)\right)^2\right] = \frac{1}{N}\left(\E_{p}\left[\frac{p^2}{q^2}\right]-d_2^2(p||q)\right)   
=\frac{1}{N}\left(d_3(p||q)-d_2^2(p||q)\right)   
\end{align}
where $d_3(p||q) = E_{p}\left[\frac{p^2}{q^2}\right]$ as in~\eqref{eq:d3-def}. It follows that:
\begin{align}
&\E_{Y_1^N}\left[\frac{1}{\frac{1}{N}\sum_{i=1}^N \frac{q_i}{p_i}}{\mathbb I}\left(\frac{1}{N}\sum_{i=1}^N \frac{q_i}{p_i} < \frac{1}{2}\right)\right]\notag\\ &\le \frac{2}{N}\left((d_3(p||q)-d_2^2(p||q))(d_2(q||p)-1)\right)^{\frac{1}{2}}+\frac{4}{N}\left(d_2(q||p)-1\right)d_2(p||q)\label{eq:less_bnd}
\end{align}
Collecting all the terms we have that:
\begin{align}
 &E_{Y_1^N}\left[\frac{N}{\sum_{i=1}^N \frac{q_i}{p_i}}\right]  \le 1+ \frac{2}{N} \left(d_2(q||p)-1\right) \notag\\&\qquad+ \frac{2}{N} \left((d_3(p||q)-d_2^2(p||q))(d_2(q||p)-1)\right)^{\frac{1}{2}}+ \frac{4}{N}\left(d_2(q||p)-1\right)d_2(p||q) \label{eq:t1-bnd}
\end{align}

Finally by using the fact that $d_2(q||p) \le \omega$ and defining
\begin{align}
\alpha(p,q) = 2(\omega-1) + 2\sqrt{\omega-1}(d_3(p||q)-d_2^2(p||q))^{\frac{1}{2}} + 4\omega \cdot d_2(p||q) \label{eq:alpha}
\end{align}
we have that:
\begin{align}
\log E_{Y_1^N}\left[\frac{N}{\sum_{i=1}^N \frac{q_i}{p_i}}\right]  \le \log\left(1+ \frac{\alpha(p,q)}{N}\right) \le \frac{\alpha(p,q)}{N} \label{eq:t1-finalbnd}
\end{align}

We will now consider the second term in~\eqref{eq:decomp}. Consider the following:
\begin{align}
&E_{Y_1^N}\left[\sum_{i=1}^N \frac{\frac{q_i}{p_i}\log\left(\frac{q_i}{p_i}\right)}{\sum_{j=1}^N \frac{q_j}{p_j}}-\frac{\sum_{i=1}^N \frac{q_i}{p_i}\log \frac{q_i}{p_i}}{N}\right] = E_{Y_1^N}\left[\frac{\frac{1}{N}\sum_{i=1}^N \frac{q_i}{p_i}\log\frac{q_i}{p_i}}{\frac{1}{N}\sum_{i=1}^N \frac{q_i}{p_i}}\left(1-\frac{\sum_{i=1}^N \frac{q_i}{p_i}}{N}\right)\right]
\end{align}

Now we consider the following:
\begin{align}
&\E\left[\left(\frac{\frac{1}{N}\sum_{i=1}^N \frac{q_i}{p_i}\log\frac{q_i}{p_i}}{\frac{1}{N}\sum_{i=1}^N \frac{q_i}{p_i}}\right)\left(1-\frac{\sum_{i=1}^N \frac{q_i}{p_i}}{N}\right)\right]  \notag\\
&=\E\left[\left(\frac{\frac{1}{N}\sum_{i=1}^N \frac{q_i}{p_i}\log\frac{q_i}{p_i}}{\frac{1}{N}\sum_{i=1}^N \frac{q_i}{p_i}}-D(q||p)\right)\left(1-\frac{\sum_{i=1}^N \frac{q_i}{p_i}}{N}\right)\right]\label{eq:zero_mean}\\
&=\E\left[\left(\frac{\frac{1}{N}\sum_{i=1}^N \frac{q_i}{p_i}\log\frac{q_i}{p_i} - D(q||p) \frac{1}{N}\sum_{i=1}^N \frac{q_i}{p_i}}{\frac{1}{N}\sum_{i=1}^N \frac{q_i}{p_i}}\right)\left(1-\frac{\sum_{i=1}^N \frac{q_i}{p_i}}{N}\right)\right]\\
&=\E\left[\left(\frac{\frac{1}{N}\sum_{i=1}^N \frac{q_i}{p_i}\log\frac{q_i}{p_i} - D(q||p) \frac{1}{N}\sum_{i=1}^N \frac{q_i}{p_i}}{\frac{1}{N}\sum_{i=1}^N \frac{q_i}{p_i}}\right)\left(1-\frac{\sum_{i=1}^N \frac{q_i}{p_i}}{N}\right){\mathbb I}\left(\frac{1}{N}\sum_{i=1}^N \frac{q_i}{p_i} \ge \frac{1}{2}\right)\right]\notag\\
&\quad + \E\left[\left(\frac{\frac{1}{N}\sum_{i=1}^N \frac{q_i}{p_i}\log\frac{q_i}{p_i} - D(q||p) \frac{1}{N}\sum_{i=1}^N \frac{q_i}{p_i}}{\frac{1}{N}\sum_{i=1}^N \frac{q_i}{p_i}}\right)\left(1-\frac{\sum_{i=1}^N \frac{q_i}{p_i}}{N}\right){\mathbb I}\left(\frac{1}{N}\sum_{i=1}^N \frac{q_i}{p_i} < \frac{1}{2}\right)\right] \label{eq:2-terms}
\end{align}
Here~\eqref{eq:zero_mean} follows from the fact that $E\left(1-\frac{\sum_{i=1}^N \frac{q_i}{p_i}}{N}\right)=0$. Using triangular inequality we have that:
\begin{align}
&\Bigg|E_{Y_1^N}\left[\sum_{i=1}^N \frac{\frac{q_i}{p_i}\log\left(\frac{q_i}{p_i}\right)}{\sum_{j=1}^N \frac{q_j}{p_j}}-\frac{\sum_{i=1}^N \frac{q_i}{p_i}\log \frac{q_i}{p_i}}{N}\right]\Bigg| \le \notag\\
&\quad \Bigg| \E\left[\left(\frac{\frac{1}{N}\sum_{i=1}^N \frac{q_i}{p_i}\log\frac{q_i}{p_i} - D(q||p) \frac{1}{N}\sum_{i=1}^N \frac{q_i}{p_i}}{\frac{1}{N}\sum_{i=1}^N \frac{q_i}{p_i}}\right)\left(1-\frac{\sum_{i=1}^N \frac{q_i}{p_i}}{N}\right){\mathbb I}\left(\frac{1}{N}\sum_{i=1}^N \frac{q_i}{p_i} \ge \frac{1}{2}\right)\right]\Bigg| \notag\\
&\qquad + \Bigg|\E\left[\left(\frac{\frac{1}{N}\sum_{i=1}^N \frac{q_i}{p_i}\log\frac{q_i}{p_i} - D(q||p) \frac{1}{N}\sum_{i=1}^N \frac{q_i}{p_i}}{\frac{1}{N}\sum_{i=1}^N \frac{q_i}{p_i}}\right)\left(1-\frac{\sum_{i=1}^N \frac{q_i}{p_i}}{N}\right){\mathbb I}\left(\frac{1}{N}\sum_{i=1}^N \frac{q_i}{p_i} < \frac{1}{2}\right)\right] \Bigg|
\end{align}
We will bound the two terms above separately. Now note that:
\begin{align}
&\Bigg| \E\left[\left(\frac{\frac{1}{N}\sum_{i=1}^N \frac{q_i}{p_i}\log\frac{q_i}{p_i} - D(q||p) \frac{1}{N}\sum_{i=1}^N \frac{q_i}{p_i}}{\frac{1}{N}\sum_{i=1}^N \frac{q_i}{p_i}}\right)\left(1-\frac{\sum_{i=1}^N \frac{q_i}{p_i}}{N}\right){\mathbb I}\left(\frac{1}{N}\sum_{i=1}^N \frac{q_i}{p_i} \ge \frac{1}{2}\right)\right]\Bigg| \notag \\
&\le 2 \E\left[\Bigg|\left({\frac{1}{N}\sum_{i=1}^N \frac{q_i}{p_i}\log\frac{q_i}{p_i} - D(q||p) \frac{1}{N}\sum_{i=1}^N \frac{q_i}{p_i}}\right)\Bigg|\Bigg|\left(1-\frac{\sum_{i=1}^N \frac{q_i}{p_i}}{N}\right)\Bigg|\right] \\
&\le 2\sqrt{\E\left[\left({\frac{1}{N}\sum_{i=1}^N \frac{q_i}{p_i}\log\frac{q_i}{p_i} - D(q||p) \frac{1}{N}\sum_{i=1}^N \frac{q_i}{p_i}}\right)^2\right]}\sqrt{\E\left[\left(1-\frac{\sum_{i=1}^N \frac{q_i}{p_i}}{N}\right)^2\right] }
\end{align} Here the last step follows from Chebyshev's inequality.
Next note that:
\begin{align}
    \E\left[\left(1-\frac{\sum_{i=1}^N \frac{q_i}{p_i}}{N}\right)^2\right] = \frac{1}{N}\left(d_2(q||p)-1\right)
\end{align}
and 
\begin{align}
&\E\left[\left({\frac{1}{N}\sum_{i=1}^N \frac{q_i}{p_i}\log\frac{q_i}{p_i} - D(q||p) \frac{1}{N}\sum_{i=1}^N \frac{q_i}{p_i}}\right)^2\right] \notag\\
&=
\E\left[\left({\frac{1}{N}\sum_{i=1}^N \frac{q_i}{p_i}\log\frac{q_i}{p_i} - D(q||p) + D(q||p) - D(q||p) \frac{1}{N}\sum_{i=1}^N \frac{q_i}{p_i}}\right)^2\right] \\
&\le 2\E\left[\left({\frac{1}{N}\sum_{i=1}^N \frac{q_i}{p_i}\log\frac{q_i}{p_i} - D(q||p)}\right)^2\right]
+ 2  \E\left[\left(D(q||p) \frac{1}{N}\sum_{i=1}^N \frac{q_i}{p_i} - D(q||p)\right)^2\right]\\
&=\frac{2}{N}\left(\E_{p}\left[\left(\frac{q}{p}\log\frac{q}{p}\right)^2\right]-D^2(q||p)\right) + \frac{2 }{N}D^2(q||p)(d_2(q||p)-1)
\end{align}
Thus we have that:
\begin{align}
    &\Bigg| \E\left[\left(\frac{\frac{1}{N}\sum_{i=1}^N \frac{q_i}{p_i}\log\frac{q_i}{p_i} - D(q||p) \frac{1}{N}\sum_{i=1}^N \frac{q_i}{p_i}}{\frac{1}{N}\sum_{i=1}^N \frac{q_i}{p_i}}\right)\left(1-\frac{\sum_{i=1}^N \frac{q_i}{p_i}}{N}\right){\mathbb I}\left(\frac{1}{N}\sum_{i=1}^N \frac{q_i}{p_i} \ge \frac{1}{2}\right)\right]\Bigg| \notag \\
    &\le \frac{2}{N}\sqrt{\left(2\left(\E_{p}\left[\left(\frac{q}{p}\log\frac{q}{p}\right)^2\right]-D^2(q||p)\right) + 2 D^2(q||p)(d_2(q||p)-1)\right)\left(d_2(q||p)-1\right)}\\
    &= \frac{2}{N}g(p,q)
\end{align}
where 
\begin{align}
g(p,q) =\sqrt{\left(2\left(\E_{p}\left[\left(\frac{q}{p}\log\frac{q}{p}\right)^2\right]-D^2(q||p)\right) + 2 D^2(q||p)(d_2(q||p)-1)\right)\left(d_2(q||p)-1\right)}\label{eq:g}
\end{align}
Furthermore using the fact that $q(y)/p(y) \le \omega$ we can show that:
\begin{align}
g(p,q) \le 2(\omega-1)\log \omega.
\end{align}
In a similar manner consider:
\begin{align}
&\Bigg|\E\left[\left(\frac{\frac{1}{N}\sum_{i=1}^N \frac{q_i}{p_i}\log\frac{q_i}{p_i} - D(q||p) \frac{1}{N}\sum_{i=1}^N \frac{q_i}{p_i}}{\frac{1}{N}\sum_{i=1}^N \frac{q_i}{p_i}}\right)\left(1-\frac{\sum_{i=1}^N \frac{q_i}{p_i}}{N}\right){\mathbb I}\left(\frac{1}{N}\sum_{i=1}^N \frac{q_i}{p_i} < \frac{1}{2}\right)\right] \Bigg| \notag\\
&\le \left( \log \omega \right) \E\left[{\mathbb I}\left(\frac{1}{N}\sum_{i=1}^N \frac{q_i}{p_i} < \frac{1}{2}\right)\right] \\
&\le \frac{\log \omega}{N}
\left(4\left(d_2(q||p)-1\right)\right)
= \frac{h(p,q)}{N}
\end{align}
where we used the bound in \eqref{eq:chebyshev} and define
\begin{align}
   h(p,q)&=\log\omega\left(4\left(d_2(q||p)-1\right)\right)\notag\\
      &\le 4 (\omega-1)\log \omega.\label{eq:h} 
\end{align}

We have established that:
\begin{align}
&E_{Y_1^N}\left[\sum_{i=1}^N \frac{\frac{q_i}{p_i}\log\left(\frac{q_i}{p_i}\right)}{\sum_{j=1}^N \frac{q_j}{p_j}}\right] \le E_{Y_1^N}\left[\frac{\sum_{i=1}^N \frac{q_i}{p_i}\log \frac{q_i}{p_i}}{N}\right] + \frac{6(\omega-1)\log \omega}{N} \\
&= E_{Y}\left[\frac{p_{Y|X}(Y|x)}{p_Y(Y)}\log\frac{p_{Y|X}(Y|x)}{p_Y(Y)}\right] + \frac{6(\omega-1)\log \omega}{N} \\
&= D(p_{Y|X}(\cdot|X=x)||p_Y(\cdot)) + \frac{6(\omega-1)\log \omega}{N}  \label{eq:t2-bnd-final}
\end{align}
Thus collecting~\eqref{eq:decomp},~\eqref{eq:t1-finalbnd} and \eqref{eq:t2-bnd-final} we have:
\begin{align}
E_{Y_1^N}\left[\sum_{i=1}^N \frac{\frac{q_i}{p_i}\log\left(N\frac{\frac{q_i}{p_i}}{\sum_{j=1}^N \frac{q_j}{p_j}}\right)}{\sum_{j=1}^N \frac{q_j}{p_j}}\right] \le
D(p_{Y|X}(\cdot|X=x)||p_Y(\cdot)) + \frac{6(\omega-1)\log \omega}{N} + \frac{\alpha(p_Y(\cdot),p_{Y|X}(\cdot))}{N}
\end{align}
where $\alpha(\cdot)$ is defined in~\eqref{eq:alpha}.
It only remains to establish the following inequality stated in~\eqref{eq:inv_ineq}:
\begin{align}
\frac{1}{\frac{1}{N} \sum_{i=1}^N \frac{q_i}{p_i}} \le \frac{1}{N}\sum_{i=1}^N \frac{p_i}{q_i}. \label{eq:inv_ineq2}
\end{align}

Now define $\beta_i= \frac{q_i}{p_i}$ as before. And observe that:
\begin{align}
\frac{1}{\frac{1}{N}\sum_{i=1}^N \frac{q_i}{p_i}}&=\left(\frac{\sum_{i=1}^N \beta_i }{N}\right)\left(\frac{\sum_{i=1}^N \beta_i \frac{p_i}{q_i}}{\sum_{i=1}^N \beta_i} \right)\left(\frac{\sum_{i=1}^N \beta_i \frac{p_i}{q_i}}{\sum_{i=1}^N \beta_i}\right)\\
&=\left(\frac{\sum_{i=1}^N \beta_i}{N}\right)\E_\beta\left[\frac{p}{q}\right]
\E_\beta\left[\frac{p}{q}\right]\\
&\le \frac{\sum_{i=1}^N \beta_i}{N} \E_\beta\left[\left(\frac{p}{q}\right)^2\right] \label{eq:cs2}\\
&= \frac{1}{N}\sum_{i=1}^N \frac{p_i}{q_i}
\end{align}
where~\eqref{eq:cs2} is a consequence of Cauchy-Schwartz inequality.

\section{Proof of Prop.~\ref{prop:entbnd} in Section~\ref{sec:alt-thm1} in this document.}
\label{app:entbnd}
Applying Hoeffding's Inequality we get that:

\begin{align}
\Pr\left(\frac{1}{N-1}\sum_{j=2}^N \frac{p_{Y|X}(Y_j|x)}{p_Y(Y_j)} \le 1-\epsilon  \right)   \le \exp\left(-2(N-1) \epsilon^2/\omega^2\right).
\end{align}
Equivalently, if
$$\Omega = \left\{y_2^N: \frac{1}{N-1}\sum_{j=2}^N \frac{p_{Y|X}(y_j|x)}{p_Y(y_j)} \le 1-\epsilon \right\}$$
then $\Pr(\Omega) \le \exp\left(-2(N-1) \epsilon^2/\omega^2\right) $.

Now revisiting~\eqref{eq:expectaton} we have
\begin{align}
&E_{Y_1^N} [N \lambda_1 \log (N \lambda_1)]  =  E_{Y_1^N} [N \lambda_1 \log (N \lambda_1)|\Omega] \Pr(\Omega) +   E_{Y_1^N} [N \lambda_1 \log (N \lambda_1)|\Omega^c] \Pr(\Omega^c)  \\
&\le  E_{Y_1^N} [N \lambda_1 \log (N \lambda_1)|\Omega] \exp\left(-2(N-1) \epsilon^2/\omega^2\right) +  E_{Y_1^N} [N \lambda_1 \log (N \lambda_1)|\Omega^c]   \label{eq:condn-exp}
\end{align}

We bound each of the two terms in~\eqref{eq:condn-exp}. First consider
\begin{align}
E_{Y_1^N} [N \lambda_1 \log (N \lambda_1)|\Omega] &= E\left[N \frac{\frac{p_{Y|X}(Y_1|x)}{p_Y(Y_1)} }{\sum_{j=1}^N \frac{p_{Y|X}(Y_j|x)}{p_Y(Y_j)}}  \log \left( N \frac{\frac{p_{Y|X}(Y_1|x)}{p_Y(Y_1)} }{\sum_{j=1}^N \frac{p_{Y|X}(Y_j|x)}{p_Y(Y_j)}} \right) \Bigg|~ \Omega \right] \\
&\le N \log N,
\end{align}
where the second step uses the fact that $\frac{P_{Y|X}(y|x)}{P_Y(y)} \ge 0$.
Furthermore we have that
\begin{align}
&E_{Y_1^N} [N \lambda_1 \log (N \lambda_1)|\Omega^c] =E\left[N \frac{\frac{p_{Y|X}(Y_1|x)}{p_Y(Y_1)} }{\sum_{j=1}^N \frac{p_{Y|X}(Y_j|x)}{p_Y(Y_j)}}  \log \left( N \frac{\frac{p_{Y|X}(Y_1|x)}{p_Y(Y_1)} }{\sum_{j=1}^N \frac{p_{Y|X}(Y_j|x)}{p_Y(Y_j)}} \right) \Bigg|~ \Omega^c \right] \\
&\le E_{Y_1 \sim P_Y(\cdot)} \left[\frac{N}{(N-1)(1-\epsilon)}\frac{p_{Y|X}(Y_1|x)}{p_Y(Y_1)} \log  \frac{N}{(N-1)(1-\epsilon)}\frac{p_{Y|X}(Y_1|x)}{p_Y(Y_1)}  \right]\\
&=\frac{N}{(N-1)(1-\epsilon)} D(p_{Y|X}(\cdot|x) || p_Y(\cdot))  + \frac{N}{(N-1)(1-\epsilon)}  \log \frac{N}{(N-1)(1-\epsilon)}  
\end{align}

\section{Proof of Prop.~\ref{prop:lem2} in this document}
\label{app:lem2}
For simplicity we will denote $\lambda_i = \lambda(Y_i) = \frac{p_{Y|X}(Y_i|X)}{p_Y(Y_i)}$ and $\beta_i = \beta(Y_i) = \frac{q_{Y|X}(Y_i|X=x)}{q_Y(Y_i)}$. For simplicity, we keep the conditioning on $Y_1=y_1$ implicit. For convenience, we define $\bN = N-1$ and consider the following normalization:
\begin{align}
 E_{Y_2^N}\left[\frac{\frac{1}{\bN}\sum_{i=1}^N \beta_i}{\left(\frac{1}{\bN}\sum_{i=1}^N \lambda_i\right)^2}\right] 
\end{align}
and let us define:
\begin{align}
    \Omega_1 = E_{Y_2^N}\left[\frac{1}{\bN} \sum_{i=1}^N \beta_i\right] = \frac{\beta_1}{\bN} + 1
\end{align}
as well as 
\begin{align}
    \Omega_2 &= E_{Y_2^N}\left[\left(\frac{1}{\bN} \sum_{i=1}^{N} \lambda_i\right)^2\right]\\
    &=\frac{1}{\bN^2}E_{Y_2^N}\left[\left( \lambda_1^2 + \sum_{i=2}^{N} \lambda_i^2 +2\lambda_1 \sum_{j=2}^N \lambda_j + 2\sum_{i=2}^N\sum_{j=i+1}^N \lambda_i \lambda_j\right)\right]\\
    &= \frac{\lambda_1^2}{\bN^2} + \frac{d_2(q||p)}{\bN}+\frac{2\lambda_1}{\bN} + \frac{\bN-1}{\bN}\\
    &= \left(1+ \frac{\lambda_1}{\bN}\right)^2 + \frac{d_2(q||p)-1}{\bN} \\
    &= 1+ \frac{\gamma}{\bN}
\end{align}

where we use the fact that $E_{Y_i}[\lambda_i]=1$ and $E_{Y_i}[\lambda_i^2]=d_2(q||p)$ and define \begin{align}\gamma = {2\lambda_1 + d_2(q||p)-1 + \frac{\lambda_1^2}{\bN}}. \label{eq:g-def}\end{align} 
Next we consider the following:
\begin{align}
 &E_{Y_2^N}\left[\frac{\frac{1}{\bN}\sum_{i=1}^N \beta_i}{\left(\frac{1}{\bN}\sum_{i=1}^N \lambda_i\right)^2}\right] -\frac{\Omega_1}{\Omega_2}\\
 &=E_{Y_2^N}\left[\frac{1}{\left(\frac{1}{\bN}\sum_{i=1}^N \lambda_i\right)^2}\left(\frac{1}{\bN}\sum_{i=1}^N \beta_i-\frac{\Omega_1}{\Omega_2}\left(\frac{1}{\bN}\sum_{i=1}^N \lambda_i\right)^2\right)\right]\label{eq:obj-1}
\end{align}
Now note that: 
\begin{align}
    E_{Y_2^N}\left[\left(\frac{1}{\bN}\sum_{i=1}^N \beta_i-\frac{\Omega_1}{\Omega_2}\left(\frac{1}{\bN}\sum_{i=1}^N \lambda_i\right)^2\right)\right]=0.
\end{align}

Thus~\eqref{eq:obj-1} is equivalent to:
\begin{align}
    E_{Y_2^N}\left[\left(\frac{1}{\left(\frac{1}{\bN}\sum_{i=1}^N \lambda_i\right)^2}-\frac{1}{c^2}\right)\left(\frac{1}{\bN}\sum_{i=1}^N \beta_i-\frac{\Omega_1}{\Omega_2}\left(\frac{1}{\bN}\sum_{i=1}^N \lambda_i\right)^2\right)\right]\label{eq:obj-2}
\end{align}
for any constant $c\neq 0$. We select \begin{align}
c = E_{Y_2^N}\left[\frac{1}{\bN}\sum_{i=1}^N \lambda_i\right]  = \frac{\lambda_1}{\bN} + 1  
\end{align}
We can express~\eqref{eq:obj-2} as follows:
\begin{align}
   &E_{Y_2^N}\left[\frac{1}{\left(\frac{1}{\bN}\sum_{i=1}^N \lambda_i\right)^2}\left(1-\frac{{\left(\frac{1}{\bN}\sum_{i=1}^N \lambda_i\right)^2}}{c^2}\right)\left(\frac{1}{\bN}\sum_{i=1}^N \beta_i-\frac{\Omega_1}{\Omega_2}\left(\frac{1}{\bN}\sum_{i=1}^N \lambda_i\right)^2\right)\right]  \notag  \\
   &=E_{Y_2^N}\left[\frac{1}{\left(\frac{1}{\bN}\sum_{i=1}^N \lambda_i\right)^2}\left(1-\frac{{\left(\frac{1}{\bN}\sum_{i=1}^N \lambda_i\right)^2}}{c^2}\right)\left(\frac{1}{\bN}\sum_{i=1}^N \beta_i-\frac{\Omega_1}{\Omega_2}\left(\frac{1}{\bN}\sum_{i=1}^N \lambda_i\right)^2\right){\mathbb I}\left(\frac{1}{\bN}\sum_{i=2}^{\bN} \lambda_i \ge \frac{1}{2}\right)\right]\notag\\
   &+E_{Y_2^N}\left[\frac{1}{\left(\frac{1}{\bN}\sum_{i=1}^N \lambda_i\right)^2}\left(1-\frac{{\left(\frac{1}{\bN}\sum_{i=1}^N \lambda_i\right)^2}}{c^2}\right)\left(\frac{1}{\bN}\sum_{i=1}^N \beta_i-\frac{\Omega_1}{\Omega_2}\left(\frac{1}{\bN}\sum_{i=1}^N \lambda_i\right)^2\right){\mathbb I}\left(\frac{1}{\bN}\sum_{i=2}^{\bN} \lambda_i < \frac{1}{2}\right)\right]
\end{align}

Now consider make use of the triangular inequality:
\begin{align}
 &\left|E_{Y_2^N}\left[\frac{\frac{1}{\bN}\sum_{i=1}^N \beta_i}{\left(\frac{1}{\bN}\sum_{i=1}^N \lambda_i\right)^2}\right] -\frac{\Omega_1}{\Omega_2}\right| \notag \\
 &\le
 \left|E_{Y_2^N}\left[\frac{1}{\left(\frac{1}{\bN}\sum_{i=1}^N \lambda_i\right)^2}\left(1-\frac{{\left(\frac{1}{\bN}\sum_{i=1}^N \lambda_i\right)^2}}{c^2}\right)\left(\frac{1}{\bN}\sum_{i=1}^N \beta_i-\frac{\Omega_1}{\Omega_2}\left(\frac{1}{\bN}\sum_{i=1}^N \lambda_i\right)^2\right){\mathbb I}\left(\frac{1}{\bN}\sum_{i=2}^{\bN} \lambda_i \ge \frac{1}{2}\right)\right]\right|\notag\\
 &+\left|E_{Y_2^N}\left[\frac{1}{\left(\frac{1}{\bN}\sum_{i=1}^N \lambda_i\right)^2}\left(1-\frac{{\left(\frac{1}{\bN}\sum_{i=1}^N \lambda_i\right)^2}}{c^2}\right)\left(\frac{1}{\bN}\sum_{i=1}^N \beta_i-\frac{\Omega_1}{\Omega_2}\left(\frac{1}{\bN}\sum_{i=1}^N \lambda_i\right)^2\right){\mathbb I}\left(\frac{1}{\bN}\sum_{i=2}^{\bN} \lambda_i < \frac{1}{2}\right)\right]\right|\label{eq:obj-3}
\end{align}

We now consider each of the two terms in~\eqref{eq:obj-3} separately:
\begin{align}
   & \left|E_{Y_2^N}\left[\frac{1}{\left(\frac{1}{\bN}\sum_{i=1}^N \lambda_i\right)^2}\left(1-\frac{{\left(\frac{1}{\bN}\sum_{i=1}^N \lambda_i\right)^2}}{c^2}\right)\left(\frac{1}{\bN}\sum_{i=1}^N \beta_i-\frac{\Omega_1}{\Omega_2}\left(\frac{1}{\bN}\sum_{i=1}^N \lambda_i\right)^2\right){\mathbb I}\left(\frac{1}{\bN}\sum_{i=2}^{\bN} \lambda_i \ge \frac{1}{2}\right)\right]\right|\notag\\
    &\le
    \frac{4}{c^2}\left|E_{Y_2^N}\left[\left(c^2-{\left(\frac{1}{\bN}\sum_{i=1}^N \lambda_i\right)^2}\right)\left(\frac{1}{\bN}\sum_{i=1}^N \beta_i-\frac{\Omega_1}{\Omega_2}\left(\frac{1}{\bN}\sum_{i=1}^N \lambda_i\right)^2\right){\mathbb I}\left(\frac{1}{\bN}\sum_{i=2}^{\bN} \lambda_i \ge \frac{1}{2}\right)\right]\right|\\
&\le   \frac{4}{c^2}E_{Y_2^N}\left[\left|\left(c^2-{\left(\frac{1}{\bN}\sum_{i=1}^N \lambda_i\right)^2}\right)\right|\left|\left(\frac{1}{\bN}\sum_{i=1}^N \beta_i-\frac{\Omega_1}{\Omega_2}\left(\frac{1}{\bN}\sum_{i=1}^N \lambda_i\right)^2\right)\right|{\mathbb I}\left(\frac{1}{\bN}\sum_{i=2}^{\bN} \lambda_i \ge \frac{1}{2}\right)\right]\\
&\le \frac{4}{c^2}E_{Y_2^N}\left[\left|\left(c^2-{\left(\frac{1}{\bN}\sum_{i=1}^N \lambda_i\right)^2}\right)\right|\left|\left(\frac{1}{\bN}\sum_{i=1}^N \beta_i-\frac{\Omega_1}{\Omega_2}\left(\frac{1}{\bN}\sum_{i=1}^N \lambda_i\right)^2\right)\right|\right]\\
&\le \frac{4}{c^2}\sqrt{E_{Y_2^N}\left[\left(c^2-{\left(\frac{1}{\bN}\sum_{i=1}^N \lambda_i\right)^2}\right)^2\right]}\sqrt{E_{Y_2^N}\left[\left(\frac{1}{\bN}\sum_{i=1}^N \beta_i-\frac{\Omega_1}{\Omega_2}\left(\frac{1}{\bN}\sum_{i=1}^N \lambda_i\right)^2\right)^2\right]}\label{eq:ode-4}
\end{align}

We now consider each of the two terms above separately.

\begin{align}
{E_{Y_2^N}\left[\left(c^2-{\left(\frac{1}{\bN}\sum_{i=1}^N \lambda_i\right)^2}\right)^2\right]} 
&={E_{Y_2^N}\left[\left(c-{\left(\frac{1}{\bN}\sum_{i=1}^N \lambda_i\right)}\right)^2\left(c+{\left(\frac{1}{\bN}\sum_{i=1}^N \lambda_i\right)}\right)^2\right]}\\
&={E_{Y_2^N}\left[\left(1-{\left(\frac{1}{\bN}\sum_{i=2}^N \lambda_i\right)}\right)^2\left(c+{\left(\frac{1}{\bN}\sum_{i=1}^N \lambda_i\right)}\right)^2\right]}\\
&\le \left(1+ \frac{N+1}{\bN}\omega\right)^2E_{Y_2^N}\left[\left(1-{\left(\frac{1}{\bN}\sum_{i=2}^N \lambda_i\right)}\right)^2\right]\\
&=\left(1+ \frac{N+1}{\bN}\omega\right)^2\frac{1}{\bN} E_{p}\left[\frac{q^2}{p^2} -1\right] \\
&=\frac{1}{\bN}\left(1+ \frac{N+1}{\bN}\omega\right)^2\left(d_2(q||p)-1\right). \label{eq:ode-41}
\end{align}

We now consider the second term in~\eqref{eq:ode-4}.
\begin{align}
&E_{Y_2^N}\left[\left(\frac{1}{\bN}\sum_{i=1}^N \beta_i-\frac{\Omega_1}{\Omega_2}\left(\frac{1}{\bN}\sum_{i=1}^N \lambda_i\right)^2\right)^2\right] \notag \\
&=E_{Y_2^N}\left[\left(\frac{1}{\bN}\sum_{i=1}^N \beta_i -\Omega_1 + \Omega_1 -\frac{\Omega_1}{\Omega_2}\left(\frac{1}{\bN}\sum_{i=1}^N \lambda_i\right)^2\right)^2\right] \\
&\le 2 E_{Y_2^N}\left[\left(\frac{1}{\bN}\sum_{i=1}^N \beta_i -\Omega_1 \right)^2\right] + 2\left(\frac{\Omega_1}{\Omega_2}\right)^2E_{Y_2^N}\left[\left(\left(\frac{1}{\bN}\sum_{i=1}^N \lambda_i\right)^2 - \Omega_2\right)^2\right]\label{eq:ode-5}
\end{align}

We consider each term in~\eqref{eq:ode-5} separately.
\begin{align}
E_{Y_2^N}\left[\left(\frac{1}{\bN}\sum_{i=1}^N \beta_i -\Omega_1 \right)^2\right] 
&=E_{Y_2^N}\left[\left(\frac{1}{\bN}\sum_{i=2}^N \beta_i -1 \right)^2\right]\\
&=\frac{1}{\bN}E_{p}\left[\left(\frac{r}{p}\right)^2-1\right]\\
&=\frac{1}{\bN}\left(d_2(r||p)-1\right) \label{eq:ode-51}
\end{align}

Next consider
\begin{align}
&E_{Y_2^N}\left[\left(\left(\frac{1}{\bN}\sum_{i=1}^N \lambda_i\right)^2 - \Omega_2\right)^2\right]=E_{Y_2^N}\left[\left(\left(\frac{1}{\bN}\sum_{i=1}^N \lambda_i\right)^2 - \left(1+ \frac{\lambda_1}{\bN}\right)^2 - \frac{d_2(q||p)-1}{\bN}\right)^2\right]\\
&\le 2E_{Y_2^N}\left[\left(\left(\frac{1}{\bN}\sum_{i=1}^N \lambda_i\right)^2 - \left(1+ \frac{\lambda_1}{\bN}\right)^2\right)^2\right] + 2\left(\frac{d_2(q||p)-1}{\bN}\right)^2
\end{align}

Now we can upper bound the first term as follows:
\begin{align}
&E_{Y_2^N}\left[\left(\left(\frac{1}{\bN}\sum_{i=1}^N \lambda_i\right)^2 - \left(1+ \frac{\lambda_1}{\bN}\right)^2\right)^2\right] \\
&=E_{Y_2^N}\left[\left(\left(\frac{1}{\bN}\sum_{i=1}^N \lambda_i\right) - \left(1+ \frac{\lambda_1}{\bN}\right)\right)^2\left(\left(\frac{1}{\bN}\sum_{i=1}^N \lambda_i\right) + \left(1+ \frac{\lambda_1}{\bN}\right)\right)^2\right] \\
&\le\left(1+ \frac{N+1}{\bN}\omega\right)^2 E_{Y_2^N}\left[\left(1-{\left(\frac{1}{\bN}\sum_{i=2}^N \lambda_i\right)}\right)^2\right]\\
&=\frac{1}{\bN}\left(1+ \frac{N+1}{\bN}\omega\right)^2\left(d_2(q||p)-1\right) \label{eq:ode-52}
\end{align}

As a result, using~\eqref{eq:ode-5},~\eqref{eq:ode-51} and~\eqref{eq:ode-52} we have:
\begin{align}
&E_{Y_2^N}\left[\left(\frac{1}{\bN}\sum_{i=1}^N \beta_i-\frac{\Omega_1}{\Omega_2}\left(\frac{1}{\bN}\sum_{i=1}^N \lambda_i\right)^2\right)^2\right] \\
&\le \frac{2}{\bN}\left(d_2(r||p)-1\right)+ \frac{4}{\bN}\left(\frac{\Omega_1}{\Omega_2}\right)^2\left\{\left(1+ \frac{N+1}{\bN}\omega\right)^2\left(d_2(q||p)-1\right) + \frac{(d_2(q||p)-1)^2}{\bN}\right\} \label{eq:ode-42}
\end{align}

Consequently using~\eqref{eq:ode-4},~\eqref{eq:ode-41} and~\eqref{eq:ode-42}, we can show that:
\begin{align} &\left|E_{Y_2^N}\left[\frac{1}{\left(\frac{1}{\bN}\sum_{i=1}^N \lambda_i\right)^2}\left(1-\frac{{\left(\frac{1}{\bN}\sum_{i=1}^N \lambda_i\right)^2}}{c^2}\right)\left(\frac{1}{\bN}\sum_{i=1}^N \beta_i-\frac{\Omega_1}{\Omega_2}\left(\frac{1}{\bN}\sum_{i=1}^N \lambda_i\right)^2\right){\mathbb I}\left(\frac{1}{\bN}\sum_{i=2}^{\bN} \lambda_i \ge \frac{1}{2}\right)\right]\right| \notag\\
&\le\frac{4}{\bN c^2} \sqrt{\left(1+ \frac{N+1}{\bN}\omega\right)^2\left(d_2(q||p)-1\right)}\notag\\
&\times\sqrt{2\left(d_2(p||r)-1\right) + 4\left(\frac{\Omega_1}{\Omega_2}\right)^2\left\{\left(1+ \frac{N+1}{\bN}\omega\right)^2\left(d_2(p||q)-1\right) + \frac{(d_2(q||p)-1)^2}{\bN}\right\}}\\
&\le 4 \frac{(\omega-1)}{\bN c^2}\left(1+ \frac{N+1}{\bN}\omega\right)\sqrt{2+ 4\left(\frac{\Omega_1}{\Omega_2}\right)^2\left\{\left(1+ \frac{N+1}{\bN}\omega\right)^2 + \frac{(\omega-1)}{\bN}\right\}}\\
&=\frac{1}{\bN}K_1(\bN)
\end{align}
where we repeatedly use the fact that $d_2(q||p) \le \omega$ and $d_2(r||p)\le \omega$.
Here we have introduced: 
\begin{align}
K_1(\bN) = 4 \frac{(\omega-1)}{ (1+ \frac{\lambda_1}{\bN})^2}\left(1+ \frac{N+1}{\bN}\omega\right)\sqrt{2+ 4\left(\frac{1+ \frac{\beta_1}{\bN}}{1+ \frac{\gamma_1}{\bN}}\right)^2\left\{\left(1+ \frac{N+1}{\bN}\omega\right)^2 + \frac{(\omega-1)}{\bN}\right\}} \\
\le 4 \frac{(\omega-1)}{ (1+ \frac{\lambda_1}{\bN})^2}\left(1+ \frac{N+1}{\bN}\omega\right)\sqrt{2+ 4\left(\frac{1+ \frac{\beta_1}{\bN}}{1+ \frac{2\lambda_1}{\bN}}\right)^2\left\{\left(1+ \frac{N+1}{\bN}\omega\right)^2 + \frac{(\omega-1)}{\bN}\right\}}, \label{eq:k1-bnd}
\end{align}where we use the fact that $\gamma_1 \ge 2 \lambda_1$ following~\eqref{eq:g-def}. Note that $K_1(\bN)=\Theta(1)$.

Now consider the other term in~\eqref{eq:obj-3}:
\begin{align}
&\left|E_{Y_2^N}\left[\frac{1}{\left(\frac{1}{\bN}\sum_{i=1}^N \lambda_i\right)^2}\left(1-\frac{{\left(\frac{1}{\bN}\sum_{i=1}^N \lambda_i\right)^2}}{c^2}\right)\left(\frac{1}{\bN}\sum_{i=1}^N \beta_i-\frac{\Omega_1}{\Omega_2}\left(\frac{1}{\bN}\sum_{i=1}^N \lambda_i\right)^2\right){\mathbb I}\left(\frac{1}{\bN}\sum_{i=2}^{\bN} \lambda_i < \frac{1}{2}\right)\right]\right|\\
&\le \left|E_{Y_2^N}\left[\frac{1}{\left(\frac{1}{\bN}\sum_{i=1}^N \lambda_i\right)^2}\left(\frac{1}{\bN}\sum_{i=1}^N \beta_i-\frac{\Omega_1}{\Omega_2}\left(\frac{1}{\bN}\sum_{i=1}^N \lambda_i\right)^2\right){\mathbb I}\left(\frac{1}{\bN}\sum_{i=2}^{\bN} \lambda_i < \frac{1}{2}\right)\right]\right|\\
&\le \left|\omega E_{Y_2^N}\left[\frac{1}{\left(\frac{1}{\bN}\sum_{i=1}^N \lambda_i\right)^2}{\mathbb I}\left(\frac{1}{\bN}\sum_{i=2}^{\bN} \lambda_i < \frac{1}{2}\right)\right]\right|\label{eq:ode-6}
\end{align}

Next, we can show that:
\begin{align}
\frac{1}{\left(\frac{1}{\bN}\sum_{i=1}^N \lambda_i\right)^2}&\le \frac{1}{\left(\frac{1}{\bN} \sum_{i=2}^N \lambda_i\right)^2}\\
&\le \frac{1}{\bN}\sum_{i=2}^N \left(\frac{p_i}{q_i}\right)^2 \label{eq:pq-bnd2}
\end{align}
The proof of~\eqref{eq:pq-bnd2} will be shown at the end of this section.
Furthermore note that $E_p[p^2/q^2] = d_3(p||q)$. Thus we can upper-bound~\eqref{eq:ode-6} as:
\begin{align}
&\le \left|\omega E_{Y_2^N}\left[\left(\frac{1}{\bN}\sum_{i=2}^{\bN} \frac{p_i^2}{q_i^2} \right){\mathbb I}\left(\frac{1}{\bN}\sum_{i=2}^{\bN} \lambda_i < \frac{1}{2}\right)\right]\right|\\
&= \omega \left|E_{Y_2^N}\left[\left(\frac{1}{\bN}\sum_{i=2}^{\bN} \frac{p_i^2}{q_i^2} -d_3(p||q) +d_3(p||q)\right){\mathbb I}\left(\frac{1}{\bN}\sum_{i=2}^{\bN} \lambda_i < \frac{1}{2}\right)\right]\right|\\
 &\le \omega \left|E_{Y_2^N}\left[\left(\frac{1}{\bN}\sum_{i=2}^{\bN} \frac{p_i^2}{q_i^2} -d_3(p||q)\right){\mathbb I}\left(\frac{1}{\bN}\sum_{i=2}^{\bN} \lambda_i < \frac{1}{2}\right)\right]\right| + \omega\left(d_3(p||q)\right)E\left[{\mathbb I}\left(\frac{1}{\bN}\sum_{i=2}^{\bN} \lambda_i < \frac{1}{2}\right)\right]\\
 &\le \omega \left|E_{Y_2^N}\left[\left(\frac{1}{\bN}\sum_{i=2}^{\bN} \frac{p_i^2}{q_i^2} -d_3(p||q)\right){\mathbb I}\left(\frac{1}{\bN}\sum_{i=2}^{\bN} \lambda_i < \frac{1}{2}\right)\right]\right| + \omega\left(d_3(p||q) \right)\frac{4}{\bN}(d_2(q||p)-1)
\end{align}

The first term above can be upper bounded using Cauchy-schwartz as follows:
\begin{align}
&\left|E_{Y_2^N}\left[\left(\frac{1}{\bN}\sum_{i=1}^{\bN} \frac{p_i^2}{q_i^2} -d_3(p||q)\right){\mathbb I}\left(\frac{1}{\bN}\sum_{i=2}^{\bN} \lambda_i < \frac{1}{2}\right)\right]\right|\\
&\le\sqrt{E_{Y_2^N}\left[\left(\frac{1}{\bN}\sum_{i=1}^{\bN} \frac{p_i^2}{q_i^2} -d_3(p||q)\right)^2\right]}\sqrt{E\left[{\mathbb I}\left(\frac{1}{\bN}\sum_{i=2}^{\bN} \lambda_i < \frac{1}{2}\right)\right]}\\
&\le \sqrt{\frac{1}{\bN}(d_5(p||q)-d_3(p||q)^2)}\sqrt{\frac{4}{\bN}(d_2(q||p)-1)}
\end{align}

It thus follows that we can express:
\begin{align}
&\left|E_{Y_2^N}\left[\frac{1}{\left(\frac{1}{\bN}\sum_{i=1}^N \lambda_i\right)^2}\left(1-\frac{{\left(\frac{1}{\bN}\sum_{i=1}^N \lambda_i\right)^2}}{c^2}\right)\left(\frac{1}{\bN}\sum_{i=1}^N \beta_i-\frac{\Omega_1}{\Omega_2}\left(\frac{1}{\bN}\sum_{i=1}^N \lambda_i\right)^2\right){\mathbb I}\left(\frac{1}{\bN}\sum_{i=2}^{\bN} \lambda_i < \frac{1}{2}\right)\right]\right|\notag\\ 
&\le \frac{2\omega}{\bN}L(\bN)
\end{align}
where 
\begin{align}
L(\bN) &= \sqrt{(d_5(p||q)-d_3(p||q)^2)}\sqrt{d_2(q||p)-1} + \left(d_3(p||q) \right)(d_2(q||p)-1) \\
&\le  \sqrt{\omega-1}\sqrt{(d_5(p||q)-d_3(p||q)^2)} + (\omega-1) d_3(p||q)  \label{eq:l-bnd}
\end{align}

Thus using~\eqref{eq:obj-3},~\eqref{eq:k1-bnd} and~\eqref{eq:l-bnd} it follows that:
\begin{align}
&\bN  E_{Y_2^N}\left[\frac{\sum_{i=1}^N \beta_i}{\left(\sum_{i=1}^N \lambda_i\right)^2}\right] \notag  \le
\frac{\frac{\beta_1}{\bN} + 1}{\left(1+ \frac{\lambda_1}{\bN}\right)^2} + \frac{1}{\bN} K_1(\bN) + \frac{2\omega}{\bN}L(p,q)
\end{align}

It remains to show
\begin{align}
\frac{1}{\left(\frac{1}{\bN} \sum_{i=2}^N \frac{q_i}{p_i}\right)^2}
&\le \frac{1}{\bN}\sum_{i=2}^N \left(\frac{p_i}{q_i}\right)^2 \label{eq:pq-bnd3}
\end{align}

Note that:
\begin{align}
\frac{1}{\left(\frac{1}{\bN} \sum_{i=2}^N \frac{q_i}{p_i}\right)^2} 
&= \frac{\sum_{i=2}^{\bN} \frac{q_i}{p_i}}{\bN} \left(\frac{\sum_{i=2}^{\bN} \frac{q_i}{p_i}\frac{p_i}{q_i} }{\sum_{i=2}^{\bN}\frac{q_i}{p_i}} \right)^3 \\
&=\frac{\sum_{i=2}^{\bN} \frac{q_i}{p_i}}{\bN} \left(E_{\lambda}\left[\frac{p}{q}\right]\right)^3 \label{eq:lam-intr}\\
&\le\frac{\sum_{i=2}^{\bN} \frac{q_i}{p_i}}{\bN} E_{\lambda}\left[\frac{p^3}{q^3}\right] \label{eq:js2} \\
&=\frac{\sum_{i=2}^{\bN} \frac{q_i}{p_i}}{\bN} \left(\frac{\sum_{i=2}^{\bN} \frac{q_i}{p_i}\frac{p_i^3}{q_i^3} }{\sum_{i=2}^{\bN}\frac{q_i}{p_i}} \right)\\
&=\frac{1}{\bN}\sum_{i=2}^{\bN}\frac{p_i^2}{q_i^2}.
\end{align}
In~\eqref{eq:lam-intr} we define $\lambda$ to be probability vector which select index $i$ with probability proportional to $q_i/p_i$. We use Jensen's inequality in~\eqref{eq:js2} since the function $f(x) = x^3$ is convex on $x\ge0$.

\section{Proof of Prop.~\ref{prop:hoeff} in this document}
\label{app:hoeff}
For $0 <\epsilon <1$, define
\begin{align}
{\mathcal E} =\left\{(Y_j)_{j=2}^N :  \frac{1}{N-1}\sum_{j=2}^N \lambda(Y_j) \ge 1-\epsilon,  \frac{1}{N-1}\sum_{j=2}^N \beta(Y_j) \le 1+\epsilon \right\},
\end{align}
then using Hoeffding's inequality and the union bound, we have that:
\begin{align}
\Pr\left({\mathcal E}^c \right) \le 2 \exp\left(-2(N-1)\epsilon^2/\omega^2\right).
\end{align}

Now observe that:
\begin{align}
&E_{Y_2^N} \left[ \frac{\sum_{j=1}^N\ \beta(Y_j) }{(\sum_{j=1}^N \lambda(Y_j) )^2  }\bigg| Y_1=y_1\right] \\
&= E_{Y_2^N} \left[ \frac{\sum_{j=1}^N\ \beta(Y_j) }{(\sum_{j=1}^N \lambda(Y_j) )^2  }\bigg| Y_1=y_1, \cE \right] \\
&\quad + E_{Y_2^N} \left[ \frac{\sum_{j=1}^N\ \beta(Y_j) }{(\sum_{j=1}^N \lambda(Y_j) )^2  }\bigg| Y_1=y_1, \cE^c \right]  \Pr(\cE^c)\\
&\le \frac{\beta(y_1) + (N-1)(1+\epsilon)}{(\lambda(y_1) + (N-1)(1-\epsilon))^2} + \frac{N \omega }{\lambda(y_1)^2} 2 e^{-(N-1)\epsilon^2/\omega^2}
\end{align}

Collecting all the terms we have that:
\begin{align}
&E_{Y_2^N}\left[\frac{\left(\sum_{j=1}^N\frac{q_{Y|X}(Y_j|x)}{p_Y(Y_j)}\right)}{\left(\sum_{j=1}^N\frac{p_{Y|X}(Y_j|x)}{p_Y(Y_j)}\right)} \bigg| Y_1=y_1, U_p=1\right]\notag\\
&\le (N-1+\lambda(y_1))\left(\frac{\beta(y_1) + (N-1)(1+\epsilon)}{(\lambda(y_1) + (N-1)(1-\epsilon))^2} + \frac{N \omega }{\lambda(y_1)^2} 2 e^{-(N-1)\epsilon^2/\omega^2}\right)
\end{align}

\section{Decoding with Neural Estimator}
\label{app:nce}
We recall that in the problem of lossy compression with side information at the decoder, the random variables $V, W$ and $T$ follow the Markov chain $T-V-W$. Following the setup and algorithm in Section~\ref{prop:SI} in the main paper,  the encoding step is relatively straightforward, given that $p_{W|V}(.|.)$ and $p_{W}(.)$ can be predefined. On the other hand,  during the decoding step, the decoder needs to compute the following quantity:
\begin{align*}
&U_q =  \arg\min_{1\le i \le N}\frac{S_i }{\frac{Q_{Y|Z}(Y_i|t,l_{U_p})}{p_Y(Y_i)}}\notag\\
&=\arg\min_{1\le i \le N}\frac{S_i }{\frac{p_{W|T}(W_i|t) {\mathbb I}(l_i= l_{U_p})}{p_W(W_i) p_{\mathsf l}(l_i)}}, 
\end{align*}
which can be hard to compute due to the presence of $p_{W|T}(W_i|t)$, especially when the distribution is unknown and complicated. As a result, the quantity $\log \frac{p_W(W_i)}{p_{W|T}(W_i|t)}$ has to be learned from the training dataset. We note that while techniques like Markov Chain Monte Carlo (MCMC) or variational inference can be employed, their usage may lead to significant time complexity or sub optimal performance due to the inherent limitations in expressing complex distributions accurately.

Instead, we construct and train a neural network $\Gamma \colon \mathcal{W}\times \mathcal{T}\to [0,1]$ to directly estimate the above ratio \citep{hermans2020likelihood, cranmer2015approximating}, by classifying whether $W,T \sim p_{W,T}(.)$ (positive samples) or $W,T \sim p_W(.)p_T(.)$ (negative samples). Following the Markov chain $T-V-W$, one can construct a positive sample by first sampling from the training set a pair of $\{T,V\}$ and then get $W\sim p_{W|V}(.)$ where $p_{W|V}(.)$ is predefined. On the other hand, to obtain negative samples, we sample $\{T,V\}$ from the training set and $W\sim p_W(.)$. Note that ratio between positive and negative samples should be 1. Furthermore, we  define  $\Gamma(W, T)= \sigma(h_\gamma(W,T))$ where $h_\gamma$ is a neural networks with parameters $\gamma$ and $\sigma$ is the sigmoid activation:
\begin{equation*}
    \sigma(x) = \frac{1}{1 + \exp({-x})}
\end{equation*}
\cite{hermans2020likelihood} shows that the logit values of the optimal classifier can be then used as a log-likelihood estimator, that is:
\begin{equation}
    h_{\gamma^*}(W,T) \approx -\log \frac{p_W(W)p_T(T)}{p_{W,T}(W, T)} =  -\log \frac{p_W(W)}{p_{W|T}(W|T)}.
\end{equation}
 which is the quantity we would like to estimate. We train our classifier using the standard cross-entropy loss with Adam optimizer. For details about the neural network architecture of each experiment (MNIST and CIFAR-10), refer Section \ref{app:add_exps} below.
 

\section{ADDITIONAL EXPERIMENT RESULTS}\label{app:add_exps}

\subsection{Synthetic Gaussian Case}
We provide details of how we compute the conditional distribution $p_{W|T}(.)$, inverse variance weighting and additional experimental results

\textbf{Calculating $p_{W|T}$.} 
 We recall the setup we are following. Assume that the source $V{\sim} \mathcal{N}(0,\sigma_V^2{=}1.0)$ and the side information $T = V + \zeta_{T|V}$ where $\zeta_{T|V} {\sim} \mathcal{N}(0, 0.01)$, i.e $p_{T|V}(.|v){=}\mathcal{N}(v, \sigma_{T|V}^2= 0.01)$. Furthermore, the encoder and decoder have access to the shared randomness $(S_i, Y_i,\ell_i)^N_i$ as described previously. The decoder must ideally output $W \sim p_{W|V}$, where $p_{W|V}(\cdot|v) {=}{\mathcal N}(v, \sigma_{W|V}^2)$. We start with the joint distribution of $V$ and $T$, which can be expressed as:
 \begin{equation*}
     \begin{pmatrix}
  V   
\\ 
T
\end{pmatrix} \sim \mathcal{N} \left(\begin{bmatrix}
0\\ 0
\end{bmatrix}, \begin{bmatrix}
\sigma^2_V & \sigma^2_V\\ 
\sigma^2_V & \sigma^2_T
\end{bmatrix}\right),
 \end{equation*}
 where $\sigma^2_T = \sigma^2_V + \sigma^2_{T|V}$. Following this, we have the conditional probability of $V$ given $T$ as:
 \begin{equation*}
     p_{V|T}(.|T=t) = \mathcal{N}\left( \frac{\sigma^2_V}{\sigma^2_T}t, \left( 1- \frac{\sigma^2_V}{\sigma^2_T}\right) \sigma^2_V\right), 
 \end{equation*}
 Using the Markov chain $T-V-W$, we have:
 \begin{align*}
     p_{W|T}(w|t)&=\int_{-\infty}^{\infty} p_{W|V}(w|v ,t )p_{V|T}(v|t) dv \\
     &=\int_{-\infty}^{\infty}  p_{W|V}(w|v )p_{V|T}(v|t) dv
 \end{align*}
As $p_{W|V}(.)$ and $p_{V|T}(.)$ are two Gaussians with fix variance, we obtain: 
\begin{equation*}
    p_{W|T}(.|t) = \mathcal{N}\left( \frac{\sigma^2_V}{\sigma^2_T}t,  \sigma^2_W- \frac{\sigma^4_V}{\sigma^2_T}\right)
\end{equation*}
where $\sigma^2_W = \sigma^2_V + \sigma^2_{W|V}$. We then use this quantity to compute the decoder index as explained in the main paper.
 
\begin{figure}[h]%
    \centering%
        \includegraphics[width=0.4\linewidth]{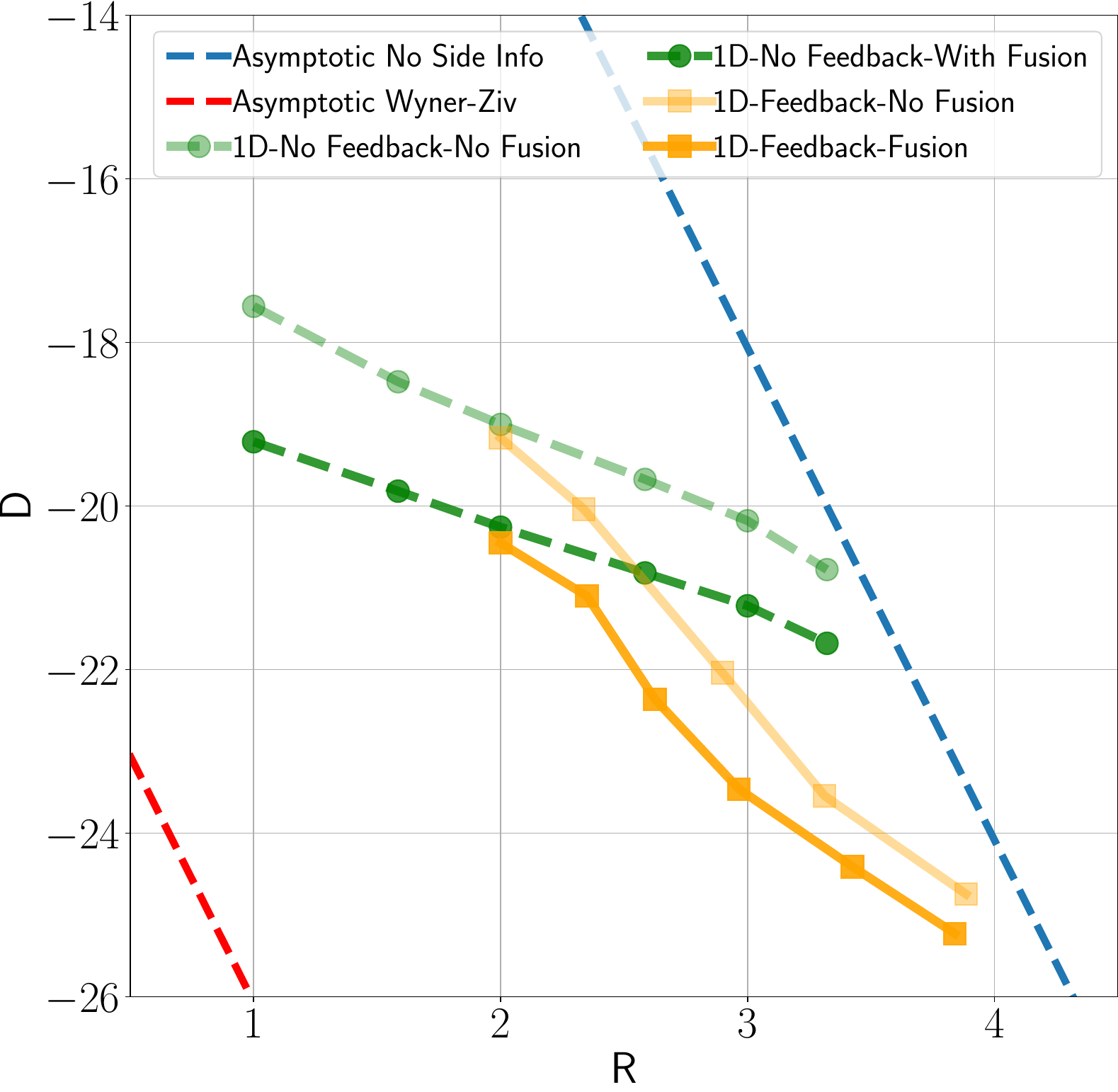}%
        \caption{Effects of inverse variance weighting (fusion) on improving the estimator accuracy.}%
    \label{fig:fusion}%
\end{figure}

\textbf{Inverse Variance Weighting.} We combine the decoder output $W_{U_q}$ with the side information $T \sim p_{T|V}$ to obtain a lower variance estimator $\hat{V}$ of $V$ by applying the inverse variance weighting fusion method proposed by \citep{graybill1959combining}, which we show its effectiveness in Figure \ref{fig:fusion}. We note that in the case without feedback, the decoder's output $W_{U_q}$ might not closely follow the target Gaussian distribution due to mismatching error, and applying inverse variance weighting might yield suboptimal results, i.e. its distortion is higher than that of using side-information alone, which is also demonstrated in Figure \ref{fig:fusion}. As such, in the main paper,  when this situation happens, we simply ignore the information from the source and only consider the side information for reconstruction. In the case where feedback is used, our inverse variance weighting estimator performs consistently well since $W_{U_q}$ now follows more closely to the target Gaussian distribution. 

\textbf{Simulation Parameters.} For all the experiments where we only compress 1 sample, we set $N=2^{15}$ and use grid-search on $L\in \{2, 4, 6, 8,10\}$, $\sigma^2_{W|V} \in \{0.01, 0.008, 0.006, 0.005, 0.003, 0.002, 0.001\}$. For the 5D case, we set $N=2^{27}$, use grid-search on $L\in \{2^{5}, 2^{6}, 2^{7}, 2^{8}, 2^{9}, 2^{10}, 2^{11}, 2^{12}, 2^{13}, 2^{14}\}$. We note that if the encoder detects a mismatch between the two indices, it can send either the rest or a part of the MSB of its index, which we refer to as $L_2$. For reference, we provide some optimal parameters we found in Table ~\ref{tab:rd_params}. Results in the main paper are averaged over 10 runs. 
\begin{table}[]
    \centering
    \caption{Rate-Distortion Parameters for 5D Gaussian Compression with Perfect Feedback.}
   \begin{tabular}{|c |c |c |c |c|c|} 
 \hline
 Dimension & $L$ & $L_2$ & $\sigma^2_{W|V}$ & Rate & Distortion (dB)\\ [0.5ex] 
 \hline
 1 & $2$ & $3$ & 0.01 & 2.133 & -20.61\\
 \hline
 1 & $2$ & $6$ &  0.008& 2.35 & -21.10\\  
 \hline
 1 & $2$ & $8$ & 0.005& 2.625 & -22.36\\  
 \hline
 1 & $2$ & $12$ & 0.003& 2.966 & -23.46\\  
 \hline
 1 & $2$ & $16$ & 0.001& 3.425 & -24.41\\  
 \hline
 5 & $2^5$ & $2^2$ & 0.01 & 1.18 & -21.78 \\
 \hline
 5 & $2^8$ & $2^{5}$ & 0.008& 1.33 & -22.33\\  
 \hline
 5 & $2^5$ & $2^{14}$ & 0.005& 1.88 & -24.49\\  
 \hline
 5 & $2^{12}$ & $2^{8}$ & 0.003& 2.65 &  -27.04 \\  
 \hline
 5 & $2^{14}$ & $2^{6}$ & 0.001& 3.06 &  -28.84\\
 \hline 
\end{tabular}

    \label{tab:rd_params}
\end{table}

\textbf{Additional Results.} 

\begin{figure}[h]%
    \centering%
        \includegraphics[width=0.4\linewidth]{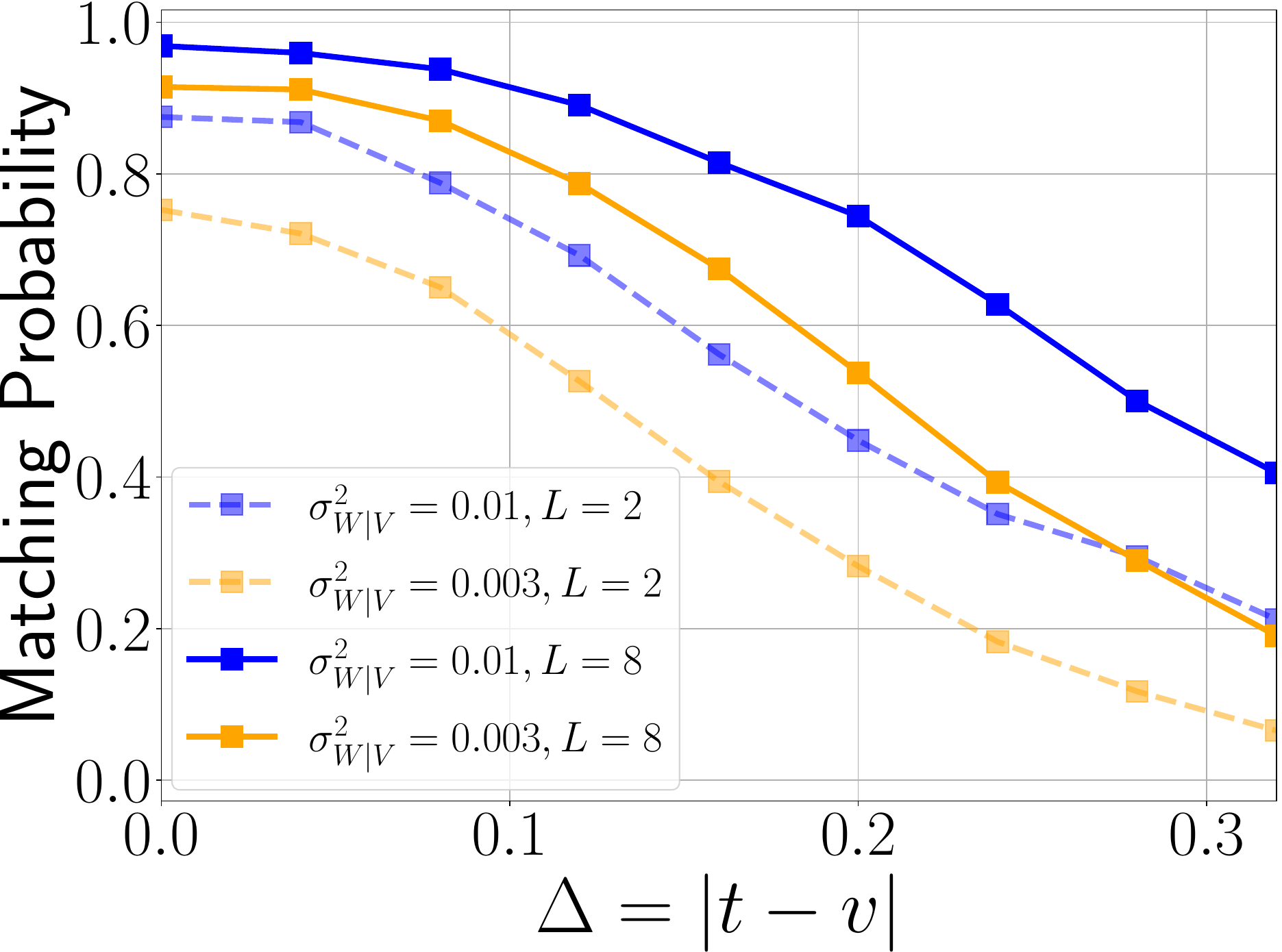}%
        
        \caption{Matching Probability w.r.t side information quality. We use the distance $\Delta=|t-v|$ to quantify the side information quality, lower $\Delta$ correspond to better quality.}%
    \label{fig:match_delta}%
\end{figure}

\emph{Matching Probability and Side Information Quality. }We show in Figure \ref{fig:match_delta} the matching probability as a function of side-information quality $\Delta$ for different $\sigma^2_{W|V}$ and $L$, where we measure the side-information quality by its associated distance to the source, i.e $\Delta=|t-v|$. To obtain the matching probability, for each $\Delta$, we sample $V\sim p_V(.)$, send the side information $T=V \pm \Delta$  to the decoder and compare $U_q$ and $U_p$. This process is simulated and averaged over 1000 runs to obtain the matching probability. We observe that the matching probability decrease with $\Delta$ and increasing $\sigma^2_{W|V}$ and $L$ consistently improves the matching rate for all $\Delta$. Finally, we note that the matching probability is not 1.0 for $\Delta=0$ due to the distribution mismatch between $p_{W|V}(.|v)$ and $p_{W|T}(.|t)$.

\emph{On Feedback Error. } We provide the feedback error (collision) due to hashing during the feedback step in Table \ref{tab:fb_error}. Specifically, we define feedback error as the probability that the encoder index and decoder index are different but have the same hash value. Also, when their hash values are different, the two indices must be different. In general, we observe that this error depends on multiple factors, such as the number of samples we are compressing, the matching probability in the first round (which depends on $L$), the number of proposals, and $\sigma^2_{W|V}$. As a simple illustrative example, consider the case when we compress 5 samples jointly with the number of proposals $N=2^{25}$, then using $4$ bits feedback is sufficient for  the encoder to obtain the exact position when $L=2^{21}$ but it is not the case when $L=2^{10}$. Overall, we found that we can adjust the parameters to obtain low feedback error in most of the cases.

\begin{table}[]
    \centering
    \caption{Feedback error with different parameters. The first column represents the feedback rate for the encoder to recover the full index. The second column represents the (hashed) non-ideal feedback rate. We measure feedback rate by bits.}
   \begin{tabular}{|c|c |c |c | c|c |c|} 
 \hline
  Feedback Rate (Ideal) & Feedback Rate (non-ideal) & Dimension & $L$ & N & $\sigma^2_{W|V}$ & Error Probability \\ [0.5ex] 
 \hline
 11 & 1 & 1 & $2^1$ & $2^{12}$ & 0.01 & 13.1\%\\ 
 \hline
 11 & 1 & 1 & $2^1$ & $2^{12}$ & 0.008 & 16.1\%\\ 
 \hline
 9 & 1 & 1 & $2^3$ & $2^{12}$ & 0.004 & 7.21\%\\ 
 \hline
 22 & 1 & 5 & $2^5$ & $2^{27}$ & 0.01 & 12.36\%\\ 
 \hline
 20 & 1 & 5 & $2^7$ & $2^{27}$ & 0.01 & 4.14\%\\ 
 \hline
 19 & 1 & 5 & $2^{8}$ & $2^{27}$ & 0.005 & 4.88\%\\
 \hline
 17 & 1 & 5 & $2^{10}$ & $2^{27}$ & 0.005 & 2.48\%\\
 \hline
 13 & 1 & 5 & $2^{14}$ & $2^{27}$& 0.001 & 1.39\%\\
 \hline
 22 & 5 & 5 & $2^5$ & $2^{27}$ & 0.01 & 1.09\%\\
 \hline
 17 & 5 & 5 & $2^{10}$ & $2^{27}$ & 0.005& 0.09\%\\  
 \hline
 13 & 5 & 5 & $2^{14}$ & $2^{27}$ & 0.001& 0.03\%\\  
 \hline
\end{tabular}

    \label{tab:fb_error}
\end{table}

\subsection{Distributed Image Compression}
We provide details on the network architecture and additional results for the distributed lossy image compression experiment on MNIST dataset \citep{lecun1998gradient}. 
\subsubsection{Network Architecture}
\textbf{Encoder-Decoder Network.} We show the architecture our $\beta-$VAE network, including the encoder network $f_e(v)$, the projection network $h(t)$, and the decoder $g(w,t)$ in the Table.\ref{mnist compressor}. Convolutional and transposed convolutional layers are denoted as ``conv'' and ``upconv'' respectively, which are accompanied by a number of filters, kernel size, stride, and padding. For ``upconv'', we have an additional parameter which is the output padding at the end. The encoder network maps an image into 2 vectors of size $4$ (total $8$D output), where the first vector represents the output mean $f_e(v)^{(1)}$ and the second vector $f_e(v)^{(2)}$ represents the output variance. Specifically, we define $p_{W|V}(.) = \mathcal{N}(f_e(v)^{(1)}, f_e(v)^{(2)})$ and use the prior distribution $p_W(.)=\mathcal{N}(0,1)$. 

At the decoder side, a projection network $h(t)$ first maps the side information image $T$ to a vector of size $128$, which is then combined with a vector of size $4$ from the encoder, resulting in a $132$D vector. This $132D$ vector is then fed into a decoder network $g(w,t)$ that outputs a reconstruction of size $28\times28$, which we denote as $\hat{V}$. 

\textbf{Loss Function} We train  our $\beta-$VAE network by optimize the following rate-distortion loss:
\begin{equation}
    \mathcal{L} = \beta (V - \hat{V})^2  -  E_V[D_{\text{KL}}(p_{W|V}(.|v)|| p_W(.))]
\end{equation}
where we vary $\beta$ for different rate-distortion tradeoff. We train each model for $30$ epochs on an NVIDIA-RTX A4500, which takes 30 minutes per model.

\begin{table}[h]
    \centering
    \caption{Encoder, project network, and  decoder for MNIST distributed image compression.} 
    \begin{tabular}[t]{ |c| }
    \multicolumn{1}{c}{\textbf{(a)}Encoder $f_e(v)$} \\
	\hline 
        Input $28\times28\times1$  \\ 
        \hline 
        conv (128:3:1:1), ReLU \\
        \hline 
        conv (128:3:2:1), ReLU \\
        \hline 
        conv (128:3:2:1), ReLU \\
        \hline 
        Flatten \\
        \hline 
        Linear (6272, 512), ReLU\\
        \hline 
        Linear (512, 8)\\
    \hline
    \end{tabular}
    \quad
    \begin{tabular}[t]{ |c| }
    \multicolumn{1}{c}{\textbf{(b)}Projection Network $h(t)$} \\
	\hline 
        Input $14\times14\times1$  \\ 
        \hline 
        conv (32:3:1:1), ReLU \\
        \hline 
        conv (64:3:2:1), ReLU \\
        \hline 
        conv (128:3:2:1), ReLU \\
        \hline 
        Flatten \\
        \hline 
        Linear (2048, 512), ReLU\\
        \hline 
        Linear (512, 128)\\
    \hline
    \end{tabular}
    \quad
    \begin{tabular}[t]{ |c| }
    \multicolumn{1}{c}{\textbf{(c)}Decoder Network $g(w,t)$} \\
	\hline 
        Input-($4 {+} 128$)  \\ 
        \hline 
        Linear-($132, 512$), ReLU  \\ 
        \hline 
        upconv (64:3:2:1:1), ReLU \\
        \hline 
        upconv (32:3:2:1:1), ReLU \\
        \hline
        upconv (1:3:1:1), Tanh \\
    \hline
    \end{tabular}
    
    \label{mnist compressor}
\end{table}
\textbf{Neural Estimator Network.} The neural estimator network in this case consists of two subnetworks. The first subnetwork will project the side-information into an embedding of size $128$ and the second subnetwork combines that $128$D embedding with the 4D embedding, either from $p_{W|V}$ or $p_W$, and outputs the probability of whether $T,W$ are from the joint or from the marginal distributions. We note that the projection network architecture is the same as the one in our $\beta$-VAE network. Finally, we note that this model is trained with 100 epochs. 

\begin{table}[h]
    \centering
    \caption{Neural Estimator Networks for Distributed Image Compression.} 
    \begin{tabular}[t]{ |c| }
    \multicolumn{1}{c}{\textbf{(a)}Projection Network} \\
	\hline 
        Input $14\times14\times1$  \\ 
        \hline 
        conv (32:3:1:1), ReLU \\
        \hline 
        conv (64:3:2:1), ReLU \\
        \hline 
        conv (128:3:2:1), ReLU \\
        \hline 
        Flatten \\
        \hline 
        Linear (2048, 512), ReLU\\
        \hline 
        Linear (512, 128)\\
    \hline
    \end{tabular}
    \begin{tabular}[t]{ |c| }
    \multicolumn{1}{c}{\textbf{(b)} Combine and Classify } \\
	\hline 
        Input $128 + 4$  \\ 
        \hline 
        Linear (132, 128), l-ReLU \\
        \hline 
        Linear (128,128), l-ReLU \\
        \hline 
        Linear (128,128), l-ReLU \\
        \hline 
        Linear (128, 1)\\
    \hline
    \end{tabular}
    \label{mnist nce}
\end{table}

\textbf{Simulation Parameters.} To train our network, we varies $\beta \in \{0.15, 0.25, 0.35,  0.45,0.55, 0.65, 0.75, 0.85, 0.95\}$. For one sample compression, we perform grid search on $N \in \{2^7, 2^8, 2^9, 2^{10}, 2^{11}\}$, $L \in \{2^4, 2^5, 2^6, 2^7, 2^8\}$. For two sample compression, we use $N \in \{2^{20}, 2^{21}, 2^{22}, 2^{23}, 2^{24}, 2^{25}\}$ and $L \in \{2^{15}, 2^{16}, 2^{17}, 2^{18}, 2^{19}, 2^{20}\}$. In both cases, we send the full MSB index in the second transmission. We provide some optimal values shown in Table \ref{tab:rd_params_mnist}. Results in the main paper are averaged over 10 runs. 

\begin{table}[]
    \centering
    \caption{Rate-Distortion Parameters for MNIST Compression with Feedback.}
   \begin{tabular}{|c |c |c |c |c|c|} 
 \hline
 Number of Samples & $L$ & $N$ & $\beta$ & Rate & Distortion (MSE)\\ [0.5ex] 
 \hline
 1 & $2^{10}$ & $2^{15}$ & 0.95 & 11.96 & 0.0488\\
 \hline
 1 & $2^{7}$ & $2^{12}$ &  0.75& 8.64& 0.0566\\  
 \hline
 1 & $2^{3}$ & $2^{8}$ & 0.25& 6.865 & 0.0635\\  
 \hline
 2 & $2^{20}$ & $2^{25}$ & 0.95 & 11.01 & 0.0489 \\
 \hline
 2 & $2^{15}$ & $2^{20}$ & 0.75& 7.79 & 0.0565\\  
 \hline
 2 & $2^{10}$ & $2^{15}$ & 0.35& 6.2 & 0.0618\\  
 \hline 
\end{tabular}
    \label{tab:rd_params_mnist}
\end{table}

\subsubsection{Additional Examples}
We provide additional examples where the decoder outputs correct/incorrect reconstructions during the first transmission. This again confirms that our neural estimator selects a semantically meaningful message from the encoder's output.
\begin{figure}[h]%
    \centering%
        \includegraphics[width=1.0\linewidth]{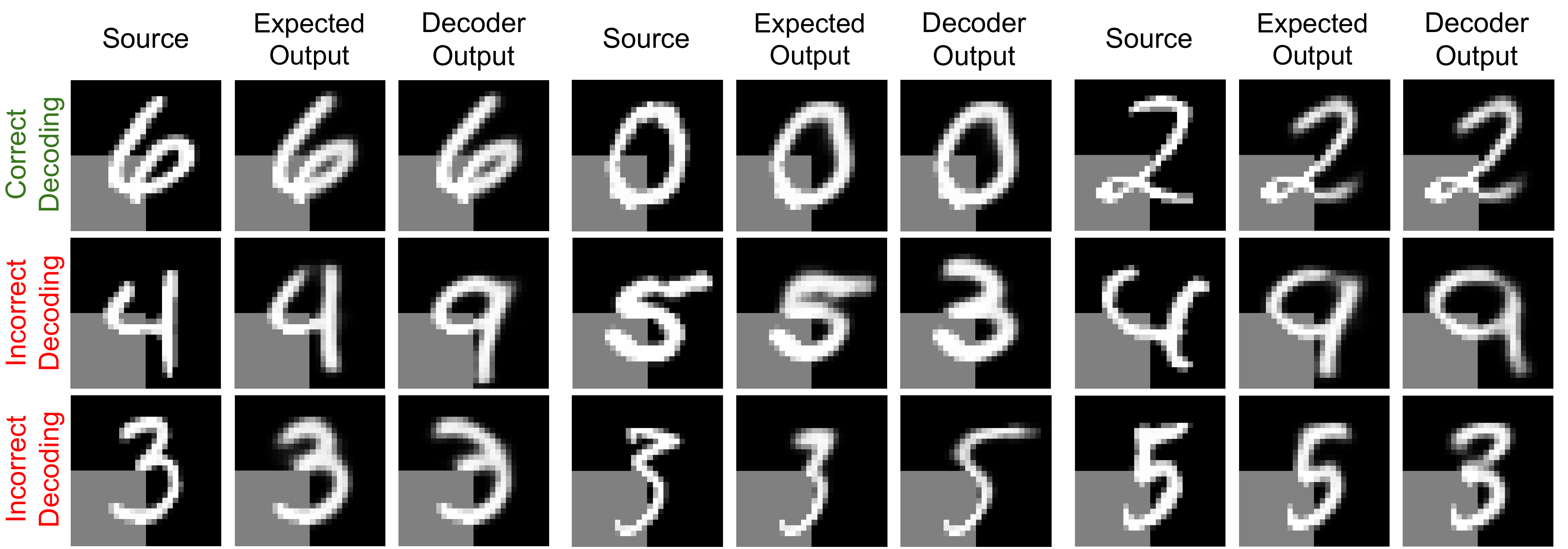}%
        
        \caption{Distributed Image Compression without Feedback. Additional Results.}%
    \label{fig:extra_mnist}%
\end{figure}

\subsubsection{Compression with limited Feedback}

\begin{figure}[t]%
    \centering%
        \includegraphics[width=0.4\linewidth]{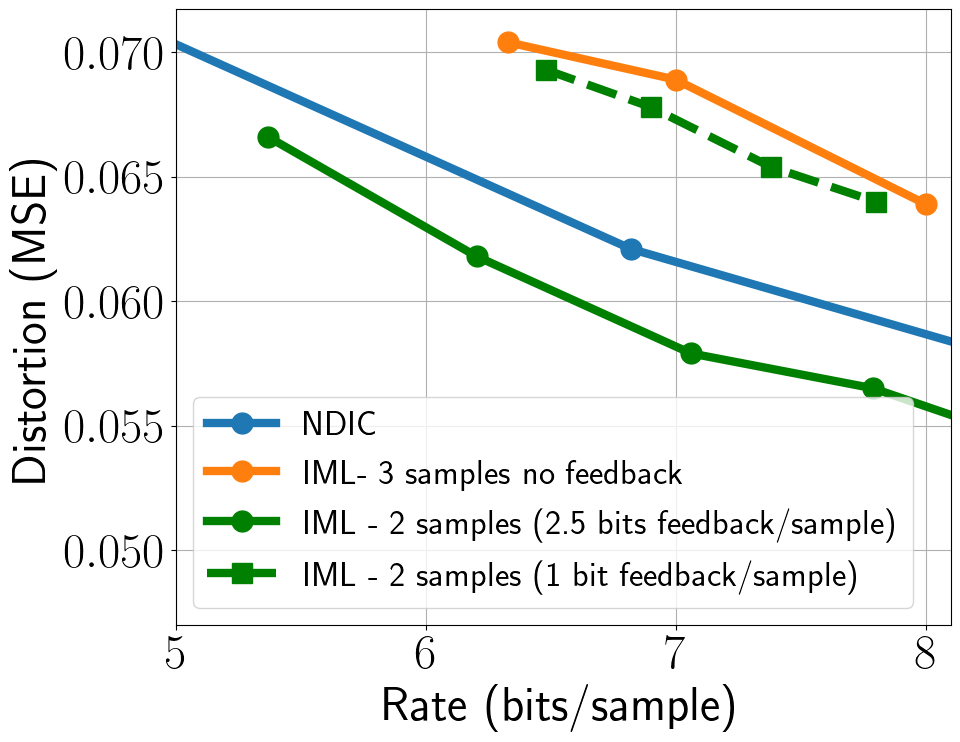}%
        
        \caption{MNIST Distributed Compression with Different Feedback Rates. Here, $2.5$ bits feedback/sample is sufficient to recover the index when two samples are jointly compressed.}%
    \label{fig:mnist_no_fb}%
\end{figure}

We provide a new experiment R-D plot for our feedback-free scheme (compressing $3$ samples together) for the MNIST experiment in Sec.~\ref{sec:mnist_exp} in  Fig.~\ref{fig:mnist_no_fb}. We also include the case where the feedback signal is imperfect. Although slightly worse than the NDIC baseline,  the latter requires engineering neural networks, involving complex loss function with several hyper-parameters.

\subsection{Vertical Federated Learning - CIFAR 10}
We provide details on network architecture and simulation parameters in the vertical federated learning experiment with CIFAR-10 \citep{krizhevsky2009learning}.

\begin{figure}[h]%
    \centering%
        \includegraphics[width=0.2\linewidth]{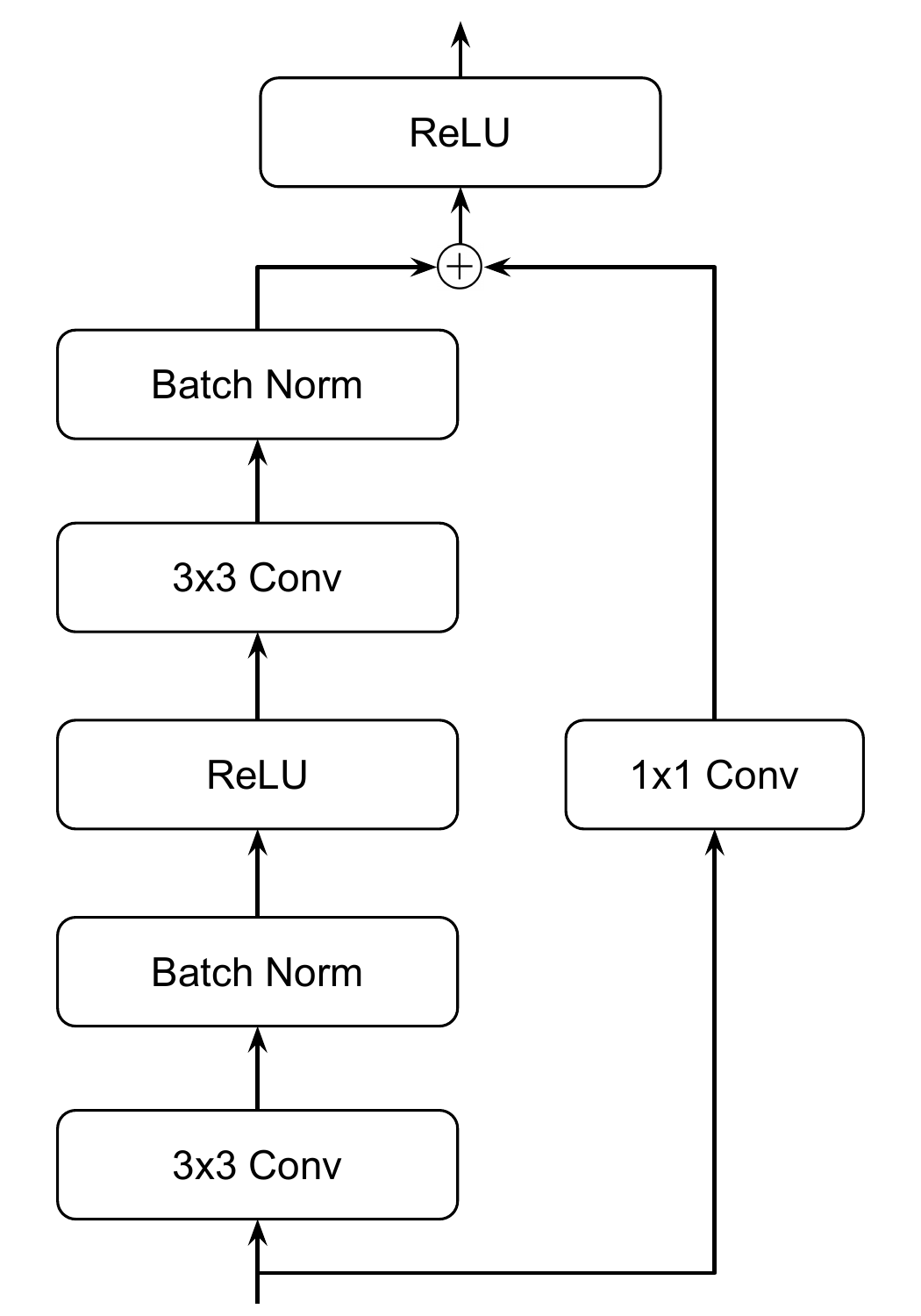}%
        
        \caption{Residual Block. In our description, each ``residual block'' is described by number of filters and stride of the 3x3 convolution operator, which we set the value of padding to 1. }%
    \label{fig:res_block}%
\end{figure}

\textbf{Network Architecture}
We present our networks, including the model at each party, the server model, and the neural estimator module in the Table \ref{cifar10-model}. We use the ``residual block'' which is shown in Figure ~\ref{fig:res_block}. Each party model in this case will project its quadrant to a 4D embedding and send them to the server model, which will output the prediction. We train the model for $100$ epochs on an NVIDIA-RTX A4500, which converges after 2 hours training.

\textbf{Loss Function. } We train our network end-to-end with a standard cross-entropy loss. We augment the dataset by applying horizontal flip and random cropping to the original image, before cropping the two quadrants (bottom-left and top-right), that would be then distributed to both parties. 

\begin{table}[h]
    \centering
    \caption{Party Model, Server Model, and Neural Estimator.} 
    \begin{tabular}[t]{ |c| }
    \multicolumn{1}{c}{\textbf{(a)}Party Model} \\
	\hline 
        Input  \\ 
        \hline 
        conv (64:3:1:1), BatchNorm2D, l-ReLU \\
        \hline 
        residual block (64:1) \\
        \hline 
        residual block (128:2) \\
        \hline 
        residual block (256:2) \\
        \hline 
        residual block (512:2) \\
        \hline
        Linear (2048, 4) \\
    \hline
    \end{tabular}
    \quad
    \begin{tabular}[t]{ |c| }
    \multicolumn{1}{c}{\textbf{(b)}Server Module} \\
	\hline 
        Input $4 \times 2$\\ 
        \hline 
        Linear (8, 128)\\
        \hline 
        Linear (128, 10)\\
    \hline
    \end{tabular}
    \quad
    \begin{tabular}[t]{ |c| }
    \multicolumn{1}{c}{\textbf{(c)}Neural Estimator} \\
	\hline 
        Input-($4 \times 2$)  \\ 
        \hline 
        Linear (8, 128), l-ReLU \\
        \hline 
        Linear (128, 128), l-ReLU \\
        \hline
        Linear (128, 128), l-ReLU \\
        \hline 
        Linear (128, 128), l-ReLU \\
        \hline 
       Linear (128, 1)
        \\
    \hline
    \end{tabular}
    \label{cifar10-model}
\end{table}

\textbf{Simulation Parameters. } We perform grid-search and show the optimal parameters in Table \ref{tab:cifar_param}. Results in the main paper are averaged over 10 runs. 

\begin{table}[h]
    \centering
    \caption{Parameters for C-VFL experiments with CIFAR-10.}
   \begin{tabular}{|c |c |c |c|c|} 
 \hline
 $L$ & $N$ & $\sigma^2_{W|V}$ & Rate & Accuracy\\ [0.5ex] 
 \hline
   $2^3$& $2^6$ & 0.07 & 4.96 & 0.764\\
 \hline
 $2^4$  & $2^8$ &  0.06 & 6.5 & 0.789\\
 \hline
  $2^5$ & $2^9$ & 0.04 & 7.4 & 0.798\\
 \hline
  $2^{9}$ & $2^{13}$ & 0.005 & 11.41 & 0.811\\
 \hline
\end{tabular}
    \label{tab:cifar_param}
\end{table}

\subsection{Breast Cancer Dataset}
The parameters for IML is shown in Table \ref{tab:extra_vfl}. We note that in this experiment, there is no neural network at the encoder side and we aim to lossily transmit the features to the decoder.

\begin{table}[h]
    \centering
    \caption{Parameters for C-VFL experiments with Breast Cancer Dataset (Our method).}
   \begin{tabular}{|c |c |c |c|c|} 
 \hline
 $L$ & $N$ & $\sigma^2_{W|V}$ & Rate & Accuracy\\ [0.5ex] 
 \hline
   $2^3$& $2^6$ & 0.07 & 4.8 & 0.9\\
 \hline
 $2^5$  & $2^{9}$ &  0.04 & 7.3 & 0.93\\
 \hline
  $2^6$ & $2^{10}$ & 0.01 & 8.6 & 0.95\\
 \hline
 $2^9$ & $2^{12}$ & 0.005 & 10.48 & 0.97\\
 \hline
\end{tabular}
    \label{tab:extra_vfl}
\end{table}




\end{document}